\documentclass{article}
\usepackage{fullpage}

\usepackage[utf8]{inputenc} % allow utf-8 input
\usepackage[T1]{fontenc}    % use 8-bit T1 fonts
\usepackage{hyperref}       % hyperlinks
\usepackage{url}            % simple URL typesetting
\usepackage{booktabs}       % professional-quality tables
\usepackage{amsfonts}       % blackboard math symbols
\usepackage{nicefrac}       % compact symbols for 1/2, etc.
\usepackage{microtype}      % microtypography

\usepackage{graphicx}
\usepackage{enumitem}
\usepackage{subfig}
\usepackage{amsmath}
\usepackage{amssymb}
\usepackage{amsthm}
\usepackage{xcolor}
\usepackage{algorithm}
\usepackage[noend]{algpseudocode}
\usepackage{xpatch}
\usepackage{natbib}

\newcommand{\numpap}{n}
\newcommand{\numrev}{m}
\newcommand{\numcat}{k}
\newcommand{\papload}{\lambda}
\newcommand{\revload}{\mu}
\newcommand{\ind}{w}
\newcommand{\hypothesis}{H}
\newcommand{\pacc}{\pi}
\newcommand{\sbpacc}{\pacc^{(\text{sb})}}
\newcommand{\dbpacc}{\pacc^{(\text{db})}}
\newcommand{\commonmatrix}{\Pi}
\newcommand{\sbmatrix}{\commonmatrix^{\text{sb}}}
\newcommand{\dbmatrix}{\Pi^{\text{db}}}

\newcommand{\sbdecision}{Y}
\newcommand{\dbdecision}{X}
\newcommand{\mean}{\pi}
\newcommand{\coef}{\beta}
\newcommand{\dbscore}{\widetilde{q}}
\newcommand{\tscore}{q^{\ast}}
\newcommand{\estcoef}{\widehat{\coef}}

\newcommand{\level}{\alpha}
\newcommand{\counting}{{\sc Counting}}
\newcommand{\permutation}{{\sc Disagreement}}
\newcommand{\indset}{\mathcal{J}}
\newcommand{\noindset}{\overline{\indset}}

\newcommand{\tomkins}{\citet{tomkins17wsdm}}
\newcommand{\shorttomkins}{Tomkins et al.}

\newcommand{\shift}{\nu}
\newcommand{\logshift}{\widetilde{\nu}}
\newcommand{\chng}{h_{\shift}}
\newcommand{\logchng}{{\textsl{g}}_{\logshift}}

\newcommand{\class}{\mathcal{F}}
\newcommand{\papset}{\mathcal{P}}
\newcommand{\sbpapset}{\mathcal{P}_{\text{SB}}}
\newcommand{\dbpapset}{\mathcal{P}_{\text{DB}}}
\newcommand{\nullshift}{f_0}

\newcommand{\commentedforneurips}[1]{}
\newcommand{\commentedforarxiv}[1]{}
\newcommand{\subclass}{\class}

\newcommand{\logregclass}{\widetilde{\class}_{\bound}}

\newcommand{\statp}{\tau}
\newcommand{\statc}{\gamma}
\newcommand{\indicator}[1]{\mathbb{I} \left[ #1 \right]}
\newcommand{\arraymaj}{U}
\newcommand{\arraymin}{V}
\newcommand{\arraymajentry}{u}
\newcommand{\arrayminentry}{v}

\newcommand{\expectation}[1]{\mathbb{E}\left[ #1 \right]}

\newcommand{\prob}[1]{\mathbb{P}\left[ #1 \right]}

\newcommand{\paprepresentation}{q}
\newcommand{\revrepresentation}{r}

\newcommand{\minigap}{\delta}
\newcommand{\procedure}{\psi}
\newcommand{\gap}{\Delta}

\newcommand{\linmodel}{\subclass_{\gap}}
\newcommand{\logmodel}{\widetilde{\subclass}_{\bound}}

\newcommand{\isarxiv}[1]{}

\newcommand{\revset}{\mathcal{R}}
\newcommand{\sbrevset}{\revset_{\text{SB}}}

\newcommand{\err}{\varepsilon}
\newcommand{\sample}{\numpap_0}
\newcommand{\sbrev}{\mathcal{R}_{\text{SB}}}
\newcommand{\dbrev}{\mathcal{R}_{\text{DB}}}
\newcommand{\reals}{\mathbb{R}}
\newcommand{\sbcoef}{\coef^{\text{(sb)}}}
\newcommand{\dbcoef}{\coef^{\text{(db)}}}

\newcommand{\simmatrix}{S}
\newcommand{\threshold}{z}

\newcommand{\noise}{\sigma}

\newcommand{\bound}{\widetilde{\Delta}}
\newcommand{\lb}{\ell}
\newcommand{\ub}{{b}}
\newcommand{\lrinc}{d_2}
\newcommand{\upinc}{d_1}

\newcommand{\constant}{c}
\newcommand{\constanta}{c}

\newcommand{\settuple}{\mathcal{T}}
\newcommand{\tuple}{t}

\newcommand{\algo}{\mathcal{A}}
\newcommand{\assignment}{A}
\newcommand{\testassignment}{A^{\ast}}

\newtheorem{theorem}{Theorem}
\newtheorem{lemma}{Lemma}

\newtheorem{corollary}{Corollary}
\newtheorem{proposition}{Proposition}
{
\theoremstyle{definition}
\newtheorem{assumption}{Assumption}
\newtheorem{problem}{Problem}

\newtheorem*{remark*}{Remark}

}

\newtheorem{manualprobleminner}{Problem}
\newenvironment{manualproblem}[1]{%
  \renewcommand\themanualprobleminner{#1}%
  \manualprobleminner
}{\endmanualprobleminner}

\graphicspath{{Figures/}}

\title{On Testing for Biases in Peer Review}

\author{\\
  Ivan Stelmakh, Nihar B. Shah and Aarti Singh\\~\\
  School of Computer Science \\ 
  Carnegie Mellon University\\
  \texttt{\{stiv,nihars,aarti\}@cs.cmu.edu} 
}
\date{}

\begin{document}

\maketitle

\begin{abstract}
We consider the issue of biases in scholarly research, specifically, in peer review. There is a long standing debate on whether exposing author identities to reviewers induces biases against certain groups, and our focus is on designing tests to detect the presence of such biases. Our starting point is a remarkable recent work by Tomkins, Zhang and Heavlin which conducted a controlled, large-scale experiment to investigate existence of biases in the peer reviewing of the WSDM conference. We present two sets of results in this paper. The first set of results is negative, and pertains to the statistical tests and the experimental setup used in the work of Tomkins et al. We show that the test employed therein does not guarantee control over false alarm probability and under correlations between relevant variables coupled with any of the following conditions, with high probability, can declare a presence of bias when it is in fact absent: (a) measurement error, (b) model mismatch, (c) reviewer calibration. Moreover, we show that the setup of their experiment may itself inflate false alarm probability if (d) bidding is performed in non-blind manner or (e) popular reviewer assignment procedure is employed.  Our second set of results is positive and is built around a novel approach to testing for biases that we propose. We present a general framework for testing for biases in (single vs. double blind) peer review. We then design hypothesis tests that under minimal assumptions guarantee control over false alarm probability and non-trivial power even under conditions (a)--(c) as well as propose an alternative experimental setup which mitigates issues (d) and (e). Finally, we show that no statistical test can improve over the non-parametric tests we consider in terms of the assumptions required to control for the false alarm probability. %\commentnihar{Our positive results can also be used to conduct semi-randomized controlled trials to test various other effects in peer review.}
 \end{abstract}

\section{Introduction}
\label{section:introduction}

Past research in social sciences indicates that humans display various biases including gender, race and age biases in many critical domains such as hiring~\citep{bertrand2004emily}, university admission~\citep{thornill18black}, bail decisions~\citep{arnold18bail} and many others. Our focus is on fairness in academia and scholarly research, and specifically, on biases in peer review. Peer review is a backbone of scholarly research and is employed by a vast majority of journals and conferences. Due to the widespread prevalence of the Matthew effect -- rich get richer and poor get poorer -- in academia~\citep{thorngate2014numbers, squazzoni2012saint}, any biases in  peer review can have far reaching consequences on career trajectories of researchers.  Specifically, we follow the long-standing debate~\citep[and references therein]{blank91effects, seeber17newcomers, snodgrass06literature, largent16blindpeerreview, okike16sbvsdb, budden08dbfemale, webb08femalebenefit, shawndra03identification} on whether the authors' identities should be hidden from reviewers or not. \emph{
The focus of this paper is on designing statistical tests to detect the presence of biases in peer review.} 

In a recent remarkable piece of work,~\tomkins{} conducted a large scale (semi-) randomized controlled trial during the peer review for the ACM International Conference on Web Search and Data Mining  (WSDM) 2017. In their experiment, the entire pool of reviewers was partitioned uniformly at random into two equal groups -- single blind and double blind -- and each paper was assigned to two reviewers from each of the groups. In this manner, the peer-review data contained both single-blind and double-blind reviews for each paper. The experiment allowed them to conduct a causal inference to test for biases, and conclude that the single-blind system induces a bias in favor of papers authored by (i) researchers from top-universities, (ii) researchers from top companies and (iii) famous authors. Interestingly, no bias against female-authored submissions was detected by their test, though a meta-analysis confirmed the presence of such bias. The conclusions of this experiment have had a significant impact. For instance, the WSDM conference itself completely switched to double-blind peer review starting 2018.

Testing for the presence of hypothesized phenomena is a common task in various branches of science including the biological, social, and physical sciences. The general approach therein is to impose a hard constraint on the probability of false alarm (claiming existence of the phenomenon when there is none; also called Type-I error) to some predefined threshold called significance level typically set as 0.05 or 0.01. The test would then aim to maximize the probability of detecting the phenomenon when it is actually present, while not violating the aforementioned hard  constraint. The present paper also follows this general approach, for the specific setting of testing for biases using single versus double blind reviewing.

\begin{figure}[t]
    \centering
    \subfloat[\label{fig:intro:type1} Bias is absent. Any valid test must have false alarm probability below 0.05. %Presence of correlations; noise in estimates of true scores by double-blind reviewers.
    ]{{\includegraphics[width=4.5cm]{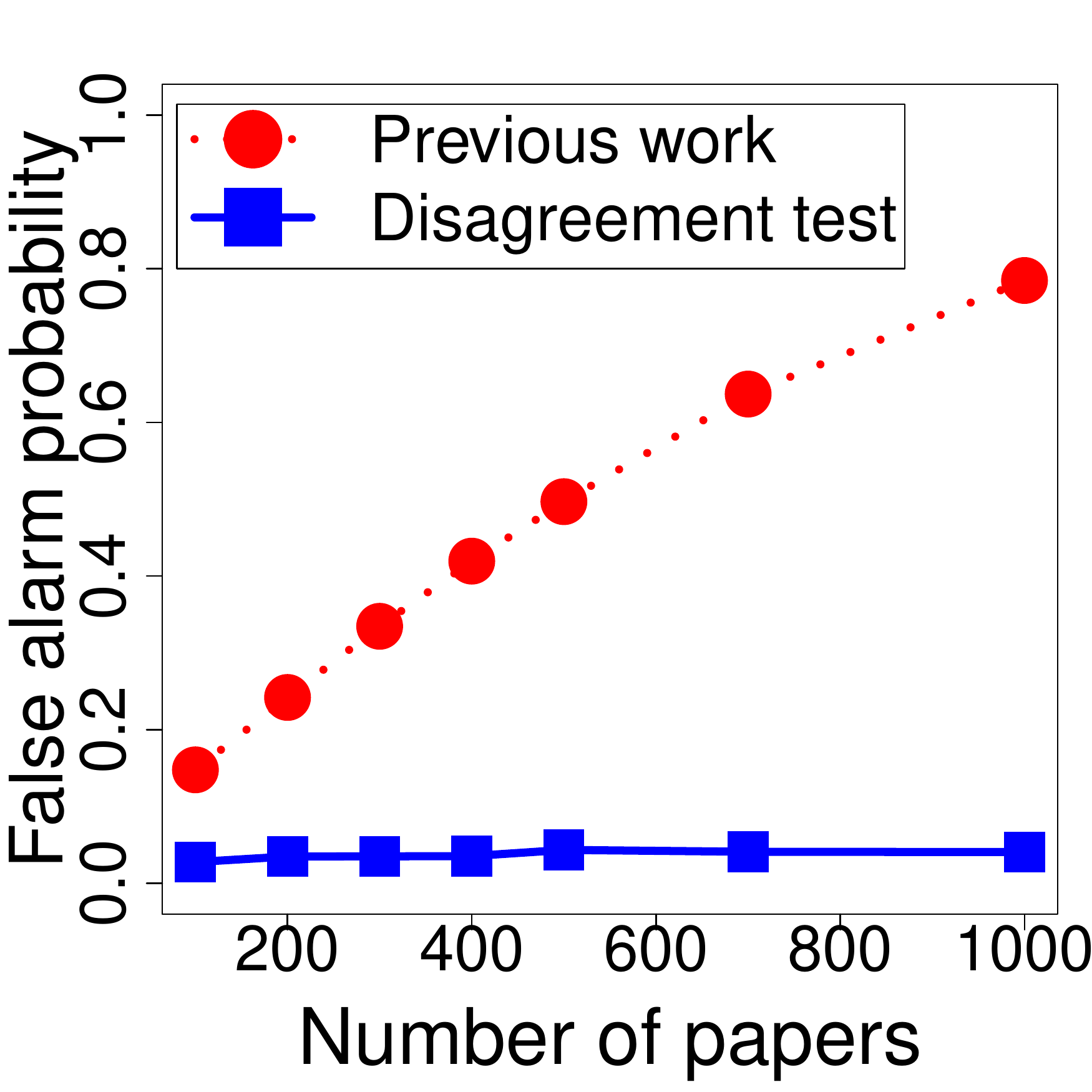}}}%
    \qquad 
    \subfloat[\label{fig:intro:type2} Bias is present. Higher probability of detection is better. %In presence of correlations double-blind reviewers estimate true scores with some noise. 
    ]{{\includegraphics[width=4.5cm]{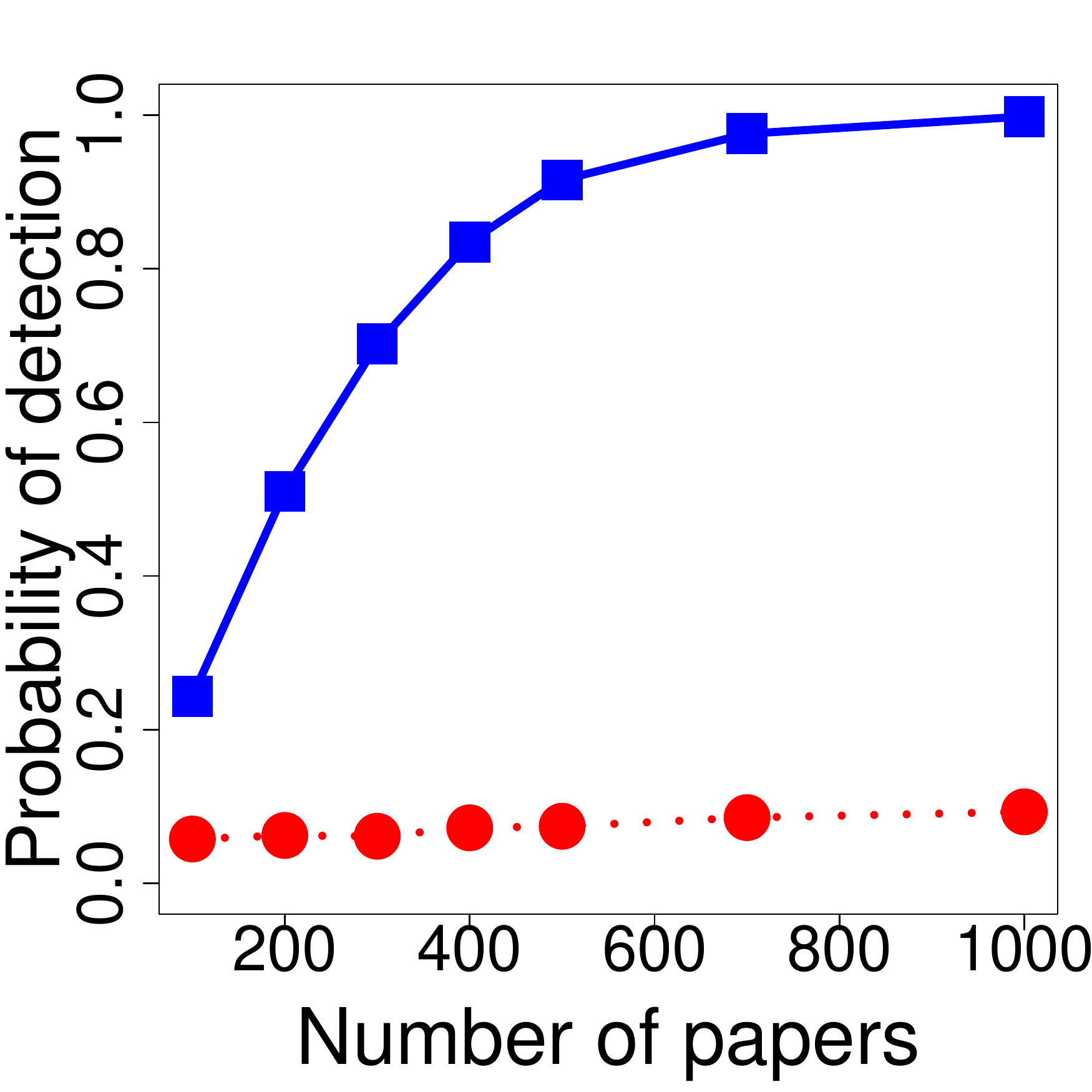}}}%
    \qquad
    \subfloat[\label{fig:intro:easy}Bias is present. Higher probability of detection is better. %There is no correlation or noise in reviewers' evaluations.
    ]{{\includegraphics[width=4.5cm]{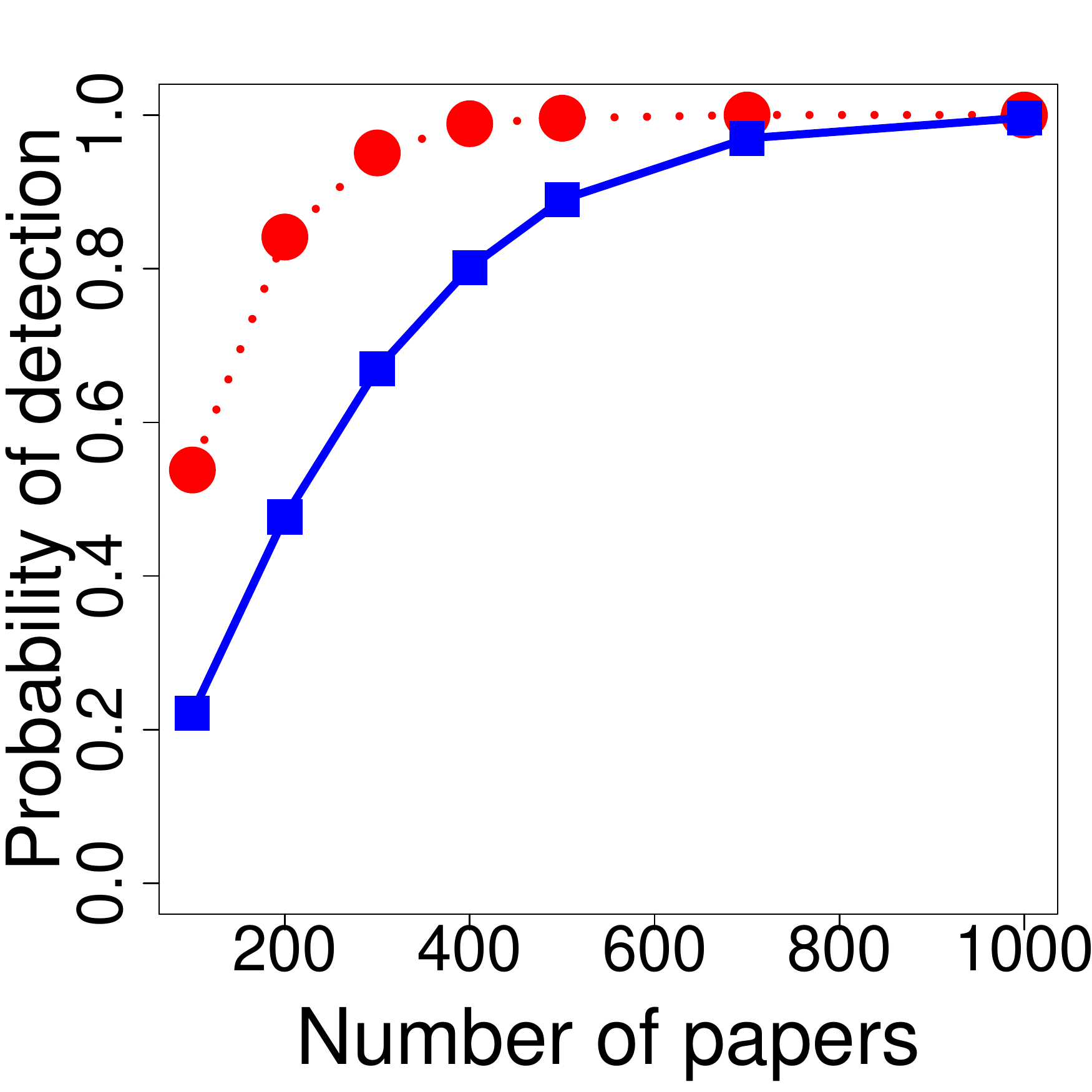}}}%
    \caption{Synthetic simulations evaluating performance of the test in \tomkins{} (``previous work'') and the test proposed in this paper (``\permutation{} test''). Subfigures (a) and (b) are in presence of correlations and noisy estimates of true scores by double-blind reviewers; subfigure (c) has zero correlations and perfect estimate of true scores by double-blind reviewers.    Details of the simulation setup are provided in Section~\ref{section:regression}. The error bars are too small to be visible.}%
    \label{fig:illustration}%
\end{figure}

\paragraph*{Contributions.} In this paper, we study the problem of detecting bias in peer review, and present two sets of results. 

\smallskip

\noindent \textbf{(1) Detailed investigation into methodology of past work} (Section~\ref{section:regression})
We first analyze the testing procedure used by~\tomkins{}, and show that under plausible conditions the statistical test employed therein does not control for false alarm probability. In other words, \emph{we show that under reasonable conditions, the test used by~\tomkins{} can, with probability as large as 0.5 or higher, declare the presence of a bias when the bias is in fact absent} (even when the test is tuned to have a false alarm error rate below 0.05). Specifically, we show that in presence of correlations that are reasonable to expect, any of the following factors breaks their false alarm probability guarantees: (a) measurement error caused by noise or subjectivity of reviewers, (b) model mismatch caused by violation of strong parametric assumptions on reviewers' behavior and (c) reviewer's calibration if she/he reviews more than one paper. Figures~\ref{fig:intro:type1} and~\ref{fig:intro:type2} illustrate the effect of measurement error on the false alarm probability and probability of detection of the test used by~\shorttomkins{} The issues we identify suggest that their test is at risk of committing Type-I error in declaring biases in their analysis.  

Moving beyond the specific test used in~\tomkins{}, we also study the effect of their experimental design, which is simply the standard peer-review procedure with an additional random partition of reviewers into single and double blind groups. We show that two factors -- (d) asymmetrical bidding procedure and (e) non-random assignment of papers to referees -- as is common in peer-review procedures today may introduce spurious correlations in the data, breaking some key independence assumptions and thereby violating the requisite guarantees on testing.

\smallskip

\noindent \textbf{(2) Novel approach to testing for biases} (Sections~\ref{section:framework} -~\ref{section:analysis}) We propose a general framework for the design of statistical tests to detect biases in this problem setting, that overcomes the aforementioned limitations. Specifically, our framework does not assume objectivity of reviewers and does not make any parametric assumptions on reviewers' behaviour. Conceptually, we propose to think of this problem as an instance of a two-sample testing problem where single-blind and double-blind reviews form two samples and the test operates on these samples. (In contrast, \tomkins{} study the problem under one-sample testing paradigm, operating on reviews of single-blind reviewers and using double-blind reviews to estimate some parameters in their parametric model).

We then design computationally-efficient hypothesis testing procedures that under minimal assumptions guarantee a provable control over the false alarm probability under various conditions, including aforementioned conditions (a) - (c). We supplement these tests with an alternative design of the experimental setup which coupled with our tests mitigates issues (d) - (e) while not restricting the choice of assignment algorithm. 

Our tests also have non-trivial power in that they have considerably higher probability of detection in hard cases where test used by~\shorttomkins{} fails, and a  power comparable to that of~\shorttomkins{} when their assumptions are exactly met. The performance of one of these tests is illustrated in Figure~\ref{fig:illustration}. Additionally, we show that assumptions required by our tests to control for the Type-I error rate are essentially minimal in that they cannot be further relaxed without making reliable testing impossible.

\medskip

We note that while the discussion in this paper focuses on testing for biases with respect to protected attributes,  our experimental setup and statistical tests are not restricted to that alone. Instead of comparing the single versus double blind settings, our work can be used to test for effects of aspects of a submission exogenous to the manuscript's content, for instance, the effects of the reviewer questionnaire or that of asking authors to provide extraneous information (such as prior submission history). Our work enables conducting such semi-randomized controlled trials while retaining the no-bias and veracity conditions~\citep{tomkins17wsdm}, not requiring additional reviews, and having rigorous guarantees on the tests.

\medskip

\noindent \textbf{Related work.} The problem of identifying biases in human decisions is commonly studied in social science and there are many works that design and conduct randomized field experiments in various settings, including resume screening~\citep{bertrand2004emily}, hiring in academia~\citep{mossracusin12malebenefit}, and peer review~\citep{blank91effects, okike16sbvsdb}. However, the conference peer review setup we consider in this work does not comprise a fully randomized control trial (i.e., the reviewers are not assigned to submissions at random) and past approaches fail due to idiosyncrasies of the peer-review process. For example, a popular approach~\citep{bertrand2004emily, mossracusin12malebenefit} is to assign author identities to (fabricated) documents (resumes, application packages or papers) uniformly at random and compare the outcomes for different categories of authors. In our setup, \emph{random assignment of author identities to real (i.e., non-fabricated) submissions} is problematic due to various logistical and ethical issues such as reviewers guessing actual authors thereby causing biases, and requrements of getting authors to agree to have their paper/name modified. Another approach~\citep{okike16sbvsdb} is to submit \emph{the same paper} to multiple reviewers in both single-blind and double-blind conditions and test for the difference in the acceptance rates between conditions. However, such an approach necessitates a considerable additional reviewing load. Other approaches include observational studies, and we refer the interested readers to~\citet{tomkins17wsdm} for a more in-depth literature review.

\smallskip 

\noindent It is important to note that in this work, we do not aim to prove or disprove the existence of biases declared in the experiment by~\citet{tomkins17wsdm}. Instead, our focus is on the theoretical validity of the statistical procedures used to conduct such experiments and more generally on principled statistical approach towards designing such experiments.

Finally, the results and tests we discuss in this work are also applicable beyond peer review, and can be used to test for biases in other domains such as admissions and hiring.

The remainder of this paper is organized as follows. In Section~\ref{section:preliminaries} we present the problem setting formally and describe the experimental setup of~\tomkins{}. In Section~\ref{section:regression} we uncover issues (a) - (e) with their test and setup and illustrate the detrimental effect of such issues through simulations. Next, in Sections~\ref{section:framework} and~\ref{section:proposal} we present a novel non-parametric approach to testing for biases and corresponding statistical tests as well as the alternative design of the experimental procedure. The detailed analysis is given in Section~\ref{section:analysis}. We conclude the paper with a discussion in Section~\ref{section:discussion}.

\section{Preliminaries}
\label{section:preliminaries}

The general peer-review setup we study for testing biases using single and double blind review is as considered in~\tomkins{}. We study a conference peer-review setup where $\numpap$ papers are submitted at once and $\numrev$ independent reviewers are available to review submissions, where $\numrev$ is assumed to be an even number. With a goal to test whether single-blind reviewing induces a bias against or in favor of some groups of authors, we consider some pre-defined set of $\numcat$ binary mutually non-exclusive properties pertaining to the author(s) of any paper to be tested for bias. For example, a property could be ``the first author is female'' or ``majority of authors are from the USA''. Each paper $j \in [\numpap]$ is then associated with $\numcat$ indicator variables $\ind^{(1)}_{j},\ldots,\ind^{(\numcat)}_{j}$, where $\ind^{(\ell)}_{j}= 1$ if paper $j$ satisfies property $\ell$ and $\ind^{(\ell)}_{j}= -1$ otherwise. For each $\ell \in [\numcat]$ we let $\indset_{\ell} \subseteq [\numpap]$ denote the set of papers that satisfy property $\ell$ and $\smash{\noindset_{\ell}= [\numpap] \backslash \indset_{\ell}}$ denote its complement.\footnote{Here, we adopt the standard notation $[\nu] = \{1, 2, \ldots, \nu\}$ for any positive integer $\nu$.}

For each  property $\ell \in [\numcat]$ we are interested in whether \emph{single-blind peer review setup induces a bias} against or in favor of papers that satisfy this property. For example, if we consider property ``the first author is female'', then we aim at testing for the bias against or in favor of papers with female first author. Note that with respect to the properties, the study is observational in that we cannot assign author identities to papers at random. Hence, the effect of confounding is unavoidable and utmost care must be taken to address presence of confounding factors.

For brevity, in the main text we consider the case of a single property of interest ($\numcat = 1$) which captures the complexity of our problem. For ease of notation we drop index $\ell$ from $\ind^{(\ell)}$ and $\indset_{\ell}$. In Appendix~\ref{appendix:morethanone} we generalize the results to $\numcat > 1$. Let us now give details of the testing procedure used by~\tomkins{}.

\smallskip

\noindent \textbf{Experimental setup of Tomkins et al.} The peer review process in their experiment is organized as follows. Reviewers are uniformly at random divided into two groups of equal sizes, corresponding to two conditions: (i) Double-Blind condition (DB) in which reviewers do not observe identities of papers' authors; and (ii) Single-Blind condition (SB) in which reviewers observe identities of the papers' authors.  Next, each paper is assigned to $\papload$ reviewers from the SB group and $\papload$ reviewers from the DB group such that each reviewer reviews at most $\revload$ submissions, where $\papload$ and $\revload$ are predefined constants.  In both conditions, if any reviewer $i \in [\numrev]$ is assigned to any paper $j \in [\numpap]$, then she/he returns a binary accept/reject recommendation and possibly a numeric score that estimates a quality of the paper as perceived by reviewer, accompanied by a textual review.

\smallskip

\noindent \textbf{Model and test used by Tomkins et al.}  We begin by introducing an idealized version of their model.  They assume a parametric, logistic model for the binary decisions made by SB reviewers. Specifically, for each paper $j \in [\numpap]$, let $\sbdecision_{1j},\ldots,\sbdecision_{\papload j}$ denote the binary accept/reject decisions given by the $\papload$ reviewers assigned to paper $j$ in the SB setup. It is assumed that  $\{\sbdecision_{rj}\}_ {r \in [\papload]}$ are independent draws from a Bernoulli random variable with an expectation $\mean_j$ satisfying
\begin{align}
\label{eqn:logregmodel}
    \log \frac{\mean_j}{1 - \mean_j} = \coef_0 + \coef_{1} \tscore_j + \coef_{2} \ind_j, 
\end{align}
where $\tscore_j$ is a ``true'' underlying score of paper $j$, $\ind_j$ is an indicator of property satisfaction and $\left\{\coef_0, \coef_1, \coef_2 \right\}$ are unknown coefficients. In words, the model says that if there is a positive (respectively negative) bias with respect to a property of interest, then the fact that paper satisfies the property increases (respectively decreases) the log-odds of the probability of recommending acceptance by  $2\coef_{2}$ as compared to the case if the same paper does not satisfy the property. The main difficulty with this model in the peer review setting lies in the fact that true scores $\{\tscore_j, j \in [\numpap] \}$ are unknown and hence standard tests for logistic regression model are not readily applicable.

In order to overcome the unavailability of true scores $\{\tscore_j, j \in [\numpap] \}$ in the  model~\eqref{eqn:logregmodel}, \tomkins{} use a plug-in estimate: they replace $\tscore_j$ with the mean $\dbscore_j$ of scores given by the DB reviewers to paper $j$, for every $j \in [\numpap]$.  Under this approximation and using $\dbscore_1,\ldots,\dbscore_\numpap$, they obtain maximum likelihood estimates of coefficients $\{\estcoef_0, \estcoef_1, \estcoef_{2} \}$ and then use the standard Wald test~\citep{weisberg05linreg} to test for significance of the coefficient $\coef_2$. 
A bias is declared present if the coefficient $\coef_2$ is found significant; the direction of the bias is determined as the sign of  $\estcoef_{2}$.

\section{Problems with the past approach}
\label{section:regression}

In this section we identify several issues that should be taken into account when testing for biases in the setup we consider. Noting that the issues themselves are general, we motivate and discuss them in context of the prior work by~\tomkins{} and investigate possible consequences of these issues through synthetic simulations. In the simulations to follow, we juxtapose algorithm by~\tomkins{} to our \permutation{} test introduced later in the paper. Complete details of all simulations are given in Appendix~\ref{appendix:simulations}.

%%%%%%%%%%%%%%%%%%%%%%%%%%%%%%%%%%%%%%%%%
\subsection{Testing procedure}
\label{section:regression:test}

We begin from the issues that are pertinent to the testing procedure used by~\citet{tomkins17wsdm}. To this end, recall that with respect to the property of interest the experiment is observational. Hence we \emph{cannot} assume independence between the indicator of property satisfaction $\ind$ and the true score $\tscore$. Moreover, a non-trivial amount of correlation between some properties is plausible. Consider for example a property ``paper has author from top univeristy''. For this property a non-trivial correlation between true scores and indicator of property satisfaction is natural to expect. While correlation itself does not cause issues, we identify three conditions which coupled with correlation can be significantly harmful.

\newcommand{\correlation}{\varphi}

\smallskip

\begin{figure}[t]%
    \centering
    \subfloat[Measurement error\label{fig:error}]{{\includegraphics[width=4.5cm]{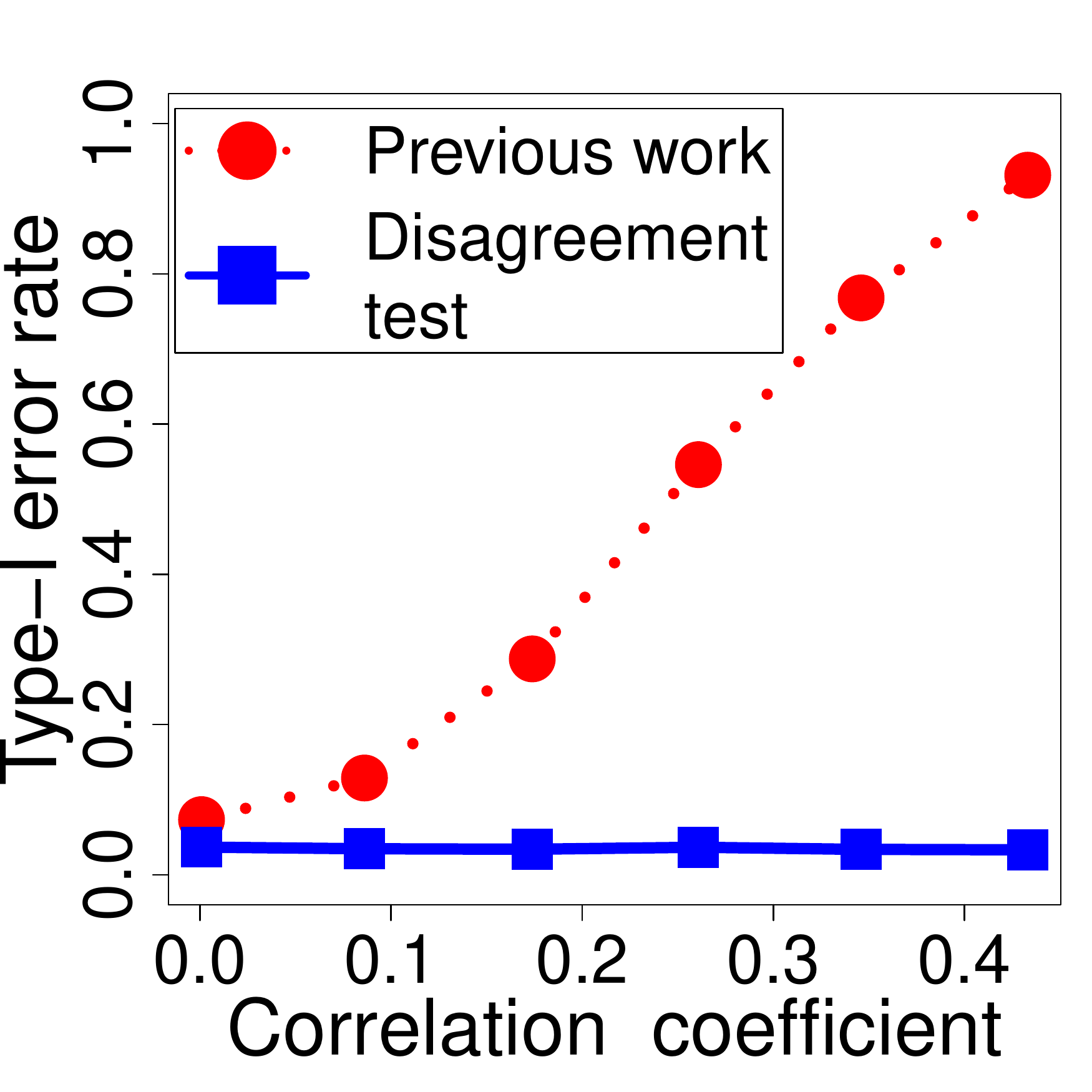}}}%
    \qquad
    \subfloat[Model mismatch\label{fig:mismatch}]{{\includegraphics[width=4.5cm]{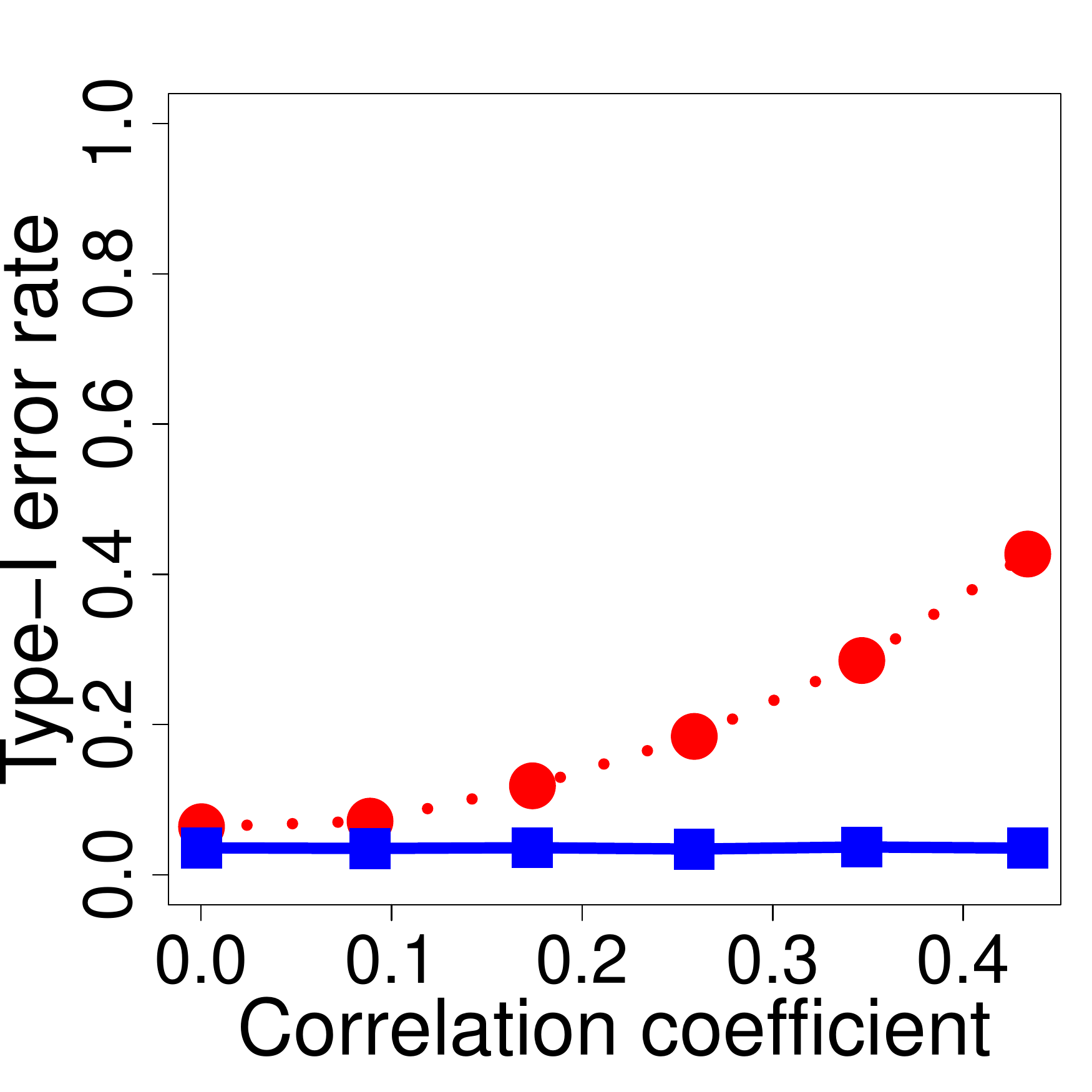}}}%
    \qquad
    \subfloat[Reviewer calibration\label{fig:revsignal}]{{\includegraphics[width=4.5cm]{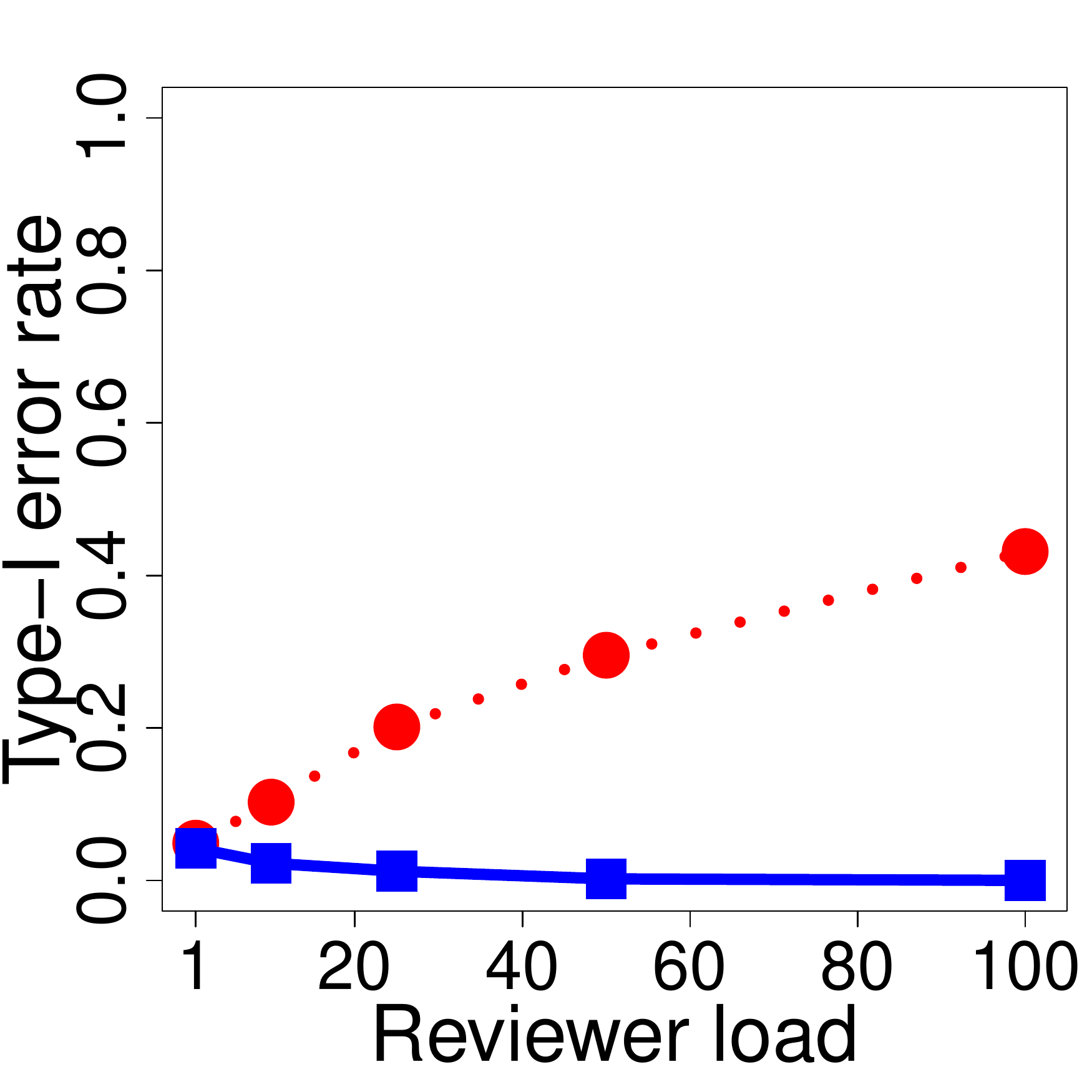}}}%
    \caption{Type-I error of the test from previous work (\citealt{tomkins17wsdm}) blows up under three different setups: bias is absent in all simulations and the tests are designed to limit the Type-I error to at most 0.05. In contrast, our \permutation{} test is robust to violations of modelling assumptions. Error bars are too small to be visible.}%
    \label{fig:tomkinsneg}%
\end{figure}

\noindent \textbf{(a) Measurement error.} \tomkins{} report low interreviewer agreement between DB reviewers which means that the estimates $\dbscore_1,\ldots,\dbscore_\numpap$  of the true scores by the DB reviewers are noisy. It is known~\citep{stefanski85errors, brunner09inflation} that noisy covariate measurement coupled with correlation between some covariates may inflate the Type-I error rate of the Wald test for logistic regression. We now investigate the impact of measurement error on the Type-I error rate of the~\shorttomkins{} test through simulations. We consider absence of any bias, and assume that model~\eqref{eqn:logregmodel} with $\coef_2 = 0$ is correct for both DB and SB reviewers. We consider DB reviewers to report noisy estimates of true scores $\tscore_j$, and vary the correlation between $\tscore$ and $\ind$. The level of noise was selected to keep correlation between the two DB reviewers assigned to each paper at the level of $0.6$, which is much better than the actual interreviewer agreement observed by~\tomkins{} (correlation 0.37). We plot the Type-I error rates in Figure~\ref{fig:error} for the test in~\tomkins{} and our proposed test, both tests are designed to restrict the Type-I error rate to 0.05.

Figure~\ref{fig:error} indicates a strong detrimental effect of measurement error on the validity of the test by~\tomkins{}. Given that interreviewer agreement in the actual WSDM conference experiment was low, the fact that some properties considered by~\shorttomkins{} may lead to correlations between $\tscore$ and $\ind$ is concerning, because it could potentially undermine the validity of their findings.

The simulations  in  Section~\ref{section:introduction} follow the setup presented here: Figures~\ref{fig:intro:type1} and~\ref{fig:intro:type2} consider measurement error with correlation fixed at 0.4 (Figure~\ref{fig:intro:type1}) and 0.6 (Figure~\ref{fig:intro:type2}) and show that (a) the negative effect of measurement error on the Type-I error rate exacerbates as sample size grows and (b) measurement error may also hinder the power of the test. Figure~\ref{fig:intro:easy} has zero correlation and no measurement error, satisfying all the assumptions of the test by~\shorttomkins{}

%%%%%%%%%%%%%%%%%%%%%%%%%%%%%%%%%%%%%%%%%%%%

\smallskip

\noindent  \textbf{(b) Model mismatch.} Model~\eqref{eqn:logregmodel} assumes a  specific parametric relationship, which may not hold in practice.  In order to check the effect of model mismatches, we consider a violation of the model~\eqref{eqn:logregmodel} and suppose that the correct model for both SB and DB reviewers is 
\begin{align*}
    \log \frac{\mean_j}{1 - \mean_j} = \coef_0 + \coef_{1} \left(\tscore_j\right)^3 + \coef_{2} \ind_{j},
\end{align*}
that is, instead of expected linear input, true scores of papers appear in the model raised to the power 3. To isolate the effect of model mismatch, we assume that true scores $\tscore_j, j \in [\numpap]$, are known exactly to the test of~\shorttomkins{} and hence abstract out the impact of the measurement error. We again consider an absence of any bias and set $\coef_2 = 0$ for both SB and DB reviewers. We then perform simulations similar to those in item (a). Figure~\ref{fig:mismatch} shows the results of the simulations.

%%%%%%%%%%%%%%%%%%%%%%%%%%%%%%%%%%%%%%%%%%%%%%%%%

\smallskip

\noindent  \textbf{(c) Reviewer calibration.} The test employed by~\citet{tomkins17wsdm} treats reviews given by the same reviewer as independent. In practice this assumption may be violated due to correlations introduced by reviewer's calibration~\citep{wang18calibration}. While some easy calibrations such as harshness/leniency can be captured by simple parametric extensions of model~\eqref{eqn:logregmodel}, more subtle patterns are beyond the scope of this model. Suppose for example that the strength of reviewers' input depends on paper's clarity --- the better the paper is written, the lower the contribution due to reviewers' calibration. Assume also that we are given a set of papers such that true score of each paper is proportional to the clarity of the paper (we formalize construction in Appendix~\ref{appendix:simulations:signal}). Coupled with the correlation between $\tscore$ and $\ind$, this pattern is sufficient to break Type-I error guarantees of the test of~\shorttomkins{} Again, to isolate the impact of reviewers' calibration, we assume that (i) true scores $\tscore_j, j \in [\numpap]$, are known to the test by~\shorttomkins{} and (ii) model~\eqref{eqn:logregmodel} is marginally correct for each reviewer, that is, each reviewer follows model~\eqref{eqn:logregmodel} for each paper she/he reviews, but her/his decisions for different papers are correlated in a specific way.

Figure~\ref{fig:revsignal} shows a result of simulations in which we vary the number of papers per reviewer, keeping correlation between $\tscore$ and $\ind$ fixed at 0.75 and the total number of papers fixed at $\numpap = 1000$. We simulate a wide range of reviewer load $\revload$ including small to medium loads of 5-15 papers typical in machine learning conferences like NeurIPS and larger loads of 40 or higher found in other smaller conferences. 

%%%%%%%%%%%%%%%%%%%%%%%%%%%%%%%%%%%%%%%%%
\subsection{Experimental setup}
\label{section:regression:setup}

\begin{figure}[t]%
    \centering
    \subfloat[Non-blind bidding\label{fig:bidding}]{{\includegraphics[width=5.5cm]{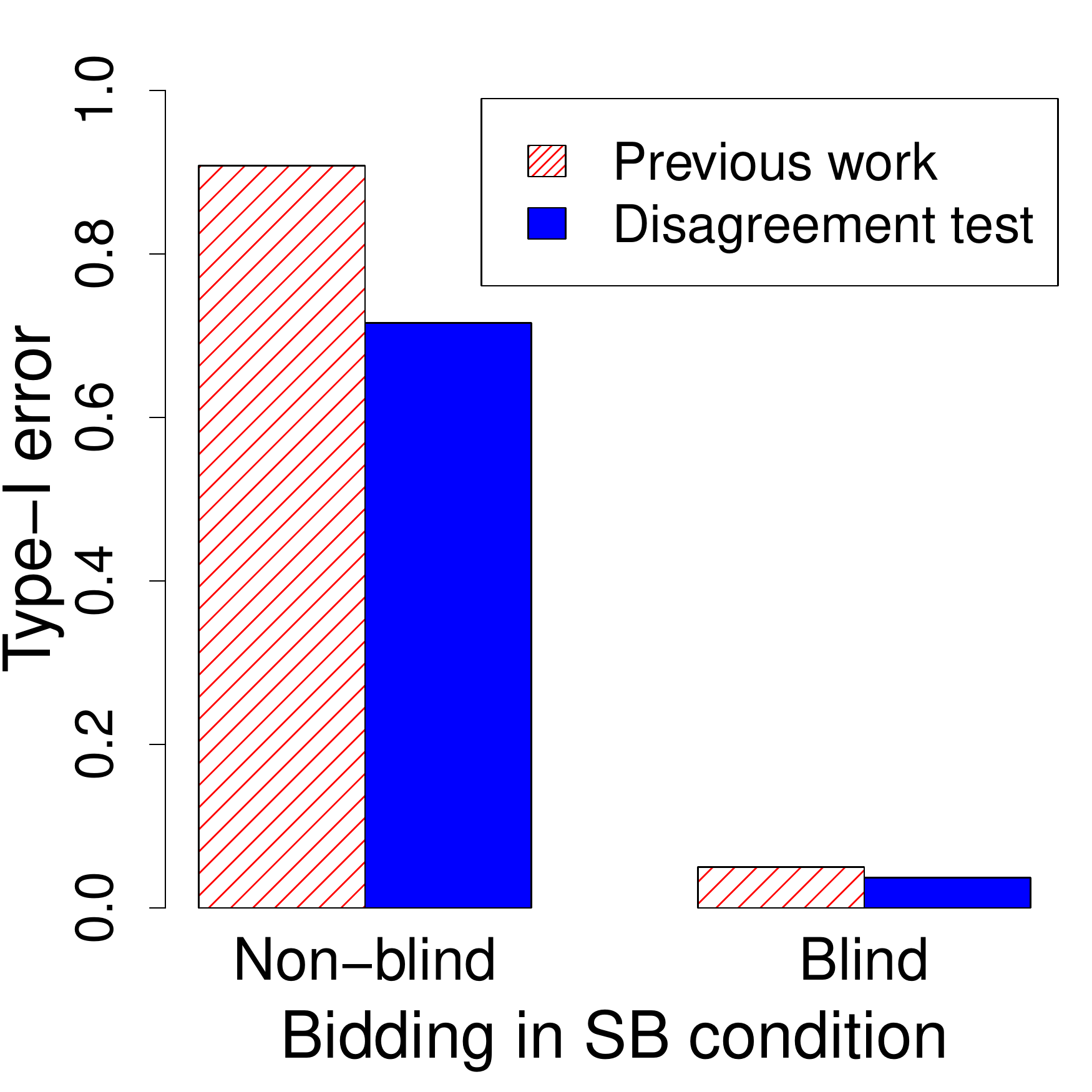}}}%
    \hspace{2cm}%
    \subfloat[Reviewer assignment\label{fig:assignment}]{{\includegraphics[width=5.5cm]{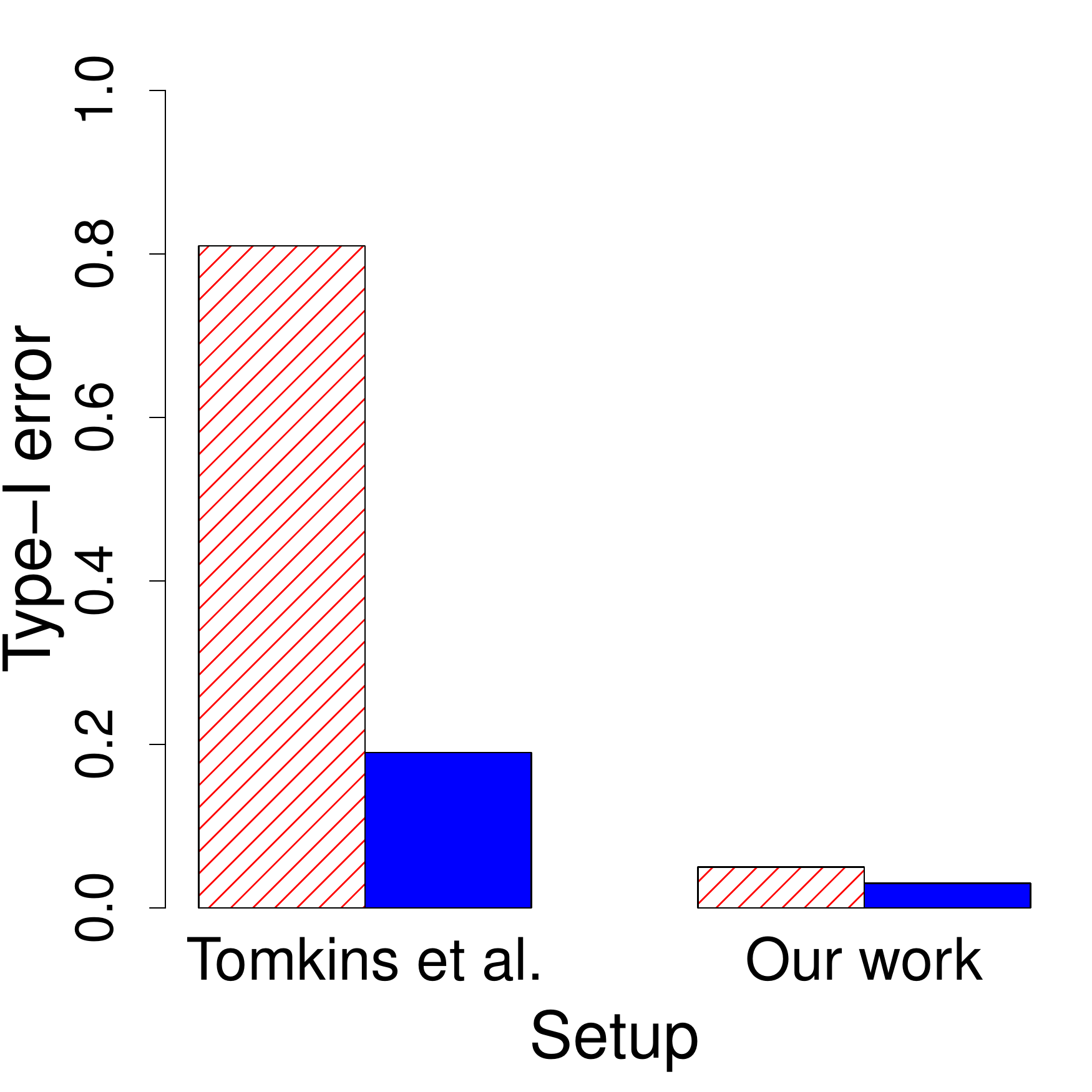}}}%
    \caption{The experimental setup from previous work (\citealt{tomkins17wsdm}) violates Type-I error guarantees of testing procedures. Bias is absent in all simulations and the tests are designed to limit the Type-I error to at most 0.05. Note that the issues which pertain to the experimental setup rather than the modelling break guarantees of both tests (leftmost columns). In contrast, our proposed setup with fully blind bidding procedure and careful management of assignment ensures Type-I error guarantees for both tests (rightmost columns). Error bars are too small to be visible.}%
    \label{fig:tomkinsneg:setup}%
\end{figure}

The issues discussed above pertain to the testing procedure and modelling assumptions made by~\tomkins{}. We now issue a commentary regarding the experimental setup considered in their work which comprises a random partition of reviewers into SB and DB groups within a standard peer review procedure. In particular, we show that the setup itself may create problems in controlling the Type-I error. 

\smallskip

\noindent \textbf{(d) Non-blind bidding.} In the experiment by~\shorttomkins{} papers are allocated to reviewers based on preferences (``bids'') declared by reviewers (reviewers could indicate that they want to review some papers and do not want to review others). Importantly, the reviewers in the SB setup also get to see author identities in the bidding stage, which may act as a confounding factor in tests for bias in the acceptance/rejection of papers. This is indeed pointed out as a caveat by \tomkins{} in their paper.

To illustrate the possible effect, consider a property of interest ``paper has a famous author'' and suppose that among all reviewers there is a subset of lenient reviewers who additionally want to read papers from top authors with the hope of reading better papers. Then in DB setup such reviewers cannot use author identity information and hence make their bidding decisions based on title and abstract only; in contrast, in SB setup these reviewers tend to bid on papers authored by top authors. Given that reviewers who preferentially bid on papers with top authors in SB condition are by coincidence lenient, the difference in bidding behavior may result in structurally different evaluations between conditions even when reviewers' evaluations are unbiased, leading to a blow-up of the Type-I error rate of any reasonable test. Figure~\ref{fig:bidding} shows a result of simulations (formal setup is in Appendix~\ref{appendix:simulations:bidding}) in which we compare non-blind and blind bidding conditions for SB reviewers and indicates a possible detrimental effect of non-blind bidding. 

\smallskip 

\noindent \textbf{(e) Reviewer assignment.} One might imagine that a natural requirement to conduct the bidding in a double blind fashion for both DB and SB reviewers would fix the issues with the setup of~\shorttomkins{} However, perhaps surprisingly, we show that even if both groups bid in a double blind fashion (or even if the bidding process is eliminated entirely), and even if the reviewers are assigned to DB or SB groups uniformly at random, the non-random assignment using algorithms such as TPMS~\citep{charlin13tpms} that assigns reviewers to papers maximizing some notion of ``similarities'' can still lead to a violation of the Type-I error guarantees. We give a formal construction in Appendix~\ref{appendix:simulations:assignment}; the intuition is as follows. Quoting~\citet{lamont2009professors}, ``evaluators often define excellence as <<what speaks to me>> which is akin to <<what is most like me>>'', that is, a similarity between a paper and a reviewer may influence the decision. That said, we construct these similarities in a careful manner: our choice ensures that despite reviewers being allocated to DB or SB conditions at random, the popular TPMS assignment algorithm with high probability constructs assignments that are in some sense structurally different between SB and DB conditions, which in turn leads to structurally different evaluations. Our construction, along with correlation between $\tscore$ and $\ind$, introduces spurious correlations in the data, thereby violating some key independence assumptions and leading to the inflation of Type-I error. 

Figure~\ref{fig:assignment} shows a result of simulations in which we compare the setup of~\tomkins{} with our proposed experimental setup introduced in Section~\ref{section:setup}. Notably, under the setup of~\shorttomkins{} even the \permutation{} test which is robust to various issues discussed in Section~\ref{section:regression:test} is unable to control for the Type-I error. In contrast, observe that under our proposed experimental setup, both the \permutation{} test and the test by~\shorttomkins{} control for the Type-I error rate at the desired level. 

Importantly, we underscore that while our experimental procedure mitigates the issues with the experimental setup of~\shorttomkins{}, their test is still \emph{susceptible to the issues we discussed in Section~\ref{section:regression:test}} even under our experimental setup. Finally, under the setup of~\shorttomkins{} the phenomenon of the Type-I guarantee violation is not restricted to the TPMS assignment and can occur in a much broader class of reviewer assignment algorithms.  

%%%%%%%%%%%%%%%%%%%%%%%%%%%%%%%%%%%%%%%%%%%%%%%%%%%%%%%%%%%%%%%%%%%%%%

\section{Novel framework to test for biases} 
\label{section:framework}

In Section~\ref{section:regression} we identified five key limitations of the approach taken by~\tomkins{}. Three of these limitations pertain to the testing procedure and the two limitations relate to the design of the experiment itself. In the next sections we design a set of tests and experimental setup with strong guarantees, and which overcome the aforementioned limitations. In this section we begin from principled definition of a bias testing problem that generalizes one made by~\shorttomkins{} and does not make any restrictive assumptions.   

At a high level, our approach to testing for biases is different from those proposed by~\shorttomkins{} in two ways. First, we relax two strict modelling assumptions: (i) instead of assuming existence of true qualities of submissions, we allow subjectivity in reviewer evaluations~\citep{kerr1977manuscript, ernst1994reviewer, bakanic1987manuscript, mahoney1977publication, lamont2009professors,noothigattu2018choosing}, and (ii) we do not assume any specific form of the relationship between a paper and its probability of acceptance by a reviewer. Instead, we allow these probabilities to be completely arbitrary and define the bias in terms of these probabilities. Second, we treat this problem conceptually differently from the work of~\shorttomkins{} The test therein treats the problem as that of one-sample testing and uses DB scores as a plugin estimate of true scores in SB model. In contrast, we approach this  problem through the lenses of two-sample testing, where SB and DB reviews form the two samples, and the goal is to test whether they belong to the same distribution. This perspective helps us to avoid a number of issues discussed in Section~\ref{section:regression}.

Formally, let $\dbmatrix \in [0, 1]^{\numrev \times \numpap}$ be a matrix whose $(i, j)^{\text{th}}$ entry, denoted as $\dbpacc_{ij}$, represents a probability that reviewer $i$ would recommend acceptance of paper $j$ if that paper is assigned to that reviewer in DB setup. Similarly, let matrix $\sbmatrix \in [0, 1]^{\numrev \times \numpap}$ be an analogous matrix in SB setup, and denote its  $(i, j)^{\text{th}}$ entry as $\sbpacc_{ij}$. 

Let $\sbrev$ be the set of reviewers allocated to the SB condition. Moreover, for each $i \in \sbrev$, let $\sbpapset(i)$ denote the set of papers assigned to reviewer $i$ and let $\sbdecision_{ij} \in \{0, 1\}$ denote the accept/reject decision given by reviewer $i$ for paper $j \in \sbpapset(i)$. We similarly define set of DB reviewers $\dbrev$ and their decisions $\left\{ \dbdecision_{ij}: i \in \dbrev,  j \in \dbpapset(i) \right\}$ . We are interested in testing for biases with respect to a property of interest. To this end, recall our notation $\indset \subseteq [\numpap]$ for the set of papers that satisfy a property of interest, and $\noindset$ as its complement. 

With this notation in place, we now define two formulations of the bias testing problem --- ``absolute'' and ``relative'': The relative bias setting is strictly more general than the absolute bias setting, but also leads to more restrictive results.\footnote{An equivalent definition of the problem from the perspective of causal inference can be found in Appendix~\ref{appendix:causal}.} Importantly, the tests we will introduce in Section~\ref{section:tests} are applicable to both formulations without additional modifications.

%%%%%%%%%%%%%%%%%%%%%%%%%%%%%%%%%%%%

\subsection{Absolute bias problem}
\label{section:absolute}

%%%%%%%%%%%%%%%%%%%%%%%%%%%%%%%%%%%%

In the absence of bias, the knowledge of authors' identities does not induce any difference in reviewers' behaviour. In the biased hypothesis, there is a positive bias in favor of papers that satisfy a property of interest: reviewers in SB condition are more lenient towards papers from $\indset$ and more harsh towards papers from $\noindset$ than they would be in DB condition. The following problem formalizes this intuition.

\begin{problem}[Absolute bias problem] \label{problem:absolute}
 Given significance level $\level \in (0, 1)$ and decisions of SB and DB reviewers, the goal is to test the following hypotheses:
\begin{align}
   & \nonumber \hypothesis_0 : \forall i \in [\numrev] \ \forall j \in [\numpap] \ \  \sbpacc_{ij} = \dbpacc_{ij} \\
   & \label{eqn:absolute:alt} \hypothesis_1 : \forall i \in [\numrev] \ \forall j \in [\numpap]   \begin{cases}
                                                \sbpacc_{ij} {\ge} \dbpacc_{ij} & \text{if } j \in \indset \\
                                                \sbpacc_{ij} {\le} \dbpacc_{ij} & \text{if } j \in \noindset ,
                                             \end{cases}
\end{align}
where at least one inequality in the alternative hypothesis~\eqref{eqn:absolute:alt} is strict.
\end{problem}

Note that one can define an alternative that represents a bias against papers from $\indset$ simply by exchanging the sets $\indset$ and $\noindset$ in~\eqref{eqn:absolute:alt}.  Our goal is to design a testing procedure that controls for Type-I error and has non-trivial power for any pair of matrices $\sbmatrix, \dbmatrix$ that fall under definition of Problem~\ref{problem:absolute}.

\smallskip

\noindent \textbf{Non-trivial power.} Informally, we say that the test has non-trivial power if for choices of $\sbmatrix$ and $\dbmatrix$ for which the presence of bias is ``obvious'', the test is able to detect the bias with probability that goes to 1 as number of papers in both $\indset$ and $\noindset$ grows to infinity. Formally, we say that matrices $\sbmatrix$ and $\dbmatrix$ satisfy alternative hypothesis~\eqref{eqn:absolute:alt} with margin $\minigap$, if all inequalities in equation~\eqref{eqn:absolute:alt} are satisfied with margin $\minigap > 0$, that is, $|\sbpacc_{ij} - \dbpacc_{ij}| > \minigap ~~\forall~ (i,j) \in [\numrev] \times [\numpap]$. Then we say that the testing procedure has non-trivial power if for any $\err > 0$ and for any $\minigap > 0$ there exists $\sample = \sample(\err, \minigap)$ such that if $\min\{|\indset|, |\noindset| \} > \sample$, then for any $\sbmatrix$ and $\dbmatrix$ that satisfy alternative hypothesis~\eqref{eqn:absolute:alt} with margin $\minigap$, the power of testing procedure is at least $1 - \err$. 

For instance, if the logistic model~\eqref{eqn:logregmodel} is correct for both SB and DB reviewers for some $\sbcoef_0 = \dbcoef_0 = \coef_0$, $\sbcoef_1 = \dbcoef_1 = \coef_1 > 0$, $\dbcoef_2 = 0$ and $|\sbcoef_2| > 0$, then the requirement of non-trivial power ensures that for any choice of true scores bounded in absolute value by a universal constant and any choice of property satisfaction indicators,  the test has power growing to 1 as  $\min\{|\indset|, |\noindset| \}$ goes to infinity.

%%%%%%%%%%%%%%%%%%%%%%%%%%%%%%%%%%%%

\subsection{Relative bias problem}
\label{section:framework:relative}

In Problem~\ref{problem:absolute} we assumed that SB (or DB) condition itself does not cause any change in reviewers' behaviour. We now consider a generalization of Problem~\ref{problem:absolute} which accommodates an additional confounding factor --- a bias in the reviewer simply due to her/his assignment in the SB or the DB group (and independent of the paper or its characteristics). For example, reviewers may not have any bias with respect to the property of interest, but just being placed in the SB condition may induce more harsh opinions than the reviewers in DB. Formally, recall the null hypothesis $\sbpacc_{ij} = \dbpacc_{ij} ~~\forall (i,j) \in  [\numrev] \times [\numpap]$ in Problem~\ref{problem:absolute}. Instead, under the null, we now allow $\sbpacc_{ij} = \nullshift(\dbpacc_{ij})$, for some non-decreasing function $\nullshift: [0, 1] \to [0, 1]$. Of course, one may not know the function $\nullshift$ and the goal of this general problem is to design a test that is guaranteed to control over Type-I error and has non-trivial power uniformly for all functions $\nullshift$ that belong to some set of non-decreasing functions $\subclass$.

\begin{problem}[Relative bias problem] \label{problem:relative} Given significance level $\level \in (0, 1)$, class of functions $\subclass$ and decisions of SB and DB reviewers, the goal is to test the following hypotheses:
\begin{align}
   & \nonumber \hypothesis_0 : \forall i \in [\numrev] \ \forall j \in [\numpap] \ \ \sbpacc_{ij} = \nullshift(\dbpacc_{ij}) \\
   & \label{eqn:relative:alt} \hypothesis_1 : \forall i \in [\numrev] \ \forall j \in [\numpap] \   \begin{cases}
                                                \sbpacc_{ij} \ge \nullshift(\dbpacc_{ij}) & \text{if } j \in \indset \\
                                                \sbpacc_{ij} \le \nullshift(\dbpacc_{ij}) & \text{if } j \notin \indset 
                                             \end{cases},
\end{align}
where $\nullshift$ is some unknown function from $\subclass$ and at least one inequality in the alternative hypothesis~\eqref{eqn:relative:alt} is strict. 
\end{problem}

For example, if the logistic model~\eqref{eqn:logregmodel} is correct for both SB and DB reviewers (with $\dbcoef_2 = 0$), but intercepts $\coef_0$ in SB and in DB conditions are allowed to be different, then the corresponding matrices $\sbmatrix$ and $\dbmatrix$ do not fall under the definition of Problem~\ref{problem:absolute}, but can be captured by Problem~\ref{problem:relative} with specific choice of $\subclass$ as we will discuss in Section~\ref{section:analysis:relative}.

The definition of non-trivial power transfers to the relative bias problem with the exception that all $\dbpacc_{ij}$ are substituted by $\nullshift(\dbpacc_{ij})$ for $\nullshift \in \subclass$. Our goal is to design a testing procedure that controls for Type-I error and has non-trivial power for any pair of matrices $\sbmatrix, \dbmatrix$ that fall under definition of Problem~\ref{problem:relative} for any function $\nullshift \in \subclass$. Ideally, we would like to achieve this goal for a set of functions $\subclass$ that contains all non-decreasing functions  $f: [0, 1] \to [0, 1]$.

%%%%%%%%%%%%%%%%%%%%%%%%%%%%%%%%%%%%

\section{Proposed solution}
\label{section:proposal}

We now introduce the proposed experimental setup as well as statistical tests we study in this work. We subsequently analyze them in the context of Problems~\ref{problem:absolute} and~\ref{problem:relative} in Section~\ref{section:analysis}. 

%%%%%%%%%%%%%%%%%%%%%%%%%%%%%%%%%%%%
\subsection{Testing procedures}
\label{section:tests}

In order to avoid correlations introduced by reviews given by the same reviewer, our tests use at most one decision per reviewer. As we discuss in Section~\ref{section:setup}, we do so by first matching reviewers into pairs, consisting of one SB and one DB reviewer who review a common paper. For the moment, assume that we are given a set of tuples $\settuple$, where each tuple $\tuple \in \settuple$ consists of a paper $j_{\tuple} \in [\numpap]$, decision of a SB reviewer for this paper $\sbdecision_{j_\tuple}$, decision of a DB reviewer for this paper $\dbdecision_{j_\tuple}$ and indicator of property satisfaction $\ind_{j_\tuple}$, with a constraint that each reviewer contributes her/his decision to at most one tuple. With this notation, we now present two tests we consider in this work. As we show subsequently, either of these tests would suffice for the absolute bias problem, but for the relative bias problem they cater to different models of reviewers' behaviour with non-intersecting areas of applicability. To provide intuition behind the tests, we define them in context of the absolute bias problem (Problem~\ref{problem:absolute}) and discuss their applicability to the relative bias problem later.

\medskip
\noindent \textbf{Disagreement-based test.} A high-level idea of the test is as follows. Consider a pair of SB and DB reviewers who disagree in their decisions for some paper. Then under the null hypothesis, the events ``SB accepts and DB  rejects'' and ``SB rejects and DB accepts'' are equally likely. In contrast, if the null hypothesis is violated, then depending on the property satisfaction and the direction of the bias, SB reviewer is more (or less) likely to vote for acceptance than her/his DB counterpart.   

{
\setcounter{algorithm}{0}
\floatname{algorithm}{Test}
\begin{algorithm}[H]
   \caption{\permutation{}}
   \label{test:permutation}
   {\bfseries Input:} Significance level $\level \in (0, 1)$ \\ 
   \hphantom{{\bfseries Input:}} Set of tuples $\settuple$, where each $\tuple \in \settuple$ is of the form  $(j_{\tuple}, \sbdecision_{j_{\tuple}}, \dbdecision_{j_{\tuple}}, \ind_{j_{\tuple}})$ for some paper $j \in [\numpap]$. \\
    \begin{enumerate}[leftmargin=*, topsep=0pt]
    
        \item \label{step:perm:init} Initialize $\arraymaj$ and $\arraymin$ to be empty arrays.
        
        \item \label{step:perm:majority} For each tuple $\tuple \in \settuple$, if $\sbdecision_{j_\tuple} \ne \dbdecision_{j_\tuple}$, append $\sbdecision_{j_\tuple}$ to 
        $
                \begin{cases}
                    \arraymaj & \text{if } \ind_{j_\tuple} = 1 \\
                    \arraymin & \text{if } \ind_{j_\tuple} = -1
                \end{cases}.
        $
        
        \item \label{step:perm:decision} Run a permutation test~\citep{fisher35permutation} at the level $\level$ to test if entries of $\arraymaj$ and $\arraymin$ are exchangeable random variables, using the test statistic:
        \begin{align*}
            \statp = \frac{1}{|\arraymaj|} \sum\limits_{r \in [|\arraymaj|]} \arraymaj_r - \frac{1}{|\arraymin|} \sum\limits_{r \in [|\arraymin|]} \arraymin_r.
        \end{align*}
        \vspace{-17pt}
        \item Reject the null if and only if the permutation test rejects the null. (If either of the arrays $\arraymin$ and $\arraymaj$ is empty, the test keeps the null.)
    \end{enumerate}
\end{algorithm}
}

We now formally present the \permutation{} test as Test~\ref{test:permutation}. In Step~\ref{step:perm:init} two empty arrays $\arraymaj$ and $\arraymin$ are initialized. Next, in Step~\ref{step:perm:majority}  we focus on pairs of SB and DB reviewers disagreeing in their decisions for a paper they both review. For each of the corresponding tuples, we add the decision of SB reviewer to the array $\arraymaj$ if a paper satisfies the property of interest and to $\arraymin$ otherwise. Finally, in Step~\ref{step:perm:decision} we define a test statistic $\statp$. According to the aforementioned intuition, under the null hypothesis $\statp$ should be close to 0, but under the alternative it should be large in absolute value. Hence, to make a decision we run a permutation test and reject the null in Step~\ref{step:perm:decision} if this test suggests that $|\statp|$ is too large for a given significance level $\level$.

\medskip

\noindent \textbf{Counting-based test.} The test is built on a simple intuition. Assume for the moment that SB setup induces a bias against papers from $\indset$ and a bias in favor of papers from $\noindset$. Then it is likely that papers from $\indset$ will receive less number of positive recommendations in SB setup as compared to DB setup. Symmetrically, for papers from $\noindset$ we expect reviewers in SB to be more lenient than their DB counterparts. In contrast, if there is no bias at all, then we expect the aforementioned differences to be small.

We now formally present the \counting{} test as Test~\ref{test:counting}. In Step~\ref{step:count:init} two empty arrays $\arraymaj$ and $\arraymin$ are created which in Step~\ref{step:count:allocation} are populated with differences between decisions of SB and DB reviewers for papers from $\indset$ and $\noindset$ respectively. Importantly, in contrast to the \permutation{} test, in the \counting{} test we do not condition on disagreeing pairs of reviewers. Noticing that mean value of entries of $\arraymaj$ (respectively $\arraymin$) measures the change of attitude towards papers from $\indset$ (respectively $\noindset$) between SB and DB conditions, in Step~\ref{step:count:diffs} we compute a test statistic $\statc$ which compares these changes. According to the aforementioned intuition, under correct null hypothesis the test statistic should be close to 0. Finally, in Step~\ref{step:count:decision} we make a decision using concentration properties of the test statistic.

{
\setcounter{algorithm}{1}
\floatname{algorithm}{Test}
  \begin{algorithm}[H]
   \caption{\counting}
   \label{test:counting}
   {\bfseries Input:} Significance level $\level \in (0, 1)$ \\ 
   \hphantom{{\bfseries Input:}} Set of tuples $\settuple$, where each $\tuple \in \settuple$ is of the form  $(j_{\tuple}, \sbdecision_{j_{\tuple}}, \dbdecision_{j_{\tuple}}, \ind_{j_{\tuple}})$ for some paper $j \in [\numpap]$.  \\
    \begin{enumerate}[leftmargin=*, topsep=0pt]
        \item \label{step:count:init} Initialize $\arraymaj$ and $\arraymin$ to be empty arrays.
        
        \item \label{step:count:allocation} For each tuple $\tuple \in \settuple$, append $(\sbdecision_{j_\tuple} - \dbdecision_{j_\tuple})$ to 
        $
                \begin{cases}
                    \arraymaj & \text{if } \ind_{j_\tuple} = 1 \\
                    \arraymin & \text{if } \ind_{j_\tuple} = -1
                \end{cases}.
        $

        \item \label{step:count:diffs} If either of the arrays $\arraymin$ and $\arraymaj$ is empty, keep the null and terminate. Otherwise, set the test statistic $\statc$ as follows:
        \begin{align}
        \label{eqn:counting_stat}
            \statc = \frac{1}{|\arraymaj|} \sum\limits_{r \in [|\arraymaj|]} \arraymaj_r - \frac{1}{|\arraymin|} \sum\limits_{r \in [|\arraymin|]} \arraymin_r.
        \end{align}
        \vspace{-7pt}
        \item \label{step:count:decision} Reject the null hypothesis if and only if
        \begin{align*}
            |\statc| > \sqrt{2 \left(|\arraymaj|^{-1} + |\arraymin|^{-1} \right) \log \nicefrac{2}{\level}}.
        \end{align*}
        \vspace{-15pt}
    \end{enumerate}
\end{algorithm} 
}

\noindent \textbf{Effect size.} In Section~\ref{section:analysis} we will establish theoretical guarantees on Type-I error control for
both \permutation{} and \counting{} tests. In addition to these guarantees, both tests provide a natural measure of the effect size:
\begin{itemize}[leftmargin=*, itemsep=0pt, topsep=1pt]
    \item \counting{}. The test statistic $\statc$ of the \counting{} test compares the within-subject differences in  acceptance rates for papers from $\indset$ and $\noindset$. Indeed, the first term in equation~\eqref{eqn:counting_stat} measures the difference between acceptance rates in SB and DB setups for papers from $\indset$. Similarly, the second term measures the same difference for papers from $\noindset$. A positive value of the test statistics then indicates that papers from $\indset$ benefit from SB review more than papers from $\noindset$. 
    
    \item \permutation{}. Slightly informally, the test statistic $\statp$ of the \permutation{} test measures the difference in acceptance rates of ``borderline'' papers from $\indset$ and $\noindset$ in the SB setup. Indeed, by conditioning on pairs of disagreeing reviewers in Step~\ref{step:perm:majority} of Test~\ref{test:permutation}, the test rules out ``clear accept'' and ``clear reject'' papers thus considering only the papers for which reviewers disagree (i.e., borderline papers).
\end{itemize}
Overall, absolute values of the test statistics $\statp$ and $\statc$ are reasonable estimates of the effect size and are in a similar vein to Cohen's $d$ and other popular effect size measures~\citep{cohen92power}.

%%%%%%%%%%%%%%%%%%%%%%%%%%%%%%We will %%%%%%%
\subsection{Setup of the experiment}
\label{section:setup}

We now propose the setup of the experiment to overcome the issues highlighted in Section~\ref{section:regression:setup} and discuss a construction of the set $\settuple$ used by the tests introduced above. At a higher level, the proposed setup has two main differences from one considered by~\tomkins{}. First, bidding is performed in blind manner by both SB and DB reviewers (Step~\ref{step:assign:bid} below). Second and more importantly, to avoid issues caused by non-random reviewers' assignment, we perform paper assignment and reviewer allocation to conditions jointly in a carefully selected manner (Steps~\ref{step:assign:selection}-\ref{step:assign:complement} below).

{
\setcounter{algorithm}{0}
\floatname{algorithm}{Procedure}
  \begin{algorithm}[H]
   \caption{Design of the experiment}
   \label{algo:experiment}
   {\bfseries Input:} Paper load $\papload \ge 1$ \\ 
   \hphantom{{\bfseries Input:}} Reviewer load $\revload \ge 1$ \\
   \hphantom{{\bfseries Input:}} Assignment algorithm $\algo$ \\
   \vspace{-15pt}
    \begin{enumerate}[leftmargin=*, topsep=5pt, itemsep=2pt]
        \item \label{step:assign:bid} Reviewers bid on papers in blind manner
        
        \item \label{step:assign:selection} Depending on the relationship between number of papers ($\numpap$) and reviewers ($\numrev$):                      
        \begin{enumerate}[noitemsep, leftmargin=8pt, topsep=0pt]
        \item If $\numrev > 2\numpap$, select $2 \numpap$ reviewers uniformly at random and use algorithm $\algo$ to assign each paper to 2 reviewers from the selected pool such that each reviewer is assigned to one paper 
        
        \item \label{step:case:example} If $\numrev < 2\numpap$, select $\numrev/2$ papers such that proportions of papers from $\indset$ and $\noindset$ are as close to each other as possible. Use algorithm $\algo$ to assign each selected paper to 2 reviewers such that each reviewer is assigned to one paper
        
        \item If $\numrev = 2 \numpap$, use algorithms $\algo$ to assign each paper to 2 reviewers such that each reviewer is assigned to one paper
        \end{enumerate}
        Denote the corresponding assignment as $\assignment^{\ast}$
        
        \item \label{step:assign:allocation} For each paper in assignment $\testassignment$, allocate one assigned reviewer to DB condition and another assigned reviewer to SB condition uniformly at random. If at this point there are reviewers who are not allocated to conditions, allocate half of them to SB and half to DB uniformly at random
        
        \item \label{step:assign:complement} Using algorithm $\algo$, complement assignment $\assignment^{\ast}$ such that each paper is assigned to $\papload$ SB and $\papload$ DB reviewers and each reviewer reviews at most $\revload$ papers. Denote the corresponding assignment as $\assignment$ and begin review process according to this assignment
    
        \item \label{step:assign:settuple} When the review process is finished, construct a set $\settuple$ as follows. For every paper $j$ from the assignment $\assignment^{\ast}$ and corresponding pair $(i_1, i_2)$ of SB and DB reviewer, add tuple $(j, \sbdecision_{i_1 j}, \dbdecision_{i_2 j}, \ind_{j})$ to the set $\settuple$
        
        \item \label{step:assign:run} Run statistical test on the set $\settuple$
        \end{enumerate}
\end{algorithm} 
}

We now formally present the experimental procedure as Procedure~\ref{algo:experiment}. It takes as input parameters of paper and reviewer loads together with any assignment algorithm that operates on similarities and/or bids. In Step~\ref{step:assign:bid} reviewers bid on the papers in blind manner, that is, using only title and abstract of submissions. Notice that in contrast to the~\shorttomkins{} setup, bidding happens even before the reviewers are allocated to SB or DB conditions. In Step~\ref{step:assign:selection} we find a partial assignment of papers to reviewers which satisfies $(\papload = 2, \revload = 1)$-load constraints. Depending on the relationship between $\numpap$ and $\numrev$, we may include only a subset of papers or reviewers in this assignment. For example, in case~\ref{step:case:example} we do not have enough reviewers to respect the one paper per reviewer constraint and hence we select subset of papers of appropriate size such that it includes approximately equal number of papers from $\indset$ and $\noindset$ and find the assignment for selected papers only. The constraint on the number of papers from $\indset$ and $\noindset$ is to ensure that the resulting set $\settuple$ is balanced which is necessary for non-trivial power. The corresponding assignment $\assignment^{\ast}$ is a building block for our tests which will use the reviews from this assignment only. Next, in Step~\ref{step:assign:allocation} reviewers are allocated to conditions in a specific manner which is crucial for our statistical guarantees. In Step~\ref{step:assign:complement} we find a full assignment $\assignment$ that is a completion of the partial assignment $\testassignment$, meaning that if reviewer $i$ was assigned to paper $j$ in assignment $\assignment^{\ast}$, she/he is also assigned to this paper in $\assignment$. Finally, in Step~\ref{step:assign:settuple} we construct a set of tuples $\settuple$ that is used by the \permutation{} and \counting{} algorithms in Step~\ref{step:assign:run}. Importantly, by construction we ensure that each reviewer contributes at most one decision to the set $\settuple$. 

As we show below, the experimental Procedure~\ref{algo:experiment} overcomes the issues with the experimental setup we discussed in Section~\ref{section:regression:setup} and leads to provable control over Type-I error for our \permutation{} and \counting{} algorithms. We underscore that (i) \permutation{} and \counting{} tests are not tied to particular experimental procedure we introduce and can be applied under the setup of~\tomkins{} with caveats discussed in Section~\ref{section:regression:setup}. For instance, in simulations (a)-(d) of Section~\ref{section:regression} the \permutation{} test was applied under the setup of~\shorttomkins{} More details on this remark are provided in Appendix~\ref{appendix:matching}; (ii) as requested by the \permutation{} and \counting{} tests, the set of tuples $\settuple$ constructed in Step~\ref{step:assign:settuple} contains at most one decision of each reviewer. This requirement allows our tests to be agnostic to reviewer calibration which may otherwise undermine Type-I error guarantees as demonstrated in Section~\ref{section:regression:test}. However, if one treats reviews given by the same reviewer as independent, thereby ignoring issues with reviewer calibration, then in Step~\ref{step:assign:settuple} of Procedure~\ref{algo:experiment} one can construct a larger set $\settuple$ by using full assignment $\assignment$ and allowing each reviewer to contribute multiple decisions to the set $\settuple$.

Finally, in addition to facilitating the experiment, in the interest of fairness in the review~\citep{tomkins17wsdm} the experimental procedure should ensure that in the eventual assignment each paper is reviewed by equal number of SB and DB reviewers. By construction, Procedure~\ref{algo:experiment} satisfies this requirement: Step~\ref{step:assign:complement} ensures that in the final assignment $\assignment$ each paper is assigned to $\papload$ SB reviewers and $\papload$ DB reviewers.

%%%%%%%%%%%%%%%%%%%%%%%%%%%%%%%%%%%%

\section{Analysis}
\label{section:analysis}

We now present the analysis of the \counting{} and \permutation{} tests in context of absolute  and relative bias problems.

%%%%%%%%%%%%%%%%%%%%%%%%%%%%%%%%%%%%%%%%%%%%%%%%%%%%%%%%%%%%%%%%%%%%

\subsection{Absolute bias problem}
\label{section:analysis:absolute}

We begin our analysis from the absolute bias problem and first formulate the main theorem of this section.

\begin{theorem}
\label{prop:absolute}
For any significance level $\level \in (0, 1)$, under the setup of the absolute bias problem (Problem~\ref{problem:absolute}), let the experiment be organized according to Procedure~\ref{algo:experiment}. Then the \permutation{} and \counting{} tests are guaranteed to control for Type-I error at the level $\level$, and also satisfy the requirement of non-trivial power.
\end{theorem}

\begin{figure}[t]
    \centering
    \subfloat[\label{fig:power:comparison:hard} DB reviewers estimate true scores with some noise in presence of correlations.
    ]{{\includegraphics[width=5.5cm]{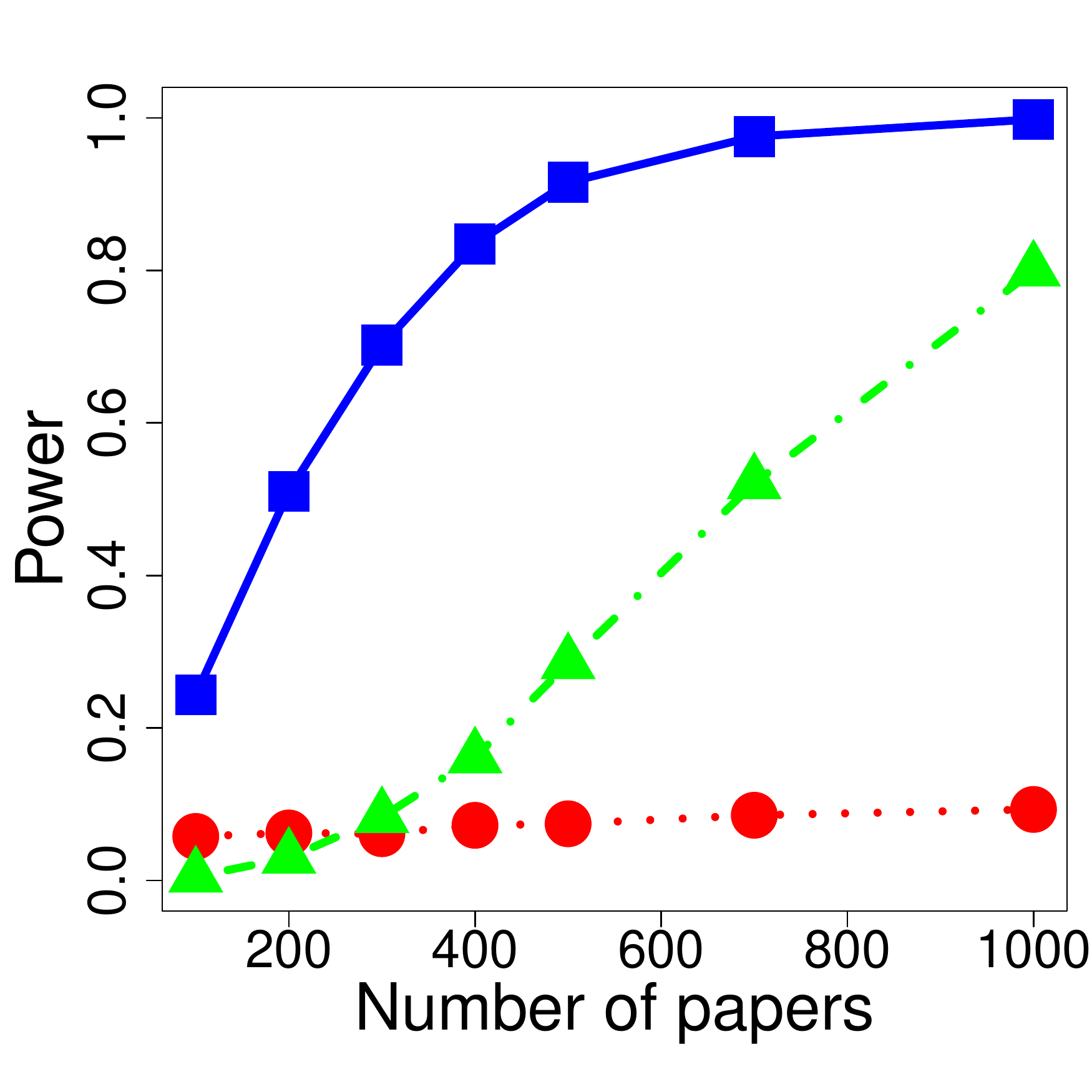}}}%
    \hspace{2cm}% 
    \subfloat[\label{fig:power:comparison:easy} DB reviewers estimate true scores without noise and correlations are absent.
    ]{{\includegraphics[width=5.5cm]{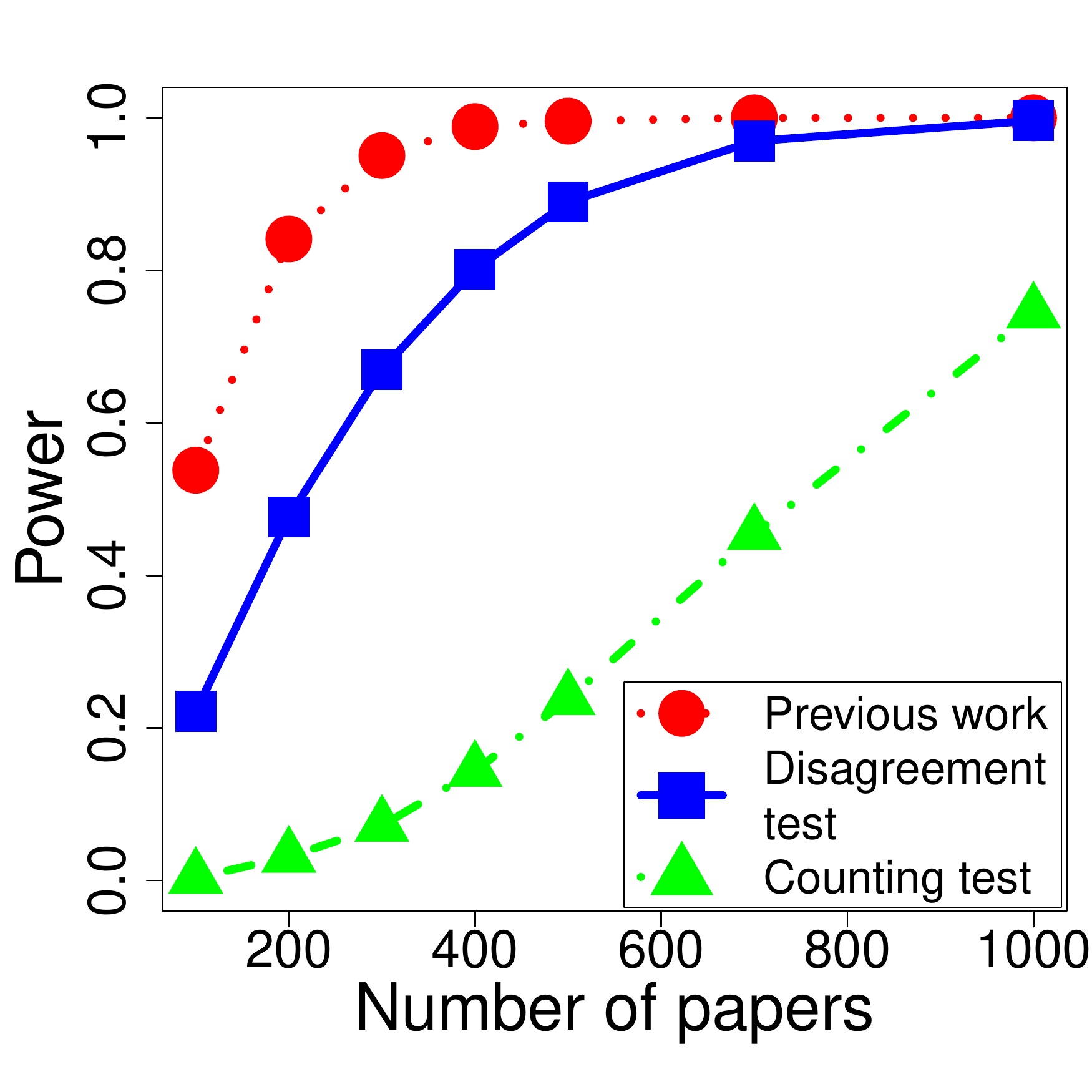}}}%
    \caption{Synthetic simulations evaluating performance of the test in \tomkins{} (``previous work'') and the tests introduced in this section (``\permutation{} and \counting{} tests'') in presence of bias when logistic model~\eqref{eqn:logregmodel} of reviewers is correct. Larger values are better. Details of the simulation setup are provided in Appendix~\ref{appendix:simulations:power}. Error bars are too small to be visible.}%
    \label{fig:power:comparison}
\end{figure}

\begin{remark*}
    1. As demonstrated in Figure~\ref{fig:power:comparison}, In practice, the \permutation{} test has a higher power as compared to the~\counting{} test and should be employed under conditions of the absolute bias problem. 
    
    2. Notice that the outcomes of the \permutation{} and \counting{} tests depend on a set $\settuple$ provided to the tests as input. That is, for two different sets $\settuple_1$ and $\settuple_2$ (that for example correspond to different assignments $\assignment^{\ast}_1$ and $\assignment^{\ast}_2$ constructed in Step~\ref{step:assign:selection} of Procedure~\ref{algo:experiment}), the outcomes of the tests might be different. Hence, one should fix a set $\settuple$ before observing reviewers' decisions to avoid chasing statistical significance.
    
    3. Finally, the \permutation{} and \counting{} tests are also applicable to the experimental procedure used by~\tomkins{} and are guaranteed to be robust to issues (a)-(c) from Section~\ref{section:regression:test}. The formal statement is given in Appendix~\ref{appendix:matching}.
\end{remark*}

\smallskip

We now discuss the issues (a)-(e) considered in Section~\ref{section:regression} in the context of our \permutation{} and \counting{} tests. 

\begin{itemize}[leftmargin=*, itemsep=1pt, topsep=0pt]
    \item \textbf{Noise.} The \permutation{} and \counting{} tests do not rely on any estimation of papers' qualities made by reviewers. Moreover, we do not even assume that there exists some objective quantity that can be estimated. Hence, our tests do not suffer from issues caused by noisy estimates of scores given by DB reviewers as illustrated by Figure~\ref{fig:error} in case of the \permutation{} test.

    \item \textbf{Model mismatch.} The only assumption we make is that under correct null hypothesis there is no difference in behavior of SB and DB reviewers. Hence, Theorem~\ref{prop:absolute} guarantees that our tests are robust to violations of specific parametric model~\eqref{eqn:logregmodel} as illustrated by Figure~\ref{fig:mismatch}.
    
    \item \textbf{Reviewer calibration.} We circumvent the detrimental effect of correlations introduced by reviewers' calibration by requiring that each reviewer contributes at most one review to the test. See Figure~\ref{fig:revsignal} for an illustration. Of course, such robustness comes at the cost of some power, but we notice that our matching procedures guarantee the use of at least a constant fraction of available data, thereby limiting reduction in the power.
    
    \item \textbf{Non-blind bidding.} The issue with bidding is straightforwardly resolved by requesting blind bidding from both SB and DB reviewers. As illustrated by Figure~\ref{fig:bidding}, we abstract out possible confoundings due to difference in bidding behaviour and ensure that the observed difference in decisions (if any) is due to bias in evaluations and not in bidding. 
    
    \item \textbf{Reviewer assignment.} The experimental design proposed in Procedure~\ref{algo:experiment} allows to execute any assignment algorithm without breaking the guarantees of the tests as demonstrated in case of TPMS assignment by Figure~\ref{fig:assignment}. The key part of the procedure that ensures such robustness are Steps~\ref{step:assign:selection} and \ref{step:assign:allocation} where we first find triples of two reviewers and a paper they review and then randomly allocate one reviewer from each triple to SB and one to DB. In this manner, we ensure that parts of the final assignment $\assignment$ that are used for testing do not exhibit any structural difference caused by non-random assignment.
\end{itemize}

We conclude the section with brief discussion of the power of the tests we introduced in this section. To provide fair comparison of tests, the simulations are performed under the setup of the experiment by~\tomkins{} and the input to the  \permutation{} and \counting{} tests is computed according to procedure described in Appendix~\ref{appendix:matching}. Figure~\ref{fig:power:comparison} contrasts the powers of the tests under the logistic model~\eqref{eqn:logregmodel} in two cases. In Figure~\ref{fig:power:comparison:hard}, DB reviewers estimate the true scores of the papers with some noise, which in presence of correlation dramatically decreases the power of the test by~\shorttomkins{} As discussed above, the \permutation{} and \counting{} tests are robust to issues caused by measurement error, as seen in Figure~\ref{fig:power:comparison:hard}. 

In contrast, Figure~\ref{fig:power:comparison:easy} considers the case when DB reviewers estimate true scores without noise, that is, true scores are known to the~\shorttomkins{} test. Although in this case their test has the highest power among 3 tests under consideration, we notice that the margin between their test and the \permutation{} test is not as large as in Figure~\ref{fig:power:comparison:hard}. Notice also that the test by~\shorttomkins{} gains power by overfitting to the strict model~\eqref{eqn:logregmodel} which leads to higher power when the model is correct, but at the cost of not being able to control over Type-I error rate under reasonable violations of the modelling assumptions as discussed in Section~\ref{section:regression}.

Finally, notice that the \counting{} test relies on the sub-Gaussian approximation to define a threshold and consequently has lower power than the \permutation{} test in both cases. While this suggests that for absolute bias problem the \permutation{} test dominates the \counting{} test, we will show in the next section that under the relative bias problem these tests are incomparable.

%%%%%%%%%%%%%%%%%%%%%%%%%%%%%%%%%%%%%%%%%%%%%%

\subsection{Relative bias problem}
\label{section:analysis:relative}

In Section~\ref{section:analysis:absolute} we showed that if under the absence of bias the behaviour of reviewers in SB and DB conditions is the same, then the \permutation{} and \counting{} tests control for Type-I error and have non-trivial power thus leading to a reliable testing procedure. In this section we relax that assumption and consider a relative bias problem with two specific choices of $\subclass$ that correspond to popular linear and logistic models. We first show that for each of these choices either the \permutation{} or the \counting{} test leads to reliable testing. We then conclude the analysis with a negative result saying that no test can control for Type-I error and have non-trivial power over both choices of $\subclass$.

Let us now introduce these two choices of $\subclass$ that we consider throughout the remaining part of this section. To this end, for each $j \in [\numpap]$ we let $\paprepresentation_j \in \reals$ denote an unknown ``representation'' of the paper $j$, where by representation we imply any function of a paper's content that defines reviewers' perception of a paper. For example, it could be that $\paprepresentation_j = \tscore_j$, where $\tscore_j$ is a true score as defined by~\shorttomkins{}

\smallskip

\noindent \textbf{Generalized linear model.} Under the generalized linear model, the SB condition itself induces a change in reviewers' evaluations, making them more harsh (or lenient). Moreover, under absence of bias the change in behaviour between SB and DB conditions is described by a constant shift in probability of acceptance for all papers, irrespective of whether they satisfy a property of interest or not. Formally, consider a fixed constant $\gap \in (0, 0.5)$. The generalized linear model assumes that (i) for every $j \in [\numpap]$  a corresponding representation  $\paprepresentation_j$ belongs to the interval $(\gap, 1 - \gap)$ and (ii) under absence of bias for every $(i, j) \in [\numrev] \times [\numpap]$ the behavior of reviewer $i$ if she/he reviews paper $j$ is described by the following parametric equations:
\begin{subequations}
\begin{align}
     \label{eqn:genlinregmodel} &\text{DB: }  \dbpacc_{ij} =  \paprepresentation_j \\ & \label{eqn:genlinregmodelsb} \text{SB: }  \sbpacc_{ij} =  \shift + \paprepresentation_j,
\end{align}
\end{subequations}
for some unknown constant $\shift \in (-\gap, \gap)$. Provided that matrix $\dbmatrix$ was generated according to the model~\eqref{eqn:genlinregmodel} of DB reviewers, the generalized linear model corresponds to an instance of a relative bias problem with a set of functions $\linmodel$ associated to a fixed constant $\gap$ and defined as
\begin{align}
\label{eqn:linmember}
    \linmodel = \left\{ \chng(t) : (\gap, 1 - \gap) \to [0, 1] \ \Big | \  \shift \in (-\gap, \gap) \right\},
\end{align}
where 
\begin{align}
\label{eqn:linf}
    \chng(t) = t + \shift, \ \ t \in (\gap, 1 - \gap). 
\end{align}
Indeed, observe that if for any  $(i, j) \in [\numpap] \times [\numrev]$ the probability of acceptance $\dbpacc_{ij}$ was generated according to the model~\eqref{eqn:genlinregmodel}, then $\sbpacc_{ij}$ defined as $\sbpacc_{ij} = \chng(\dbpacc_{ij})$ satisfies model~\eqref{eqn:genlinregmodelsb}. 

\smallskip

\noindent \textbf{Generalized logistic model.} Similar to the generalized linear model, under the generalized logistic model the SB condition also induces a change in reviewers' evaluations, but the change now is described as a constant shift in space of log-odds of the acceptance probabilities. Formally, consider a fixed constant $\bound > 0$. The generalized logistic model assumes that (i) for every $j \in [\numpap]$ a corresponding representation  $\paprepresentation_j$ belongs to the interval $(-\bound, \bound)$ and (ii) under absence of bias for every $(i, j) \in [\numrev] \times [\numpap]$ behaviour of reviewer $i$ if she/he reviews paper $j$ is described by the following parametric equations:
\begin{subequations}
\begin{align}
    \label{eqn:genlogregmodel} &\text{DB:} \log \frac{\dbpacc_{ij}}{1 - \dbpacc_{ij}} = \coef_0 + \coef_1 \paprepresentation_j \\
    \label{eqn:genlogregmodelsb} &\text{SB:} \log \frac{\sbpacc_{ij}}{1 - \sbpacc_{ij}} = \coef_0 + \logshift + \coef_1 \paprepresentation_j,
\end{align}
\end{subequations}
for some unknown constant $\logshift \in (-\bound, \bound)$, where unknown coefficients $\coef_0$ and $\coef_1$ are also bounded in absolute value by $\bound$ and $\coef_1 > 0$. Provided that matrix $\dbmatrix$ is generated according to the model~\eqref{eqn:genlogregmodel} of DB reviewers, one can verify that the generalized logistic model corresponds to an instance of a relative bias problem with a set of functions $\logregclass$ associated to a fixed constant $\bound$ and defined as
\begin{align}
\label{eqn:logmember}
    \logregclass = \left\{ \logchng(t) : [0, 1] \to [0, 1] \ \Big | \  \logshift \in (-\bound, \bound) \right\},
\end{align}
where
\begin{align}
\label{eqn:logf}
    \logchng(t) = \frac{t e^{\logshift}}{1 - t + te^{\logshift}}, \ \ t \in [0, 1]. 
\end{align}
Indeed, observe that if for some $(i, j) \in [\numpap] \times [\numrev]$ the probability that reviewer $i$ accepts paper $j$ in DB setup $\dbpacc_{ij}$ is generated according to the model~\eqref{eqn:genlogregmodel}, then setting $\sbpacc_{ij} = \logchng(\dbpacc_{ij})$ we ensure that $\sbpacc_{ij}$ satisfies model~\eqref{eqn:genlogregmodelsb}.

\smallskip

The models we defined follow an objective parametric approach assumed by~\citet{tomkins17wsdm} with two differences: (i) we do not assume that $\paprepresentation_j$ has a known meaning or that it can be measured (for instance, it may be that $\paprepresentation_j = \tscore_j$, or that $\paprepresentation_j = (\tscore_j)^3$, or $\paprepresentation_j$ may be a complex function of the content of the paper) and (ii) we do not assume that the bias is described by a linear shift in space of probabilities or log-odds, and instead consider a non-parametric definition of the bias as specified in the alternative hypothesis~\eqref{eqn:relative:alt}.

\subsubsection{Positive results}
\label{section:analysis:relative:positive}

We now show that the \counting{} and \permutation{} tests lead to reliable testing under the generalized linear and generalized logistic models respectively. Before we formulate the main result of this section, let us provide some intuition behind the models and the corresponding tests.

\smallskip

\noindent{\textbf{Generalized linear model.}} A natural strategy to test for biases under the generalized linear model is to estimate the shift in reviewers' behaviour on papers that belong to $\indset$ and to $\noindset$ separately and then compare the estimates. In fact, the \counting{} testing procedure introduced above as Test~\ref{test:counting} follows this strategy and, as guaranteed by Theorem~\ref{prop:relative}, leads to reliable testing under the generalized linear model.

\smallskip

\noindent{\textbf{Generalized logistic model.}} Intuitively, estimating a constant shift in reviewers' behavior as done by the \counting{} test may not be the optimal strategy under the generalized logistic model, as under absence of bias the change of behaviour between SB and DB conditions is given by a constant shift in \emph{log-odds} space and not in \emph{probability} space. However, it turns out that the \permutation{} test is able to capture the constant shift in \emph{log-odds} space and hence we still can perform reliable testing under this model, using the \permutation{} algorithm.

\begin{theorem}
\label{prop:relative}
For any significance level $\level \in (0, 1)$, let the experiment be organized according to Procedure~\ref{algo:experiment}. Then
\begin{enumerate}[label=(\alph*), itemsep=1pt]
    \item \label{prop:relative:counting} Under the generalized linear model with any $\gap \in (0, 0.5)$, the \counting{} test is guaranteed to control for Type-I error at the level $\level$, and also satisfies the requirement of non-trivial power.
    
    \item \label{prop:relative:permutation} Under the generalized logistic model with any $\bound > 0$, the \permutation{} test is guaranteed to control for Type-I error at the level $\level$, and also satisfies the requirement of non-trivial power.
\end{enumerate} 
\end{theorem}
\begin{remark*}

1. Result of Theorem~\ref{prop:relative}(a) holds even for a subjective version of the generalized linear model in which for each $(i, j) \in [\numpap] \times [\numrev]$ we substitute $\paprepresentation_j$ with $\paprepresentation_{ij}$ (that is, different values across reviewers) in equations~\eqref{eqn:genlinregmodel} and~\eqref{eqn:genlinregmodelsb}, thereby accounting for subjectivity of reviewers. 

2. If the logistic model~\eqref{eqn:logregmodel} assumed by~\shorttomkins{} is correct for both SB and DB reviewers with possibly different intercepts, then Theorem~\ref{prop:relative}(b) ensures that the \permutation{} test provably controls for the Type-I error and can detect a bias with probability that goes to 1 as sample size grows, without requiring knowledge (neither exact nor approximate) of papers' scores $\tscore_1,\ldots,\tscore_\numpap$.

3. Notice that the \counting{} and \permutation{} tests do not require the knowledge of $\gap$ and $\bound$ parameters to control for Type-I error and satisfy the requirement of non-trivial power.

\end{remark*}

\begin{figure}[t]
    \centering
    \subfloat[\label{fig:relative:linear} Instance of generalized linear model.
    ]{{\includegraphics[width=5.5cm]{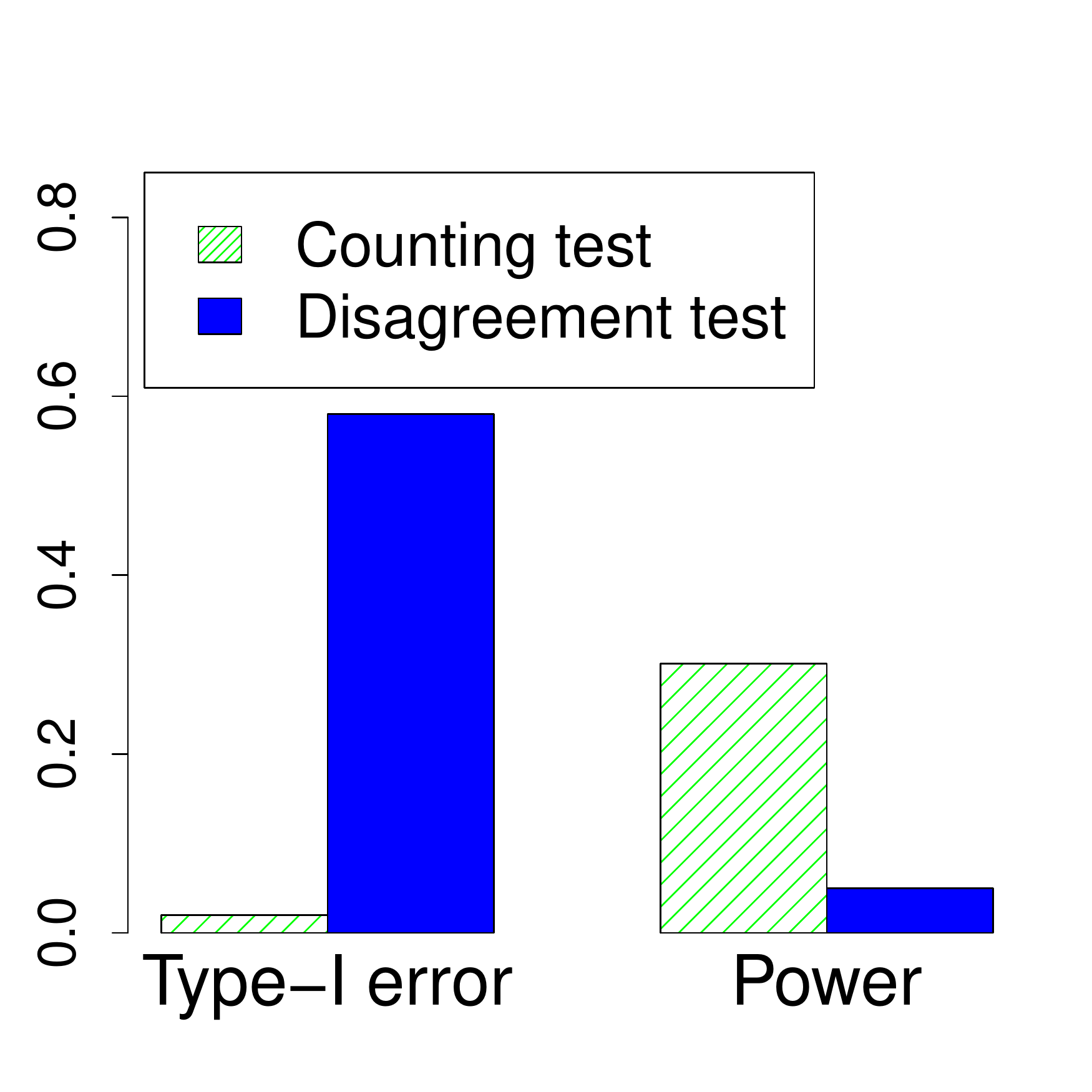}}}%
    \hspace{2cm}% 
    \subfloat[\label{fig:relative:logistic} Instance of generalized logistic model.
    ]{{\includegraphics[width=5.5cm]{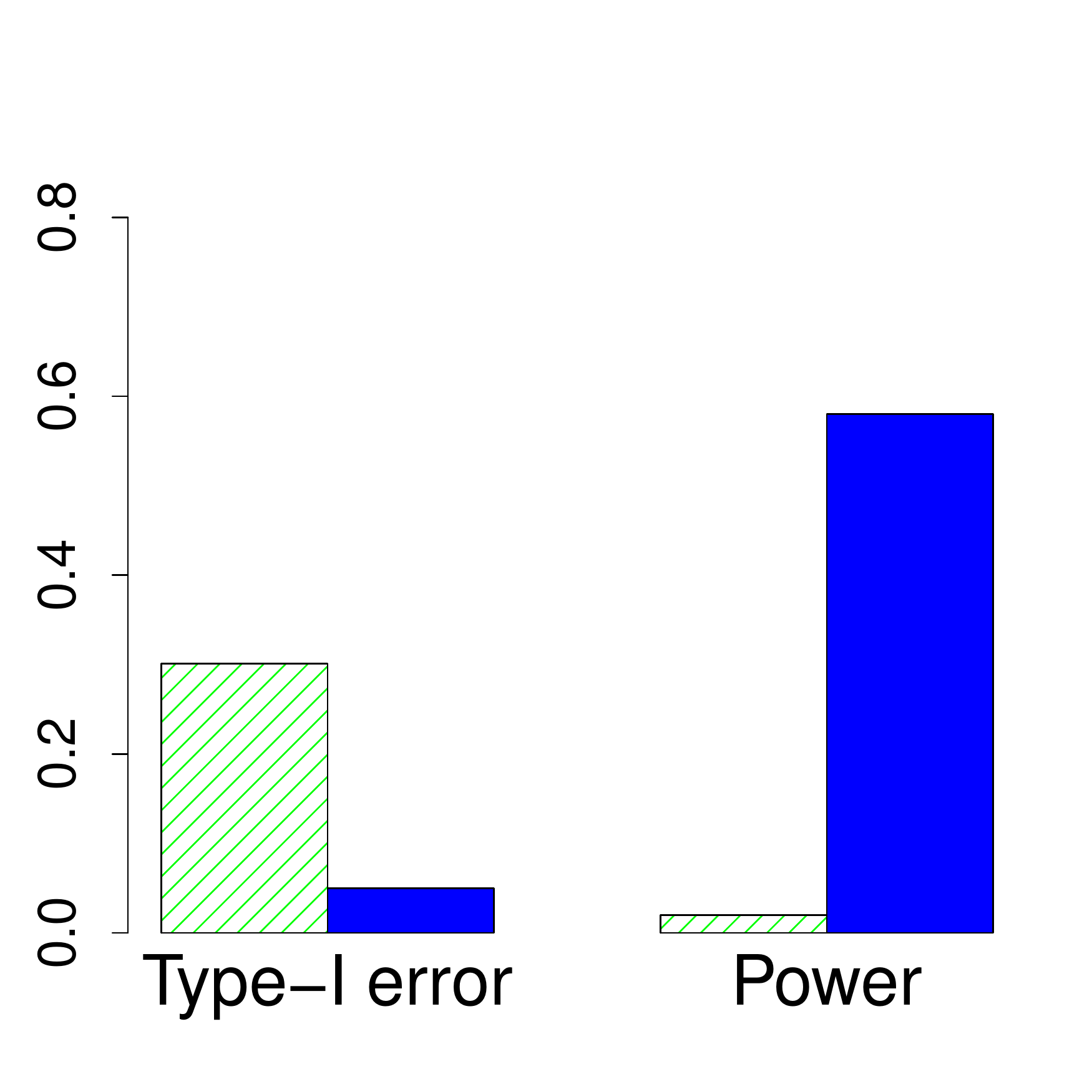}}}%
    \caption{Synthetic simulations evaluating performance of the \permutation{} and \counting{} tests under setup of the relative bias problem when generalized linear (Figure~\ref{fig:relative:linear}) or generalized logistic (Figure~\ref{fig:relative:logistic}) models is correct. For each model presence and absence of bias are simulated. Details of the simulation setup are provided in Appendix~\ref{appendix:simulations:relative}. Error bars are too small to be visible.}%
    \label{fig:relative}%
\end{figure}

Figure~\ref{fig:relative} compares the performance of the \permutation{} and \counting{} tests under specific instances of the generalized linear and logistic models, illustrating the results of Theorem~\ref{prop:relative}. Figure~\ref{fig:relative:linear} shows that the \counting{} test controls for Type-I error and has a non-trivial power under the generalized linear model. Notice that the \permutation{} test does not control for Type-I error in this instance, implying that Theorem~\ref{prop:relative}(b) cannot be extended to guarantee the Type-I error control under the generalized linear model as well. Symmetrically, Figure~\ref{fig:relative:logistic} shows that while the \permutation{} test leads to reliable testing under the generalized logistic model, the \counting{} test is unable to control for Type-I error under this model. Hence, we conclude that under the setup of the relative bias problem, none of the tests dominates another, each leading to reliable testing under the corresponding model.

We see in Figure~\ref{fig:relative} that neither the \permutation{} nor the \counting{} test is suitable for \emph{both} generalized linear and logistic models. In the next section we show that this is not a drawback of these specific tests, but rather a manifestation of a more general impossibility result. 

%%%%%%%%%%%%%%%%%%%%%%%%%%%%%%%%%%%%%%%%%%%%%%

\subsubsection{Negative result}
\label{section:analysis:relative:negative}
We conclude our analysis with a negative result that limits the complexity of the class $\subclass$ for which reliable testing in the relative bias problem is possible. Let us first state the main result of this section.

\begin{theorem}
\label{prop:richness}
        Suppose that there exist two functions $g, h \in \subclass$  and some $0 \le x_1 < x_2 \le 1$ such that $g(x_1) < h(x_1)$ and $g(x_2) > h(x_2)$.
        Suppose also that there exists a testing procedure $\procedure$ operating on decisions of SB and DB reviewers that for any given $\level \in (0, 1)$ keeps Type-I error below $\level$ for all matrices $\sbmatrix, \dbmatrix$  that satisfy the null hypothesis of Problem~\ref{problem:relative} specified by some function $\nullshift \in \subclass$. Then the testing procedure $\procedure$ cannot satisfy the requirement of non-trivial power.
\end{theorem}

The intuition behind Theorem~\ref{prop:richness} is as follows. If the class $\subclass$ contains ``too many'' functions, then some matrices $\sbmatrix$ and $\dbmatrix$ satisfy the null hypothesis of Problem~\ref{problem:relative} defined by $\nullshift \in \subclass$ and simultaneously satisfy the alternative hypothesis~\eqref{eqn:relative:alt} defined by some other function $\nullshift' \in \subclass$ with margin $\minigap > 0$. Hence, any testing procedure must either have high Type-I error rate or sacrifice the non-trivial power requirement over the class of functions $\subclass$, implying that reliable testing is impossible.

Using Theorem~\ref{prop:richness} we can deduce that one can hope to control for the Type-I error rate and simultaneously have a non-trivial power only when functions contained in $\subclass$ are pointwise totally ordered, that is, for any two functions $f, g \in \subclass$ it must be the case that either $f(x) \ge g(x)$ for all $x \in [0, 1]$ or $f(x) \le g(x)$ for all $x \in [0, 1]$.

Let us now illustrate the consequences of Theorem~\ref{prop:richness} for the generalized linear and logistic models. 

\begin{corollary}
\label{prop:joint}
For any significance level $\level \in (0, 1)$, let $\procedure_1$ and $\procedure_2$ be any testing procedures which operate on decisions of SB and DB reviewers. Suppose that under the generalized linear model with any $\gap \in (0, 0.5)$, procedure $\procedure_1$ controls for the Type-I error at the level $\level$ and satisfies the non-trivial power requirement. Suppose also that under the generalized logistic model with any $\bound > 0$, procedure $\procedure_2$ controls for the Type-I error at the level $\level$ and satisfies the non-trivial power requirement. Then 
\begin{enumerate}[label=(\alph*), itemsep=1pt]
\item Under the generalized logistic model with any $\bound > 0$, procedure $\procedure_1$ incurs a Type-I error rate strictly greater than $\level$.

\item Under the generalized linear model with any $\gap \in (0, 0.5)$, procedure $\procedure_2$ incurs a Type-I error rate strictly greater than $\level$.
\end{enumerate}
\end{corollary}

Corollary~\ref{prop:joint} shows that there does not exist a testing procedure that controls over Type-I error and has non-trivial power under both generalized linear and generalized logistic models. As a result, one needs to design different procedures for these models as we did with the \permutation{} and \counting{} tests. In case of the \permutation{} and \counting{} tests, Corollary~\ref{prop:joint} is illustrated by Figure~\ref{fig:relative}.

In Appendix~\ref{appendix:logreg} we discuss another application of Theorem~\ref{prop:richness} in context of the generalized logistic model that also suggests that generality of the \permutation{} test cannot be further increased.

%%%%%%%%%%%%%%%%%%%%%%%%%%%%%%%%%%%%%%%%%%%%%%%%%
\section{Discussion}
\label{section:discussion}
Peer review is the backbone of academia but faces a number of challenges of unfairness, biases, and inefficiency. This work contributes to the growing literature~\citep{shah2017design, kang18peerread, gao19rebuttal, wang18calibration, stelmakh2018pr4a, kobren19localfairness, noothigattu2018choosing,balietti2016peer,xu2018strategyproof,fiez2019super} in the domain of addressing these challenges in peer review, by designing a principled method to test for biases.   We show that under various conditions the approach used by the prior work of~\shorttomkins{} does not control the Type-I error rate. We underscore that we do not aim at confirming or disproving the presence of biases found in that work, but our focus is on the validity of testing methods. With this goal in mind, we propose a principled approach to testing for biases and design two statistical procedures that coupled with our novel experimental setup provably control for the Type-I error rate. Additionally, these procedures have non-trivial power under essentially a single assumption of no difference in the behavior of SB and DB reviewers when the bias is absent. We then show that this assumption cannot be relaxed in general and that to accommodate the aforementioned difference in behavior one needs to make some modelling assumptions, as we demonstrated with our tests and generalizations of popular linear and logistic models. 

We presented the \permutation{} and \counting{} tests in the context of peer review. However, we underscore that one can adapt our experimental setup (Procedure~\ref{algo:experiment}) to use our testing procedures (Test~\ref{test:permutation} and Test~\ref{test:counting}) in other applications. These applications include peer grading, university admission, and hiring where some protected attributes might be available to reviewers. 

There are several open problems suggested by our work. The first direction is associated with the statistical power of the testing procedures we propose. In this work, we show that our tests have power that going to one under certain conditions on the alternative. It is of interest to establish a bound on the statistical power of our tests in a finite sample setting and compare it with an upper bound on the maximum power that can be achieved by any computationally-efficient testing procedure. 

The second direction is related to the design of the experimental procedure. To accommodate tests for biases, one needs to deviate from the standard peer-review pipeline, thus introducing a trade-off between the quality of the peer-review process and the accuracy of the testing. Quantification of such a trade-off may help to design a better setup and understand the cost of the experiment in terms of the peer-review quality. In this work, we designed a procedure that leads to the desired accuracy, but is suboptimal in terms of the TPMS objective. In contrast, the optimal TPMS assignment would not allow to perform reliable testing. Hence, an open problem is to design an experimental procedure that accommodates our statistical tests and subject to this maximizes the quality of the assignment in terms of the TPMS objective.

%%%%%%%%%%%%%%%%%%%%%%%%%%%%%%%%%%%%%%%%%%%%%%

\bibliographystyle{apalike}
\bibliography{bibtex}

\bigskip

\noindent {\Large{\textbf{Appendix}}}
\appendix

\bigskip

\noindent We provide supplementary materials and additional discussion.

%%%%%%%%%%%%%%%%%%%%%%%%%%%%%%%%%%%%%%%%%%%%%%%%%%%%%%%%%%%%%

\section{More than one property of interest}
\label{appendix:morethanone}

Throughout the main body of the paper, we considered the case of a single property of interest. We now generalize some of the results to the case of more than one property of interest. First of all, we recall some notation. Let $\numcat$ be the total number of properties, then for each paper $j \in [\numpap]$, we let variables $\ind_j^{(1)}, \ind_j^{(2)}, \ldots, \ind_j^{(\numcat)}$ indicate whether or not the paper satisfies the corresponding property. For each property $\ell$, the set $\indset_{\ell} \subseteq [\numpap]$ contains papers that satisfy a property $\ell$ with $\noindset_{\ell}$ being its complement.

We argue that when $\numcat > 1$, one needs to think about the bias testing problem as of an instance of the relative bias problem as defined in Section~\ref{section:framework:relative}. Indeed, consider for example the case of two properties of interest ($\numcat = 2$) and assume that we are interested in testing for biases with respect to the first property. Then even if there is no bias with respect to this property, the behavior of reviewers between SB and DB conditions might be different due to possible biases with respect to the second property.

\newcommand{\vectind}{\mathbf{\ind}}

The negative result of Theorem~\ref{prop:richness} we established in Section~\ref{section:analysis:relative:negative} also applies to the case of multiple properties, implying that reliable testing is possible only under some restrictions on the difference in reviewers' behavior between SB and DB conditions under the absence of bias. Following the relative bias problem defined as Problem~\ref{problem:relative}, we now generalize it for $\numcat > 1$. To this end, we consider a problem of testing for biases with respect to the property $\ell \in [\numcat]$ and introduce an additional piece of notation. For each paper $j \in [\numpap]$, let $\vectind_j$ denote a vector of indicators of property satisfaction: $\vectind_j = (\ind^{(1)}_j, \ind^{(2)}_j, \ldots, \ind^{(\numcat)}_j)$ and let $\vectind_j^{(-\ell)}$ denote the same vector but with $\ell^{\text{th}}$ component omitted, that is, $\vectind_j^{(-\ell)} = (\ind^{(1)}_j, \ldots, \ind^{(\ell - 1)}_j, \ind^{(\ell + 1)}_j, \ldots, \ind^{(\numcat)}_j)$. 

Following the definition of the relative bias problem (Problem~\ref{problem:relative}), the set $\subclass_{\ell}$ contains functions that under the absence of bias with respect to the property $\ell$, specify the difference in behavior between DB and SB conditions. In case of a single property of interest, $\subclass$ was a subset of all non-decreasing functions $f: [0, 1] \to [0, 1]$. However, when $\numcat > 1$, even under the absence of the bias with respect to the property $\ell$, the change in reviewers' behavior between DB and SB conditions may be influenced by whether the paper satisfies properties other than $\ell$, due to possible biases with respect to these properties. Hence, under the absence of bias with respect to the property $\ell$, the change of behavior between SB and DB conditions is described as follows:
\begin{align*}
    \forall (i, j) \in [\numpap] \times [\numrev] : \ \sbpacc_{ij} = \nullshift(\dbpacc_{ij}, \vectind_j^{(-\ell)}),
\end{align*}
where function $\nullshift \in \subclass_{\ell}$ is non-decreasing in its first argument. Thus, the set $\subclass_{\ell}$ is a subset of all functions with domain $[0, 1] \times \{0, 1 \}^{\numcat - 1}$ which are non-decreasing in their first argument, that is,
\begin{align*}
    \subclass_{\ell} \subseteq \left\{f: [0, 1] \times \{0, 1 \}^{\numcat - 1} \to [0, 1] \Big | f \text{ is non-decreasing in its first argument} \right\}.
\end{align*}
Having defined the necessary notation, we are ready to introduce the relative bias problem in case of multiple properties of interest.

\begin{problem}[Relative bias problem for multiple properties.] \label{problem:relative:general} Given significance level $\level \in (0, 1)$, the property of interest $\ell$, the class of functions $\subclass_{\ell}$ and decisions of SB and DB reviewers, the goal is to test the following hypotheses:
\begin{align}
   & \nonumber \hypothesis_0 : \forall i \in [\numrev] \ \forall j \in [\numpap] \ \ \sbpacc_{ij} = \nullshift(\dbpacc_{ij}, \vectind_j^{(-\ell)}) \\
   & \label{eqn:relative:general:alt} \hypothesis_1 : \forall i \in [\numrev] \ \forall j \in [\numpap] \   \begin{cases}
                                                \sbpacc_{ij} \ge \nullshift(\dbpacc_{ij}, \vectind_j^{(-\ell)}) & \text{if } j \in \indset_{\ell} \\
                                                \sbpacc_{ij} \le \nullshift(\dbpacc_{ij}, \vectind_j^{(-\ell)}) & \text{if } j \notin \indset_{\ell} 
                                             \end{cases},
\end{align}
for some unknown $\nullshift \in \subclass_{\ell}$, and where at least one inequality in the alternative hypothesis~\eqref{eqn:relative:general:alt} is strict. 
\end{problem}

Intuitively, under the null hypothesis of absence of bias with respect to the property $\ell$, the change of reviewers' behaviour between SB and DB conditions is determined by (i) bias introduced by SB condition itself which is independent of papers' authorship information and (ii) bias with respect to properties other than the property $\ell$. The generalized linear and logistic models can be formulated in case of multiple properties of interest as follows.

\smallskip

\noindent{\textbf{Generalized linear model.}} Given a fixed constant $\gap \in (0, 0.5)$, we follow the case of a single property and assume that each paper $j \in [\numpap]$ has some unknown representation $\paprepresentation_j \in (\gap, 1 - \gap)$. The generalized linear model assumes that for each $(i, j) \in [\numrev] \times [\numpap]$, the behaviour of reviewer $i$ if she/he reviews paper $j$ is described by the following parametric equations:
\begin{subequations}
\begin{align}
\label{eqn:glinmp:db}    &\text{DB: }  \dbpacc_{ij} =  \paprepresentation_j \\   \label{eqn:glinmp:sb} &\text{SB: }  \sbpacc_{ij} =  \paprepresentation_j + \sbcoef_0 + \sum\limits_{\ell \in [\numcat]} \sbcoef_{\ell} \ind_{j}^{(\ell)},
\end{align}
\end{subequations}
where unknown coefficients are such that $|\sbcoef_0| + \sum\limits_{\ell= 1}^{\numcat} |\sbcoef_{\ell}| < \gap$. Under the generalized linear model, a bias with respect to the property $\ell$ is present whenever $\sbcoef_{\ell} \ne 0$.

\smallskip

\noindent{\textbf{Generalized logistic model.}} Given a fixed constant $\bound > 0$, the generalized logistic model assumes that (i) for every $j \in [\numpap]$, a corresponding representation $\paprepresentation_j$ belongs to the interval $(-\bound, \bound)$ and (ii) for each $(i, j) \in [\numrev]\times[\numpap]$, the behaviour of reviewer $i$ if she/he reviews paper $j$ is described by the following parametric equations:
\begin{subequations}
\begin{align}
     \label{eqn:glogmp:db} &\text{DB: } \ \log \frac{\dbpacc_{ij}}{1 - \dbpacc_{ij}} = \dbcoef_0 + \coef_1 \paprepresentation_j \\     \label{eqn:glogmp:sb}
 &\text{SB: } \ \log \frac{\sbpacc_{ij}}{1 - \sbpacc_{ij}} = \sbcoef_0 + \coef_1 \paprepresentation_j + \sum\limits_{\ell \in [k]} \sbcoef_{\ell + 1} \ind_{j}^{(\ell)},
\end{align}
\end{subequations}
where all coefficients are bounded in absolute value by $\bound$ and $\coef_1 > 0$.  Under the generalized logistic model, a bias with respect to the property $\ell$ is present whenever $\sbcoef_{\ell+1} \ne 0$.

\smallskip

\begin{remark*}
1. First, provided that matrix $\dbmatrix$ is generated according to the one of the introduced models, one can define a set of functions $\subclass_{\ell}$ that puts the corresponding model in the context of the relative bias problem for multiple properties of interest defined as Problem~\ref{problem:relative:general}.

2. The goal under each of the models introduced above is to test the significance of the coefficient in equation describing the behavior of SB reviewer that corresponds to the indicator of the property of interest. For example, if we are interested in testing for biases with respect to the property $\ell$ under the generalized logistic model, then we want to test the significance of the coefficient $\sbcoef_{\ell + 1}$ in equation~\eqref{eqn:glogmp:sb}. 

3. Notice that in case of multiple properties, the models we introduced above describe reviewers' behaviour both under the absence of bias and under the presence of bias. In this way they allow simultaneous testing for biases with respect to many properties of interest.
\end{remark*}

Observe that the relationships that describe the behaviour of SB reviewers in models~\eqref{eqn:glinmp:sb} and~\eqref{eqn:glogmp:sb} are reminiscent of the linear regression and logistic regression models respectively. As mentioned above, one cannot fit decisions of SB reviewers to these models using existing methods, because one of the covariates (paper representation $\paprepresentation$) is unknown. In their work, \shorttomkins{} employed DB reviewers to estimate this unknown covariate and used these estimates to fit the logistic model. As we discussed in Section~\ref{section:regression}, this approach leads to an unreliable testing procedure under various realistic conditions.

We now show that using ideas of the \permutation{} and \counting{} tests, one can use decisions of both SB and DB reviewers to eliminate the unknown covariate from the model, thereby enabling standard tools without the need to estimate any covariate.

\begin{proposition}
\label{prop:clearq}
    Let reviewers $i$ and $i'$ be assigned to paper $j$ in SB and DB setups correspondingly, then

\begin{subequations}
    \begin{enumerate}[itemsep=1pt, label=(\alph*)]
    \item Under the generalized linear model the expectation of the quantity $\sbdecision_{ij} - \dbdecision_{i'j}$ follows the linear model~\eqref{eqn:glinmp:sb} with $\paprepresentation_j = 0$:
    \begin{align}
    \label{eqn:glinmanipulated}
        \expectation{\sbdecision_{ij} - \dbdecision_{i'j}} = \sbcoef_0 + \sum\limits_{\ell \in [\numcat]} \sbcoef_{\ell} \ind_{j}^{(\ell)}.
    \end{align}
    
    \item Under the generalized logistic model the expectation of the quantity $\sbdecision_{ij} | \left(\sbdecision_{ij} \ne \dbdecision_{i'j}\right)$ follows the logistic model~\eqref{eqn:glogmp:sb} with $\paprepresentation_j = 0$:
       \begin{align}
        \label{eqn:glogmanipulated}
        \log \frac{\expectation{\sbdecision_{ij} | \left(\sbdecision_{ij} \ne \dbdecision_{i'j}\right)}}{1 - \expectation{\sbdecision_{ij} | \left(\sbdecision_{ij} \ne \dbdecision_{i'j}\right)}} = \coef_0 + \sum\limits_{\ell \in [k]} \sbcoef_{\ell + 1} \ind_{j}^{(\ell)},
    \end{align}
    \end{enumerate}
    \end{subequations}
    where $\coef_0 = \sbcoef_0 - \dbcoef_0$.
\end{proposition}

Proposition~\ref{prop:clearq} provides a mean to eliminate the unknown covariate from the models of SB decisions~\eqref{eqn:glinmp:sb} and~\eqref{eqn:glogmp:sb} using decisions of DB reviewers. For example, in case of the generalized logistic model it relies on the core idea of the \permutation{} test and suggests conditioning on pairs (SB reviewer, DB reviewer) such that reviewers disagree in their decisions for some paper. After conditioning, decisions of SB reviewers follow model~\eqref{eqn:glogmanipulated} with all covariates known and hence standard test for logistic regression can be applied to evaluate significance of coefficients.

Proposition~\ref{prop:clearq} also allows to avoid using noisy measurements and hence any test for significance of the coefficients applied to the models~\eqref{eqn:glinmanipulated} and~\eqref{eqn:glogmanipulated} will not be susceptible to issues caused by the use of noisy measurements and misspecification of the meaning of $\paprepresentation$ (issues (a) and (b) from Section~\ref{section:regression}). If one restricts each reviewer to input at most one decision to the testing procedure, then issue (c) will also be mitigated.

%%%%%%%%%%%%%%%%%%%%%%%%%%%%%%%%%%%%%%%%%%%%%%%

\section{Our tests under the setup of~\shorttomkins{}}
\label{appendix:matching}

\newcommand{\sbassignment}{\assignment_{\text{SB}}}
\newcommand{\dbassignment}{\assignment_{\text{DB}}}

In this section we give additional comments on the applicability of our testing procedures to the setup of~\tomkins{}. To this end, recall that our tests take as input the set of tuples $\settuple$ such that (i) each tuple $\tuple \in \settuple$ is of the form  $\tuple = (j_{\tuple}, \sbdecision_{j_{\tuple}}, \dbdecision_{j_{\tuple}}, \ind_{j_t})$, where $j_{\tuple}$ is a corresponding paper, $\sbdecision_{j_t}, \dbdecision_{j_t}$ are decisions of SB and DB reviewers for this paper and $\ind_{j_t}$ equals $1$ if $j_{\tuple} \in \indset$ and $-1$ otherwise; (ii) each reviewer contributes at most one decision to the set $\settuple$. Potentially our tests can be coupled with any experimental procedure as long as this procedure enables a construction of such a set $\settuple$. However, one needs to understand that while Procedure~\ref{algo:experiment} is robust to issues we discussed in Section~\ref{section:regression:setup}, other experimental setups may lead to an inflation of the Type-I error. 

In the experiment conducted by~\tomkins{}, reviewers were split into two groups (SB and DB) uniformly at random at the very beginning of the experiment. Then two assignments $\sbassignment$ and $\dbassignment$ were computed separately for each group of reviewers. As discussed in Section~\ref{section:regression:setup}, even if both groups of reviewers bid in a blind manner, the design of~\shorttomkins{} may lead to inflated Type-I error. Nonetheless, we now show that even under the setup of~\shorttomkins, one can employ the \permutation{} and~\counting{} tests to fix issues (a)-(c) with the testing procedure discussed in Section~\ref{section:regression:test}.

\subsection{Matching algorithms}

Let us first introduce two matching procedures that construct an input for our tests under the setup of~\tomkins{}. Given assignment of papers to reviewers in both SB and DB conditions, we discuss two choices of matching algorithms depending on the relationship between parameters $\papload$ (required number of reviewers per paper in each condition) and $\revload$ (maximum number of papers per reviewer). Notice that our goal is not to maximize the size of $\settuple$, but instead to maximize the minimum of the number of papers from $\indset$ included in the set $\settuple$ and the number of papers from $\noindset$ included in the set $\settuple$. This is because our statistical tests need decisions for papers from both $\indset$ and $\noindset$ to maximize their power. Depending on the relationship between $\papload$ and $\revload$ we can solve this problem either exactly or approximately.

\smallskip

\noindent \underline{Case 1 ($\papload \ge \revload$).} In this case, each paper can be matched to 1 SB reviewer and 1 DB reviewer by finding two separate maximum matchings (papers to SB reviewers and papers to DB reviewers) using the Hungarian matching algorithm. We formally present the matching procedure as Algorithm~\ref{alg:exact}.

\newcommand{\matching}{\mathcal{M}}

{
\setcounter{algorithm}{0}
\floatname{algorithm}{Algorithm}
\begin{algorithm}[H]
   \caption{Exact matching algorithm}
   \label{alg:exact}
   {\bfseries Input:} Assignments $\sbassignment$, $\dbassignment$ of SB and DB reviewers to papers, respectively.\\
    \begin{enumerate}[leftmargin=*, topsep=0pt]
    
        \item \label{step:construction}Construct a graph $G$ that consists of 3 layers:
        \begin{itemize}[noitemsep, leftmargin=*]
            \item \textbf{Layer 1.} One node for each SB reviewer
            \item \textbf{Layer 2.} One node for each paper
            \item \textbf{Layer 3.} One node for each DB reviewer
        \end{itemize}
        and add edges between reviewers and papers according to assignments $\sbassignment$ and $\dbassignment$. Set $\settuple = \emptyset.$
        
        \item \label{step:exact:hungary} Using the Hungarian matching algorithm with uniform tie-breaking find matchings $\matching_{\text{SB}}$ and $\matching_{\text{DB}}$ where $\matching_{\text{SB}}$ (respectively $\matching_{\text{DB}}$)  is a maximum 1-1 matching between SB (respectively DB) reviewers and papers (each reviewer is matched to at most 1 paper and each paper is matched to at most 1 reviewer). 
    
        \item Leave in graph $G$ only those edges that correspond to matched pairs in $\matching_{\text{SB}}$ and $\matching_{\text{DB}}$.
    
        \item \label{step:exact:final} For any triple of (SB reviewer $i_1$, paper $j$, DB reviewer $i_2$) such that there is a path from a node that corresponds to reviewer $i_1$ to a node that corresponds to reviewer $i_2$ through a node that corresponds to paper $j$, add $t = (j, \sbdecision_{i_1 j}, \dbdecision_{i_2 j}, \ind_j)$ to $\settuple$.
        
        \item Return $\settuple$.
    \end{enumerate}
\end{algorithm}
}

\begin{lemma}
\label{lemma:exact}
   For any assignments of SB and DB referees to papers that satisfy ($\papload, \revload$)-load constraints with $\papload \ge \revload$,  the matching procedure in Algorithm~\ref{alg:exact} is guaranteed to construct a set of tuples $\settuple$ such that for each paper $j \in [\numpap]$ there is one tuple that corresponds to this paper.
\end{lemma}

\noindent \underline{Case 2 ($\papload < \revload$).} In this case we cannot use the above idea, because there does not exist a matching such that each paper is matched to one SB and one DB reviewer, subject to a constraint that each reviewer is matched with at most one paper. While solving the exact optimization problem in this case might be hard, a simple greedy procedure constructs a sufficiently large matching  for the \permutation{} and \counting{} tests to satisfy the non-trivial power requirement. The iterative greedy procedure in each iteration matches one paper from $\noindset$ and one paper from $\indset$ to 1 SB and 1 DB reviewer and removes those reviewers from subsequent interations to maintain the constraint that each reviewer contributes at most one decision to the set $\settuple$. We formally introduce the greedy procedure as Algorithm~\ref{alg:greedy}.

\begin{lemma}
\label{lemma:greedy}
    For any assignments of SB and DB referees to papers that satisfy ($\papload, \revload$)-load constraints, the matching procedure in Algorithm~\ref{alg:greedy} is guaranteed to construct a set of tuples $\settuple$ that for large enough $\min\{|\indset|, |\noindset| \}$ contains at least $\constant \min\{|\indset|, |\noindset| \}$ tuples corresponding to papers from $\indset$ and at least $\constant \min\{|\indset|, |\noindset|\}$ tuples corresponding to papers from $\noindset$, where $\constant$ is a constant that may depend only on $\papload$ and $\revload$.
\end{lemma}
\begin{remark*}
    1. If the set $\settuple$ constructed by the Algorithm~\ref{alg:exact} is such that there exist reviewers who do not contribute any of their decisions to this set, then one can run Algorithm~\ref{alg:greedy} on assignments of these reviewers to papers and obtain the set $\settuple'$. Next, consider the updated set $\settuple^{\ast} = \settuple \cup \settuple'$ and observe that each reviewer contributes at most one decision to this set. 
    
    2. By construction both matching algorithms introduced in this section include at most one decision per reviewer in a set of tuples $\settuple$.
\end{remark*}

{
\begin{algorithm}[H]
   \caption{Greedy matching algorithm}
   \label{alg:greedy}
   {\bfseries Input:} Assignments $\sbassignment$, $\dbassignment$ of SB and DB reviewers to papers, respectively.\\
    \begin{enumerate}[leftmargin=*, topsep=0pt]
    
        \item Construct a graph $G$ that cosists of 3 layers:
        \begin{itemize}[noitemsep, leftmargin=*]
            \item \textbf{Layer 1.} One node for each SB reviewer
            \item \textbf{Layer 2.} One node for each paper
            \item \textbf{Layer 3.} One node for each DB reviewer
        \end{itemize}
        and add edges between reviewers and papers according to assignments $\sbassignment$ and $\dbassignment$. Set $\settuple = \emptyset$.
        
        \item \label{step:greedy:find} Find a triple (SB reviewer $i_1$, paper $j \in \indset$, DB reviewer $i_2$) such that there is a path in graph $G$ from a node corresponding to SB reviewer to a node corresponding to DB reviewer through a node corresponding to a paper. If there are many such triples, break ties uniformly at random. If such a triple exists, define $\tuple_1 = (j, \sbdecision_{i_1 j}, \dbdecision_{i_2 j}, \ind_j)$, otherwise set $\tuple_1 = \emptyset$.
        
        \item \label{step:greedy:find2} Find a triple (SB reviewer $i_1' \ne i_1$, paper $j' \in \noindset$, DB reviewer $i_2' \ne i_2$) such that there is a path in graph $G$ from a node corresponding to the SB reviewer to a node corresponding to the DB reviewer through a node corresponding to the paper. If there are many such triples, break ties uniformly at random. If such a triple exists, define $\tuple_2 = (j', \sbdecision_{i_1' j'}, \dbdecision_{i_2' j'}, \ind_{j'})$, otherwise set $\tuple_2 = \emptyset$.
        
        \item \label{step:greedy:iterate} Update $\settuple = \settuple \cup \left\{\tuple_1, \tuple_2 \right\}$. If both $\tuple_1$ and $\tuple_2$ are empty, return $\settuple$. Otherwise delete reviewers $i_1, i_1', i_2, i_2'$ from the graph $G$ together with the corresponding edges and go to Step~\ref{step:greedy:find}.
    \end{enumerate}
\end{algorithm}
}

Overall, let $\algo$ denote a procedure that takes assignments $\sbassignment$ and $\dbassignment$ as input and depending on the relationship between $\papload$ and $\revload$ calls Algorithm~\ref{alg:exact} or Algorithm~\ref{alg:greedy} to construct the set $\settuple$.

\subsection{Guarantees}

Having defined a procedure to construct input for the \permutation{} and~\counting{} tests, we are now ready to formulate corresponding theoretical guarantees. Recalling that the experimental setup of~\shorttomkins{} itself breaks the Type-I error guarantees, we abstract out these issues by assuming that instead of TPMS assignment algorithm or any other algorithm that computes assignments of papers to SB and DB reviewers, the assignment is selected uniformly at random from the set of all assignments satisfying $(\papload, \revload)-$constraints. For brevity, we only show the result for the absolute bias problem, but analogue of Theorem~\ref{prop:relative} also holds.

\begin{proposition}
\label{prop:tomkins}
For any given significance level $\level \in (0, 1)$, under the setup of the absolute bias problem (Problem~\ref{problem:absolute}), let experiment be organized according to the procedure of~\shorttomkins{} with random assignment. Then the \permutation{} and \counting{} tests coupled with procedure $\algo$ (Algorithm~\ref{alg:exact} and Algorithm~\ref{alg:greedy}) are guaranteed to control for the Type-I error rate at the level $\level$ and also satisfy the requirement of non-trivial power.
\end{proposition}

\begin{remark*}

1. From theoretical standpoint, the requirement of random assignment can be substituted with the following conditions, under which any assignment algorithm can be used: (i) reviewers in both conditions bid blindly and (ii) reviewers' evaluations are independent of similarities. That is, rows of matrices $\sbmatrix$ and $\dbmatrix$ are assigned to reviewers uniformly at random after the assignment is computed.

2. Proposition~\ref{prop:tomkins} ensures that the \counting{} and \permutation{} tests, coupled with matching algorithms we introduced above, are robust to issues (a)-(c) discussed in Section~\ref{section:regression:test}. However, in practice even our robust tests may still be susceptible to issues caused by the experimental setup of~\shorttomkins{}
\end{remark*}

%%%%%%%%%%%%%%%%%%%%%%%%%%%%%%%%%%%%%%%%%%%%%%%%

\section{Additional impossibility result for the generalized logistic model}
\label{appendix:logreg}

In this section we formulate an additional impossibility result that highlights the generality of the~\permutation{} test.

As we demonstrated in Theorem~\ref{prop:relative}, the \permutation{} test leads to reliable testing under the generalized logistic model. Recalling the definition of papers representations $\paprepresentation_j, j \in [\numpap]$, let us now consider an extended version of the generalized logistic model which is given by the following parametric equations describing the behaviour of DB and SB reviewers under the absence of bias
\begin{subequations}
\begin{align}
    \label{eqn:extlogregdb} &\text{DB:} \ \log \frac{\dbpacc_{ij}}{1 - \dbpacc_{ij}} = \dbcoef_0 + \dbcoef_1 \paprepresentation_j \\  \label{eqn:extlogregsb} &\text{SB:} \ \log \frac{\sbpacc_{ij}}{1 - \sbpacc_{ij}} = \sbcoef_0 + \sbcoef_1 \paprepresentation_j,
\end{align}
\end{subequations}
where parameters $\sbcoef_{0}, \sbcoef_{1} > 0, \dbcoef_{0} , \dbcoef_{1} > 0$ as well as papers' scores $\paprepresentation_j, j \in [\numpap],$ are bounded in absolute value  but otherwise are allowed to be arbitrary. This extended version specializes to the standard generalized logistic model if $\sbcoef_1 = \dbcoef_1$. In words, under the extended version, the reviewers in SB and DB conditions under the absence of bias may have not only different intercepts, but also different coefficients in front of $\paprepresentation$. The presence of bias is then defined as a violation of model~\eqref{eqn:extlogregsb} where the direction of violation is different for papers from $\indset$ and $\noindset$.

Unfortunately, as we show in Corollary~\ref{prop:extgll}, reliable testing under this extension of the generalized logistic model is not possible. 

\begin{corollary}
\label{prop:extgll}
For any significance level $\level \in (0, 1)$, consider the extension of the generalized logistic model given by equations~\eqref{eqn:extlogregdb} and~\eqref{eqn:extlogregsb}. Then no test operating on decisions of SB and DB reviewers can control for the Type-I error rate at the level $\level$ and simultaneously satisfy the non-trivial power requirement.
\end{corollary}

The negative result of Corollary~\ref{prop:extgll} applies to the \permutation{} test, implying that even our testing procedure which is robust to various issues discussed in Section~\ref{section:regression} cannot handle the extension of the generalized logistic model specified by equations~\eqref{eqn:extlogregdb} and~\eqref{eqn:extlogregsb}.

%%%%%%%%%%%%%%%%%%%%%%%%%%%%%%%%%%%%%%%%%%%%%%%%%%

\section{Causal inference viewpoint}
\label{appendix:causal}

\newcommand{\causalf}{\xi}
\newcommand{\causalsb}{\eta}
\newcommand{\causaldb}{\zeta}

When testing for biases in peer review, we aim at discovering a causal relationship between paper's authorship information and reviewers' perception of the paper. In this section, we describe a causal model under which we approach the problem, and provide an equivalent formulation of the problem from the causal inference viewpoint.

Recall that a decision of reviewer $i$ for paper $j$ if this reviewer is assigned to this paper in SB setup is denoted as $\sbdecision_{ij}$ and is a Bernoulli random variable with expectation $\sbpacc_{ij}$. Our ultimate goal is to evaluate whether the indicator variable $\ind_j \in \{-1, 1 \}$ which encodes the property satisfaction has \emph{causal impact} on the decisions of SB reviewers. To this end, we assume that for each reviewer $i \in [\numrev]$ and for each paper $j \in [\numpap]$, probability $\sbpacc_{ij}$ can be expressed as:
\begin{align}
\label{eqn:sb_prob}
    \sbpacc_{ij} = \causalf(\revrepresentation_i, \paprepresentation_j, \ind_j),
\end{align}
for some unknown function $\causalf$ with co-domain $[0, 1]$, where $\paprepresentation_j$ is an anonymized content of a paper and $\revrepresentation_i$ is an arbitrary complex representation of a reviewer. That is, we assume that decisions of SB reviewers are determined by the paper content, reviewer identity and, possibly, authorship information.

In this notation, we can state a canonical formulation of the bias testing problem:
\begin{manualproblem}{$1'$}[Canonical formulation of the bias testing problem] \label{problem:causal}
\normalfont Given significance level $\level \in (0, 1)$, and decisions of SB reviewers that are distributed according to equation~\eqref{eqn:sb_prob}, the goal is to test the following hypotheses:
\begin{align}
   & \nonumber \hypothesis_0 : \forall i \in [\numrev] \ \forall j \in [\numpap] \ \  \causalf(\revrepresentation_i, \paprepresentation_j, 1) = \causalf(\revrepresentation_i, \paprepresentation_j, -1)\\
   & \label{eqn:absolute:alt_c} \hypothesis_1 : \forall i \in [\numrev] \ \forall j \in [\numpap]  \ \
   \causalf(\revrepresentation_i, \paprepresentation_j, 1) \ge \causalf(\revrepresentation_i, \paprepresentation_j, -1)
\end{align}
where at least for one pair $(i, j) \in [\numrev] \times [\numpap]$ the inequality in the alternative hypothesis~\eqref{eqn:absolute:alt_c} is strict.
\end{manualproblem}

Unfortunately, the formulation of Problem~\ref{problem:causal} is too challenging and cannot be tested without further assumptions in the peer-review setup for the following reasons:
\begin{enumerate}[topsep=0pt, noitemsep]
    \item Fully randomized controlled experiments cannot be performed in peer-review settings, because we cannot randomize indicators $\ind_j, j \in [\numpap]$, that is, we cannot randomize authors of the papers
    \item To facilitate an observational study without further assumptions on function $\causalf$, we need to have many papers with the same content but with different authors which is also impossible
\end{enumerate}

\citet{tomkins17wsdm} attempted to circumvent the aforementioned challenges by making the following assumptions: (i) for each paper $j \in [\numpap]$, representation $\paprepresentation_j$ is simply a true score of a paper $\tscore_j \in \reals$; (ii) function $\causalf$ follows logistic model~\eqref{eqn:logregmodel}; (iii) function $\causalf$ is independent of reviewer identity $\revrepresentation$; and (iv) double blind reviewers can estimate true scores of submissions with reasonable accuracy. As we discussed in Section~\ref{section:regression}, even if assumptions (i)--(iv) are satisfied, the test used by~\citet{tomkins17wsdm} is at risk of violating its Type-I error guarantees unless DB reviewers estimate true scores of submissions without any noise (which is not the case in conference settings). Of course, further violations of the assumptions exacerbate the issues.

\smallskip

In contrast, in our work we attempt the problem making fundamentally different assumptions. Without loss of generality, we denote the probability of reviewer $i$ recommending acceptance for paper $j$ in DB condition as:
\begin{align}
\label{eqn:db_prob}
    \dbpacc_{ij} = \causalf(\revrepresentation_i, \paprepresentation_j, 0),
\end{align}
where the last argument of the function $\causalf$ is censored, indicating that DB reviewers do not have access to the authorship information. In this notation, our assumption is formulated as follows:
\begin{assumption}
\label{assumption:1}
Under absence of a bias, the behaviour of reviewers does not change between SB and DB conditions, that is, for any reviewer representation $\revrepresentation$, for any paper representation $\paprepresentation$ and for any value of the indicator $\ind \in \{-1, 1\}$, we have
\begin{align*}
    \causalf(\revrepresentation, \paprepresentation, \ind) = \causalf(\revrepresentation, \paprepresentation, 0).
\end{align*}
\end{assumption}
Under Assumption~\ref{assumption:1}, the presence of bias is defined as a deviation of reviewers in SB condition from their behavior in DB condition such that the direction of the deviation is determined by the value of the indicator $\ind$. Given that,  the canonical formulation of the bias testing problem (Problem~\ref{problem:causal}) corresponds to the absolute bias problem (Problem~\ref{problem:absolute}).

Observe that Assumption~\ref{assumption:1} does not restrict the generality of representations $\paprepresentation$ and $\revrepresentation$ and also does not make strong parametric assumptions about function $\causalf$. Instead, it essentially postulates that the condition in which a reviewer is put does not serve as a confounder, that is, under the absence of bias, the probability that reviewer $i$ votes to accept paper $j$ is independent of whether reviewer $i$ reviews paper $j$ in the SB or DB condition.

\medskip

To accommodate an additional confounding factor --- a distributional shift due to assignment of a reviewer in the SB or DB condition which is independent of papers' characteristics --- we substitute Assumption~\ref{assumption:1} with its less restrictive version.
\begin{assumption}
\label{assumption:2}
Under the absence of bias, the behaviour of any reviewer $i$ in SB condition is connected to the behaviour of that reviewer in DB condition through a linking function $\nullshift$, that is, for any reviewer representation $\revrepresentation$, for any paper representation $\paprepresentation$ and for any value of the indicator $\ind \in \{-1, 1\}$, we have
\begin{align}
\label{eqn:link}
    \causalf(\revrepresentation, \paprepresentation, \ind) = \nullshift(\causalf(\revrepresentation, \paprepresentation, 0)),
\end{align}
where $\nullshift$ is an (unknown) member of a (known) family $\subclass$ of monotonic functions acting from $[0, 1]$ to $[0, 1]$. 
\end{assumption}

First, observe that if we restrict $\subclass$ to be a singleton containing only the identity function, then Assumption~\ref{assumption:2} reduces to Assumption~\ref{assumption:1}. However, richer choices of family $\subclass$ allow to incorporate various models of confoundings due to the setup. Second, if we again define the presence of bias as a deviation from~\eqref{eqn:link}, where the direction of the deviation is determined by indicator $\ind$, then the canonical formulation of the bias testing problem (Problem~\ref{problem:causal}) reduces to the relative bias problem (Problem~\ref{problem:relative}).

In this section we have formulated two assumptions that allow us to perform causal inference and lead to the absolute and relative bias testing problems defined in Section~\ref{section:framework}. With this formulation, in Section~\ref{section:proposal} we introduce two statistical procedures to test for biases in the peer-review setup and provide their theoretical analysis in Section~\ref{section:analysis}.

%%%%%%%%%%%%%%%%%%%%%%%%%%%%%%%%%%%%%%%%%%%%%%%%

\section{Setup for simulations}
\label{appendix:simulations}

In this section we describe setup for simulations we conducted in this work. Notice that in contrast to the test of~\tomkins{} which operates on accept/reject decisions of SB reviewers and scores provided by DB reviewers, the tests we introduce in this work operate on decisions of both SB and DB reviewers. Hence, to compare tests we need to specify (i) models of DB/SB reviewers' decisions and (ii) models of DB reviewers' scores. All simulations are run for 5000 iterations.

%%%%%%%%%%%%%%%%%%%%%%%%%%%%%%%%%%%%%%%%%%%%%%%%

\subsection{Simulations in Section~\ref{section:regression}}
\label{appendix:simulations:issue}

We now provide necessary details for the simulations in Section~\ref{section:regression}.

\subsubsection{Measurement error (Figure~\ref{fig:error})}
\label{appendix:simulations:noise}

For this simulation we consider the following model of SB and DB reviewers: 
\begin{subequations}
\begin{align}
      \label{eqn:noise:db} \text{DB:} \quad &\log \frac{\dbpacc_j}{1 - \dbpacc_{j}} =  \coef_0 + \coef_1 \tscore_j \\
      \label{eqn:noise:sb} \text{SB:} \quad &\log \frac{\sbpacc_{ij}}{1 - \sbpacc_{ij}} = \coef_0 + \coef_1 \tscore_j + \coef_2 \ind_j,
\end{align}
\end{subequations}
that is, model~\eqref{eqn:logregmodel} is correct and reviews given by the same reviewer for different papers are independent. Notice that under this model all reviewers are identical and hence issues with the setup do not manifest in this case.

We set $\numrev = 2\numpap = 1000$ and $\revload = \papload = 2$. At each iteration we independently sample true scores of papers $\tscore_j, j \in [\numpap]$, from uniform distribution $\mathcal{U}[-2, 2]$ and assume that mean scores by two DB reviewers assigned to a paper $j \in [\numpap]$ estimates true score $\tscore_j$ with some Gaussian noise ($\sigma = 0.7$). We then sample values of $\ind_{j}, j \in [\numpap]$, such that correlation between $\tscore$ and $\ind$ equals $\varphi$ for values of $\varphi$ between $0$ and $0.5$. To this end, we let each paper $j \in [\numpap]$ with the score $\tscore_j < 0$ have $\ind_j = 1$ with probability $0.5 - \gamma$ and $\ind_j = -1$ otherwise. Similarly, each paper $j \in [\numpap]$ with the score $ \tscore_j \ge 0$ has $\ind_j = 1$ with probability $0.5 + \gamma$ and $\ind_j = -1$ otherwise. We then vary the value of $\gamma \in (0, 0.5)$ to achieve the necessary correlation. Finally, using models~\eqref{eqn:noise:db} and~\eqref{eqn:noise:sb} with $\coef_0 = 1$, $\coef_1 = 2$ and $\beta_2 = 0$ (no bias condition) we sample decisions of SB and DB reviewers and run the \permutation{} test and the test used by~\tomkins{}, setting the significance level to be $\level = 0.05$. We then compute a Type-I error as a fraction of iterations in which the null hypothesis ($\beta_2 = 0$) was rejected.

%%%%%%%%%%%%%%%%%%%%%%%%%%%%%%%%%%%%%%%%%%%%%%%%

\subsubsection{Model mismatch (Figure~\ref{fig:mismatch})}

For this simulation we consider a violation of model~\eqref{eqn:logregmodel} and the following model of SB and DB reviewers with $\coef_2 = 0$ (no-bias condition): 

\begin{align*}
    \text{DB:} \quad & \log \frac{\dbpacc_j}{1 - \dbpacc_{j}} =  \coef_0 + \coef_1 (\tscore_j)^3 \\
    \text{SB:} \quad & \log \frac{\sbpacc_{ij}}{1 - \sbpacc_{ij}} = \coef_0 + \coef_1 (\tscore_j)^3 + \coef_2 \ind_j. 
\end{align*}

To abstract out the effect of measurement error, in this section we assume that the true scores $\tscore_j, j \in [\numpap],$ are known, but the test used by~\tomkins{} fits the model defined by equation~\eqref{eqn:noise:sb}. Besides the change of correct model and availability of true scores $\{\tscore_j, j \in [\numpap]\}$, the simulations follow scenario we described in Appendix~\ref{appendix:simulations:noise}.

%%%%%%%%%%%%%%%%%%%%%%%%%%%%%%%%%%%%%%%%%%%%%%%%

\subsubsection{Reviewer calibration (Figure~\ref{fig:revsignal})}
\label{appendix:simulations:signal}

\newcommand{\leniency}{\ell}
\newcommand{\uniform}{\mathcal{U}}
\newcommand{\clarity}{\zeta}

In this simulation we model the effect of correlations introduced by reviewer calibration. More concretely, we construct a model of reviewer calibration under which the test by~\tomkins{} fails to control for the Type-I error rate. In this section we assume that true scores of submissions are proportional to the clarity of the writing. We then sample clarity scores $\clarity_j, j \in [\numpap],$ from uniform distribution $\uniform[-1, 1]$ and  define $\tscore_j = \clarity_j$ for each $j \in [\numpap]$. Eventually, we consider the following model of reviewer. For each $i \in [\numrev]$ and for each $j \in [\numpap]$:
\begin{align*}
    \text{DB:} \quad & \dbpacc_{ij} =  \dbpacc_{j} + \leniency_i \times \indicator{\clarity_j < 0.5}  \\
    \text{SB:} \quad & \sbpacc_{ij} =  \sbpacc_{j} + \leniency_i \times \indicator{\clarity_j < 0.5}, 
\end{align*}
where $\leniency_i$ is reviewers' leniency which equals $0.4$ with probability $0.5$ and $-0.4$ otherwise and $\dbpacc_j, \sbpacc_j$ are defined by equations~\eqref{eqn:noise:db} and~\eqref{eqn:noise:sb} with $\coef_0 = 0, \coef_1 = 0.25$ and $\coef_2 = 0$ (no bias condition). Parameters are selected to ensure that $0 \le \dbpacc_{ij}, \sbpacc_{ij} \le 1$.

In words, the above model says that for papers with high quality of writing ($\clarity > 0.5$) reviewers understand their content well and follow models~\eqref{eqn:noise:db} and~\eqref{eqn:noise:sb} exactly, but for papers with lower writing quality their leniency parameter influences their decision. Notice that under this model it is natural to expect that estimates of the true scores provided by DB reviewers are also influenced by their leniency and hence are noisy. However, to isolate the effect of reviewer identity we assume that the test used by~\tomkins{} knows true scores $\tscore_j, j \in [\numpap],$ exactly. Additionally, notice that marginally each reviewer follows the model defined by equations~\eqref{eqn:noise:db} and~\eqref{eqn:noise:sb}, and hence when $\revload = 1$, the test by~\tomkins{} has control over the Type-I error for any correlation between $\tscore$ and $\ind$.

In this section we consider an extreme pattern of correlations between $\tscore$ and $\ind$. Concretely, we assume that for any paper $j \in [\numpap]$, we have $\ind_j = 1$ if and only if $\tscore_j > 0.5$ and $\ind_j = -1$ otherwise.  Notice that in practice such strong dependence is unlikely to happen, but we underscore that in practice the test by~\tomkins{} also does not have access to noiseless true scores which will cause measurement errors and hence will exacerbate the issue. 

We then perform simulations as discussed above having $\numpap = 1000$ and $\papload = 1$ fixed and varying the number of papers per reviewer and using the modification of the Wald test with factor variable for each reviewer added (reviewer-depedent intercept).

%%%%%%%%%%%%%%%%%%%%%%%%%%%%%%%%%%%%%%%%%%%

\subsubsection{Non-blind bidding (Figure~\ref{fig:bidding})}
\label{appendix:simulations:bidding}

\newcommand{\bid}{b}

Formalizing the intuition we mentioned in Section~\ref{section:regression:setup}, we consider a setting with $\numpap = 1000, \numrev = 2000, \papload = \revload = 1$ and consider a property of interest ``paper has a famous author''. Suppose that during the bidding procedure each reviewer $i \in [\numrev]$ gives a score $\bid_{ij} \in \{-1, 0, 1\}$ to each paper $j \in [\numpap]$, where $\bid_{ij} = 1$ means that reviewer wants to review the paper, $\bid_{ij} = -1$ means that reviewer does not want to review the paper and $\bid_{ij} = 0$ is an intermediate between $\bid_{ij} = 1$ and $\bid_{ij} = -1$. Given the bids, the assignment is computed maximizing the total sum of the bids. Namely, for all $(i, j) \in [\numrev] \times [\numpap]$ let a binary indicator $A_{ij}$ equal $1$ if reviewer $i$ is assigned to paper $j$ and 0 otherwise and let $\sbrevset \subset [\numrev]$ be the set of reviewers allocated to SB condition. Then the assignment of SB reviewers to papers is computed maximizing the following objective subject to the standard $(\papload, \revload)$-load constraints.
\begin{align*}
    \sum\limits_{i \in \sbrevset} \sum\limits_{j \in [\numpap]} A_{ij} \bid_{ij}.
\end{align*}
The same objective is used to assign DB reviewers to papers. Next, we suppose that for each paper $j \in [\numpap]$ there is a true score $\tscore_j \in [0, 0.9]$ and that all reviewers belong to one of the following personality types:
\begin{itemize}[leftmargin=*, itemsep=3pt]
    \item {\bfseries Type A:} Lenient reviewers who accept each paper $j \in [\numpap]$ assigned to them with probability $\tscore_j + 0.1$ and want to read papers from top authors. If bidding is blind, they do not have any information about author identity and bid 0 on each paper, but if bidding is non-blind, then for each paper $j \in \indset$ reviewer $i$ of type A places a bid $\bid_{ij} = 1$ and for each paper $j \in \noindset$ she/he places a bid $\bid_{ij} = -1$.  
    
    \item {\bfseries Type B:} Accurate reviewers who accept each paper $j \in [\numpap]$ assigned to them with probability $\tscore_j$ and do not mind reviewing any paper. Independent of whether bidding is blind or not, reviewer $i$ of type B places a bid $\bid_{ij} = 0$ on each paper $j \in [\numpap]$.
\end{itemize}

Notice that evaluations of reviewers of both types are unbiased --- the probability of acceptance is not determined by author identities. The type of each reviewer is determined independently: reviewer $i \in [\numrev]$ is of type A with probability $0.3$ and of type B with probability $0.7$. Independently, each paper $j \in [\numpap]$ belongs to $\indset$ with probability $0.3$ and to $\noindset$ with probability 0.7.  

Having defined the setup, in each iteration we independently sample true scores of submissions from $\uniform[0, 0.9]$ (no correlation with indicator $\ind$) and compute two bidding matrices: (i) when SB reviewers observe author identities during bidding and (ii) when bidding is blind for both SB and DB reviewers. For each bidding matrix we compute assignments of SB and DB reviewers to papers and pass observed decisions to the \permutation{} test and the test used by~\tomkins{}. For the test of~\shorttomkins{}, we assume that true scores $\tscore_j, j \in [\numpap]$, are known exactly.

%%%%%%%%%%%%%%%%%%%%%%%%%%%%%%%%%%%%%%%%%%%

\subsubsection{Non-random assignment (Figure~\ref{fig:assignment})}
\label{appendix:simulations:assignment}

In this section we construct a similarity matrix $\simmatrix$ and formalize the dependence of reviewer's perception of a paper on similarity between paper and reviewer that leads to the effect demonstrated in Figure~\ref{fig:assignment}. We notice that the construction we provide here is artificial and serves as a proof of concept for our claim that non-random assignment may violate some key independence assumptions of statistical tests even if it is not based on reviewers' bids. While in practice we do not expect to observe such specific similarity matrices, we can still observe some more subtle manifestations of issues caused by non-randomness of the assignment.

First, in this section we assume that assignment is performed using the TPMS algorithms~\citep{charlin13tpms}, that is, given similarity matrix $\simmatrix$ between reviewers and papers, each paper is assigned to $\papload$ reviewers in a way that each reviewer is assigned to at most $\revload$ papers such that total sum similarity of the assignment is maximized. 

Second, consider a similarity matrix $\simmatrix$, defined as follows. For each reviewer $i \in [\numrev]$ and for each paper $j \in [\numpap]$:
\begin{align}
\label{eqn:simmatrix}
    \simmatrix_{ij} = (\numrev + 1 - i) \times (\numpap + 1 - j).
\end{align}
Given that reviewers are allocated to conditions at random, similarity matrices $\simmatrix_{\text{SB}}$ (SB condition) and $\simmatrix_{\text{DB}}$ (DB condition) are constructed by random division of rows of $\simmatrix$ into two groups of equal size and stacking them into $\simmatrix_{\text{SB}}$ and $\simmatrix_{\text{DB}}$ correspondingly.

Third, we assume that each reviewer $i \in [\numrev]$ has some value of threshold $\threshold_i$ such that if reviewer $i$ is assigned to paper $j \in [\numpap]$ in either of setups, reviewer accepts the paper with probability $\mean_{ij}$ given by:
\begin{align}
    \label{eqn:acceptance}
    \mean_{ij} = \begin{cases}
                    0.9 & \text{if } \simmatrix_{ij} \ge \threshold_i \\
                    \tscore_j & \text{if } \simmatrix_{ij} < \threshold_i,
                 \end{cases}
\end{align}
where $\tscore_j \in [0, 0.9]$ is a true score of paper $j$. We also assume that reviewer $i$ in DB condition returns $\mean_{ij}$ as an estimate of $\tscore_j$. 

Fourth, for every reviewer $i$ we set a value of threshold as follows:
\begin{align}
\label{eqn:threshold}
    \threshold_i = (\numrev +1 - i) \times (\numpap - \left[ \nicefrac{(i-1)}{2} \right]),
\end{align}
where $\left[ x \right]$ is the integral part of $x$.

Fifth and finally, we assume that true scores $\tscore$ are independently sampled from $\uniform[0, 0.9]$  and sample indicators $\ind$ such that they are correlated with $\tscore$, fixing the value of correlation $\varphi = 0.45$. We also set $\revload = \papload = 1$ and $\numrev = 2 \numpap = 1000$. Now we allocate half of reviewers to SB condition and half to DB condition uniformly at random. We then compare the performance of the \permutation{} test and the test by~\shorttomkins{} under (i) experimental setup of~\shorttomkins{} and (ii) our experimental setup.

The intuition behind our construction of matrix $\simmatrix$ in equation~\eqref{eqn:simmatrix} is that for any square submatrix of $\simmatrix$, the TPMS algorithm with parameters $\revload = \papload = 1$ will compute an assignment that corresponds to the diagonal of this submatrix.  Coupled with specific choice of thresholds~\eqref{eqn:threshold}, probabilities of acceptance~\eqref{eqn:acceptance} and correlation between $\tscore$ and $\ind$ at the level of 0.45, this choice of similarity matrix ensures that under the setup of~\shorttomkins{}, with non-zero probability most of SB reviewers will receive papers with similarities above the corresponding threshold and most of DB reviewers will receive papers with similarities below the corresponding threshold or vice versa. Hence, the assignments will be structurally different and, as demonstrated by Figure~\ref{fig:assignment}, this difference will be confused with bias by both~\tomkins{} and \permutation{} tests. In contrast, under our proposed setup the assignments of SB and DB reviewers to papers do not exhibit any structural difference and hence do not break the Type-I error guarantees of the tests.

%%%%%%%%%%%%%%%%%%%%%%%%%%%%%%%%%%%%%%%%%%%%%%%%

\subsection{Simulations in Section~\ref{section:introduction} and Section~\ref{section:analysis:absolute}}
\label{appendix:simulations:power}

The simulations in Section~\ref{section:introduction} and Section~\ref{section:analysis:absolute} were performed under the model of reviewers in~\eqref{eqn:noise:db} and~\eqref{eqn:noise:sb} following the setup described in Appendix~\ref{appendix:simulations:noise} with small differences. Instead of varying the value of correlation $\correlation$ between $\tscore$ and $\ind$, we fix the value of $\correlation$ and vary the number of papers $\numpap$. Moreover, we independently assign papers to the sets $\indset$ and $\noindset$ as follows: each paper $j$ such that $\tscore_j < 0$ belongs to the set $\indset$ with probability $0.5 - \gamma$ and otherwise belongs to the set $\noindset$, similarly, each paper $j$ with $\tscore_j > 0$ belongs to the set $\indset$ with probability $0.5 + \gamma$ and otherwise belongs to the set $\noindset$. The value of $\gamma$ is selected to achieve the required level of correlation $\correlation$ between $\tscore$ and $\ind$.
\begin{itemize}[leftmargin=*, itemsep=1pt]
    \item For Figure~\ref{fig:intro:type1} we set $\varphi = 0.4$ and perform simulations under $\coef_0 = 1, \coef_1 = 2, \coef_2 = 0$ (no bias), $\papload = 2, \revload = 1$, where true scores are sampled from $\uniform[-1, 1]$. We see that for the test used by~\tomkins{} a violation of Type-I error guarantees caused by measurement error coupled with correlations (see Appendix~\ref{appendix:simulations:noise} for details) exacerbates as sample size grows. 
    
    \item For Figure~\ref{fig:intro:type2} we set $\varphi = 0.6$ and perform simulations under $\coef_0 = 1, \coef_1 = 2, \coef_2 = -0.35$ (bias against papers that satisfy the property), $\papload = 2, \revload = 1$, where true scores are sampled from $\uniform[-0.5, 0.5]$. We see that in this case measurement error has strong harmful impact on the power of the test used by~\tomkins{}.
    
    \item For Figure~\ref{fig:intro:easy} we set $\varphi = 0$ and additionally assume that DB reviewers estimate true scores with no noise. In this case all parametric assumptions made by~\tomkins{} are satisfied. We then perform simulations under $\coef_0 = 1, \coef_1 = 2, \coef_2 = 0.35$ (bias in favour of papers that satisfy the property), $\papload = 2, \revload = 1$, where true scores are sampled from $\uniform[-1, 1]$.

    \item Simulations in Section~\ref{section:analysis:absolute} follow the simulations in Figure~\ref{fig:intro:type2} and Figure~\ref{fig:intro:easy} with the exception that the \counting{} test is added for comparison.

\end{itemize}

%%%%%%%%%%%%%%%%%%%%%%%%%%%%%%%%%%%%%%%%%%%%%%%%

\subsection{Simulations in Section~\ref{section:analysis:relative}}
\label{appendix:simulations:relative}

In this section we illustrate that the \counting{} test designed to control for Type-I error under the generalized linear model does not lead to reliable testing under the generalized logistic model under which the \permutation{} test is suitable, and vice versa. To this end, we design two instances of the relative bias problem under the generalized \emph{linear} model --- instance (i) with presence of bias and instance (ii) with absence of bias. Our construction ensures that the resulting matrices $\sbmatrix$ and $\dbmatrix$ \emph{simultaneously also} fall in the relative bias problem under the generalized \emph{logistic} model with the exception that instance (i) corresponds to absence of bias under the generalized logistic model and instance (ii) corresponds to the presence of bias under this model. 

\smallskip

\noindent \underline{Instance (i)} Under the generalized linear model, for each paper $j \in \indset$ let $\tscore_j = 0.7$ and for each paper $j \in \noindset$ let $\tscore_j = 0.5$. Additionally, let $\shift = 0.175$. This choice of parameters defines matrices $\sbmatrix_1$ and $\dbmatrix_1$ that are generated according to the equations~\eqref{eqn:genlinregmodel} and~\eqref{eqn:genlinregmodelsb} and fall under the null hypothesis of no bias. 

\smallskip

\noindent \underline{Instance (ii)} Under the generalized linear model, for each paper $j \in \indset$ let $\tscore_j = 0.65$ and for each paper $j \in \noindset$ let $\tscore_j = 0.25$. Now let matrix $\dbmatrix_2$ be defined according to the model~\eqref{eqn:genlinregmodel} and matrix $\sbmatrix_2$ be defined as follows:
	\begin{align}
	\label{eqn:tmpundefined}
		\sbpacc_{ij} = \begin{cases}
 								\tscore_j + \shift_1 & \text{if} \ \ind_j = 1 \\
 								\tscore_j + \shift_2 & \text{if} \ \ind_j = -1 \\
 							\end{cases},
	\end{align}
	where we carefully select $\shift_2 > \shift_1$ as explained below. This choice of parameters leads to a correct alternative hypothesis of presence of bias against papers that satisfy the property of interest.
	
	We now simulate reviewers decisions with $\papload = 2, \revload = 1, \numpap = 1000, \numrev = 4000$, independently allocating each paper to $\indset$ with probability $0.5$ and to $\noindset$ otherwise. We then apply the \counting{} and \permutation{} tests for each of these instances, and present the results in Figure~\ref{fig:relative:linear}. Instance (i) allows to compare Type-I error rates, and instance (ii) allows to compare powers of the tests under the generalized linear model.

\medskip

One can verify that the instances we constructed above under the generalized \emph{linear} model also fall under the generalized \emph{logistic} model for some specific choice of parameters. Indeed, consider an instance of the generalized logistic model specified by parameters $\coef_0 = -2.5 \log \nicefrac{7}{3}, \coef_1 = 5 \log \nicefrac{7}{3}$ and $\logshift = 1$. Then a straightforward verification shows that matrix $\dbmatrix_1$ satisfies equation~\eqref{eqn:genlogregmodel} which specifies the behavior of DB reviewers under the generalized logistic model. Next, observe that for each reviewer $i \in [\numrev]$ and for each paper $j \in \indset$ the corresponding entry of the matrix $\sbmatrix_1$ is \emph{larger} than prescribed by the model of SB reviewers under the absence of bias~\eqref{eqn:genlogregmodelsb}. Similarly, for each paper $j \in \noindset$ the corresponding entry of the matrix $\sbmatrix_1$ is \emph{smaller} than it should be if the bias is absent~\eqref{eqn:genlogregmodelsb}. Hence, the pair of matrices $\dbmatrix_1, \sbmatrix_1$ satisfies the \emph{alternative} hypothesis under the generalized logistic model. 

Conversely, consider an instance of the generalized logistic model specified by parameters $\coef_0 = \log \nicefrac{1}{3} - 0.625 \log \nicefrac{39}{7}, \coef_1 = 2.5 \log \nicefrac{39}{7}$ and $\logshift = 1.5$. Then a straightforward verification shows that matrix $\dbmatrix_2$ satisfies equation~\eqref{eqn:genlogregmodel} which specifies the behavior of DB reviewer under the generalized logistic model. Recall that at this point we didn't specify how we selected values $\shift_1, \shift_2$ in equation~\eqref{eqn:tmpundefined}. In fact, we selected these values such that entries of the matrix $\sbmatrix_2$ satisfy equation~\eqref{eqn:genlogregmodelsb} which specifies the behavior of SB reviewers under the generalized logistic model when the bias is absent. Namely, we set 
\begin{align*}
    \shift_1 &= -0.65 + \left(1 + \exp\left\{-\beta_0 - \logshift - 0.65 \beta_1 \right\} \right)^{-1} \\
    \shift_2 &= -0.25 + \left(1 + \exp\left\{-\beta_0 - \logshift - 0.25 \beta_1 \right\} \right)^{-1}
\end{align*}
As a result, the pair of matrices $\dbmatrix_2, \sbmatrix_2$ satisfies the \emph{null} hypothesis under the generalized logistic model.  

Finally, the power of the \counting{} test in Figure~\ref{fig:relative:linear} becomes the Type-I error rate under the instance of  the generalized logistic model with $\coef_0 = \log \nicefrac{1}{3} - 0.625 \log \nicefrac{39}{7}, \coef_1 = 2.5 \log \nicefrac{39}{7}$ and $\logshift = 1$. Similarly, the Type-I error of the \counting{} test in Figure~\ref{fig:relative:linear} becomes its power under the instance of the generalized logistic model with $\coef_0 = -2.5 \log \nicefrac{7}{3}, \coef_1 = 5 \log \nicefrac{7}{3}$ and $\logshift = 1$. The same applies to the \permutation{} test and eventually we obtain Figure~\ref{fig:relative:logistic} by simply exchanging the bars in Figure~\ref{fig:relative:linear}.

\smallskip

%%%%%%%%%%%%%%%%%%%%%%%%%%%%%%%%%%%%%%%%%%%%%%%%%%%%%%%%%%%%%%%%

%%%%%%%%%%%%%%%%%%%%%%%%%%%%%%%%%%%%%%%%%%%%%%%%%%%%%%%%%%%%%%%%%%%%%%%%%%%%%%%%%%%%%%%%%%%%%%

\section{Proofs of main results}
\label{appendix:proofs}

In this section we give proofs of our main results.

%%%%%%%%%%%%%%%%%%%%%%%%%%%%%%%%%%%%%%%%%%%%%%%
\subsection{Proof of Theorem~\ref{prop:absolute}}
\label{appendix:proof:absolute}

We prove Theorem~\ref{prop:absolute} in two steps. First, we show the result for the \permutation{} test and then for the \counting{} test. Before we delve into proofs, let us make two observations that we use in this section.

\smallskip

\noindent{\textbf{Observations:}}
\vspace{-7pt}
\begin{enumerate}[itemsep=3pt, label={{\Alph*}}, leftmargin=*]
    \item \label{observation:symmetry} For every paper $j \in [\numpap]$,  if in assignment $\testassignment$ (Step~\ref{step:assign:selection} of Procedure~\ref{algo:experiment}) paper $j$ is attributed to reviewers $i_1$ and $i_2$, then the events ``reviewer $i_1$ is allocated to SB and reviewer $i_2$ is allocated to DB'' and ``reviewer $i_1$ is allocated to DB and reviewer $i_2$ is allocated to SB'' are mutually exclusive and happen with probability 0.5 each. This is ensured by Step~\ref{step:assign:allocation} of Procedure~\ref{algo:experiment} where reviewers are allocated to conditions.
    
    \item \label{observation:size} By construction of Procedure~\ref{algo:experiment}, at least $\constant \min\{|\indset|, |\noindset| \}$ papers each from sets $\indset$ and $\noindset$ appear in assignment $\testassignment$ for some constant $\constant$ that depends only on parameters $\papload$ and $\revload$. Indeed, in cases (a) and (c) of Step~\ref{step:assign:selection}, all papers are included into assignment $\testassignment$ and hence our claim holds with $\constant = 1$. In case (b) $\frac{\numrev}{2} \ge \frac{\papload}{\revload} \numpap \ge 2\frac{\papload}{\revload} \min\{|\indset|, |\noindset| \}$ papers are selected and hence our claim holds with $\constant = \frac{\papload}{\revload}$.
\end{enumerate}

%%%%%%%%%%%%%%%%%%%%%%%%%%%%%%%%%%%%%%%%%%%%%%%%

\subsubsection{Proof for \permutation{} test}

The proof of Theorem~\ref{prop:absolute} for the \permutation{} test consists of two parts. First, we show that under the null hypothesis defined in Problem~\ref{problem:absolute}, for any matrices $\dbmatrix$ and $\sbmatrix (= \dbmatrix)$ and for any assignment $\testassignment$ constructed by Procedure~\ref{algo:experiment} in Step~\ref{step:assign:selection}, the test rejects the null with probability at most $\level$. Second, we show that if the number of papers in both $\indset$ and $\noindset$ is large enough, then the \permutation{} test satisfies the requirement of non-trivial power.

We prove both parts conditioned on the assignment $\testassignment$. The unconditional statement of the theorem then follows from the law of total probability.

\bigskip
%%%%%%%%%%%%%%%%%%%%%%%%%%%%%%%%%%%%%%%%%%%%%%%

{
\noindent \textbf{Control over Type-I error}

\noindent Let $\dbmatrix$ and $\sbmatrix (= \dbmatrix)$ be arbitrary matrices that fall under the definition of null hypothesis in Problem~\ref{problem:absolute}. Consider arrays $\arraymaj$ and $\arraymin$ constructed in Step~\ref{step:perm:majority} of the \permutation{} test from the set of tuples $\settuple$ passed to the test by Procedure~\ref{algo:experiment}. If any of them is empty, the test keeps the null and hence does not commit the Type-I error. Now without loss of generality assume that both $\arraymaj$ and $\arraymin$ are non-empty.
}

The idea of the proof is to show that under the null hypothesis, entries of arrays $\arraymaj$ and $\arraymin$ are mutually independent and identically distributed. Assume for the moment that it is indeed the case. Then entries of arrays $\arraymaj$ and $\arraymin$ are exchangeable random variables and hence the permutation test with statistic $\statp$ defined in Step~\ref{step:perm:decision} of Test~\ref{test:permutation} is guaranteed to provide control over the Type-I error rate for any given significance level $\level \in (0, 1)$ and hence the result for Type-I error control follows.

Consider any  entry $\arraymajentry$ of array $\arraymaj$. Then $u$ is a decision of SB reviewer for some paper $j_{\tuple} \in \indset$, where $\tuple$ is a tuple that corresponds to $\arraymajentry$. Corresponding SB and DB reviewers disagree in their decisions, that is, $\sbdecision_{j_{\tuple}} \ne \dbdecision_{j_{\tuple}}$. Recalling Observation~\ref{observation:symmetry}, we deduce that conditioned on assignment $\testassignment$, the symmetry of the null hypothesis guarantees that
\begin{align}
\label{eqn:conditioning}
    \sbdecision_{j_{\tuple}} | \left(\sbdecision_{j_{\tuple}} \ne \dbdecision_{j_{\tuple}} \right) \ \sim \ \text{Bernoulli}(0.5).
\end{align}
Indeed, given that both $\sbdecision_{j_{\tuple}}$ and $\dbdecision_{j_{\tuple}}$ are Bernoulli random variables, one can verify that
\begin{align*}
    \prob{\sbdecision_{j_{\tuple}} = 1, \dbdecision_{j_{\tuple}} = 0} = \prob{\sbdecision_{j_{\tuple}} = 0, \dbdecision_{j_{\tuple}} = 1},
\end{align*}
which coupled with the definition of condition probability implies~\eqref{eqn:conditioning}.

Hence, entries of array $\arraymaj$ are Bernoulli random variables with expectation $0.5$. Provided that each reviewer contributes at most one decision to $\settuple$, entries of $\arraymaj$ are also independent. The same argument applies to entries of array $\arraymin$ and hence we have shown that under the null hypothesis entries of $\arraymaj$ and $\arraymin$ are independent Bernoulli random variables with probability of success $0.5$ and thus are exchangeable. 

%%%%%%%%%%%%%%%%%%%%%%%%%%%%%%%%%%%%%%%%%%%%%%%%%%%%%%%%%%

\medskip

\noindent \textbf{Non-trivial power}

\noindent Consider any fixed choice of $\minigap > 0$ and $\err > 0$ in the definition of non-trivial power. The goal now is to show that there exists $\sample = \sample(\err, \minigap)$ such that if $\min\{|\indset|, |\noindset| \} > \sample$, then for any matrices $\dbmatrix$ and $\sbmatrix$ that satisfy the alternative hypothesis in Problem~\ref{problem:absolute} with margin $\minigap$, the \permutation{} test coupled with Procedure~\ref{algo:experiment} is guaranteed to reject the null hypothesis with probability at least $1 - \err$. Throughout the proof we use $\constant$ to denote a universal constant and allow its value to change from line to line due to multiplications by some other universal constants. Recall that problem parameters $\papload, \revload$ and $\level$ are treated as constants. For concreteness, throughout the proof we assume that the bias is in favor of papers from $\indset$. The same argument can be repeated in case of bias against papers from $\indset$. 

\smallskip

\noindent \underline{Step 1. Cardinality of $\arraymaj$ and $\arraymin$.}

\noindent Let us first show that arrays $\arraymaj$ and $\arraymin$ will with high probability contain order $\sample$ elements. To this end, recall that for tuple $t \in \settuple$ we add $\sbdecision_{j_{\tuple}}$ to $\arraymaj$ if (i) $\ind_{j_{\tuple}} = 1$ and (ii) $\sbdecision_{j_\tuple} \ne \dbdecision_{j_\tuple}$. Observation~\ref{observation:size} ensures that $\settuple$ will contain at least $\constant\sample$ tuples that correspond to papers from $\indset$. Consider any such tuple, and let $(j_{\tuple}, i_1, i_2)$ be a corresponding paper and two reviewers assigned to this paper in assignment $\testassignment$. Then conditioned on assignment $\testassignment$, $\prob{\sbdecision_{j_{\tuple}} \ne \dbdecision_{j_{\tuple}}}$ is lower bounded by:
\begin{align*}
    \prob{\sbdecision_{j_{\tuple}} \ne \dbdecision_{j_{\tuple}}} &= \frac{1}{2} \left(\sbpacc_{i_{1} j_{\tuple}} (1 - \dbpacc_{i_2 j_{\tuple}}) + \dbpacc_{i_{2} j_{\tuple}} (1 - \sbpacc_{i_1 j_{\tuple}}) \right) + \frac{1}{2} \left(\sbpacc_{i_{2} j_{\tuple}} (1 - \dbpacc_{i_1 j_{\tuple}}) + \dbpacc_{i_{1} j_{\tuple}} (1 - \sbpacc_{i_2  j_{\tuple}}) \right) \\ &\overset{{}}{\ge}  \frac{1}{2} \left(\sbpacc_{i_{1} j_{\tuple}} (1 - \dbpacc_{i_2 j_{\tuple}}) \right) + \frac{1}{2}\left( \sbpacc_{i_{2} j_{\tuple}} (1 - \dbpacc_{i_1 j_{\tuple}})\right) \\ & \overset{(i)}{\ge} \frac{1}{2} \left(\minigap^2 + \minigap^2 \right) = \minigap^2,
\end{align*}
where inequality $(i)$ follows from the fact that for any reviewer $i \in [\numrev]$ and for any paper $j \in [\numpap]$ we have $\minigap \le \sbpacc_{ij} \le 1$ and $0 \le \dbpacc_{ij} \le 1 - \minigap$ by the definition of non-trivial power requirement.

The same argument applies to tuples $t \in \settuple$ that correspond to papers from $\noindset$. Hence, we conclude that for any tuple $\tuple \in \settuple$ we are guaranteed that $\sbdecision_{j_{\tuple}} \ne \dbdecision_{j_{\tuple}}$ with probability at least $\minigap^2$.

Now notice that $|\arraymaj| = \sum\limits_{{t \in \settuple: \ \ind_{j_\tuple} = 1}} \indicator{\sbdecision_{j_{\tuple}} \ne \dbdecision_{j_{\tuple}}}$ and hence $\expectation{|\arraymaj|} \ge \constant \sample \minigap^2$. Applying Hoeffding's inequality, we can also derive that for large enough $\sample$ with probability at least $1 - \frac{\err}{4}$ we have
\begin{align*}
    |\arraymaj| > \constanta \sample \minigap^2.
\end{align*}
The same argument applies to $\arraymin$ and hence we conclude that with probability at least $1 - \frac{\err}{2}$ we have
\begin{align}
\label{eqn:cardinality}
     |\arraymaj| > \constanta \sample \minigap^2 \ \ \text{and} \ \ |\arraymin| > \constanta \sample \minigap^2.
\end{align}

\smallskip

\noindent \underline{Step 2. Distribution.}

\newcommand{\ptmp}{p}
\newcommand{\qtmp}{q}
\newcommand{\dpp}{\gamma_1}
\newcommand{\dqq}{\gamma_2}

\noindent Now we describe the distribution of components of $\arraymaj$ and $\arraymin$. By construction, the entries of these arrays are independent, so it suffices to study a single component. Consider an entry $\arraymajentry$ of array $\arraymaj$ and let $(j, i_1, i_2)$ be a corresponding paper and two reviewers assigned to this paper in assignment $\testassignment$. For brevity, denote $\ptmp = \sbpacc_{i_1 j} \in (\minigap, 1]$, $\qtmp = \dbpacc_{i_2 j} \in [0, 1 - \minigap)$, $\dpp = \ptmp - \dbpacc_{i_1 j}$ and $\dqq = \sbpacc_{i_2 j} - \qtmp$, where $\gamma_1 > \minigap$ and $\gamma_2 > \minigap$ by definition of non-trivial power requirement. Then, we can derive the following chain of bounds:
\begin{align*}
    2\prob{\arraymajentry = 1} - 1 &= 2\prob{\sbdecision_{j} = 1 | \sbdecision_{j} \ne \dbdecision_{j}} - 1 \\ &=  \frac{\ptmp(1 - \qtmp)}{\ptmp(1 - \qtmp) + \qtmp(1 - \ptmp)} + \frac{(\qtmp + \dqq)(1 - \ptmp + \dpp)}{(\qtmp + \dqq)(1 - \ptmp + \dpp) + (\ptmp - \dpp)(1 - \qtmp - \dqq)} - 1 \\& \overset{(i)}{\ge} \frac{\ptmp(1 - \qtmp)}{\ptmp(1 - \qtmp) + \qtmp(1 - \ptmp)} + \frac{(\qtmp + \minigap)(1 - \ptmp + \minigap)}{(\qtmp + \minigap)(1 - \ptmp + \minigap) + (\ptmp - \minigap)(1 - \qtmp - \minigap)} - 1 \\&= \frac{1}{2} \left( \frac{\ptmp - \qtmp}{\ptmp + \qtmp - 2 \ptmp \qtmp} - \frac{\ptmp - \qtmp - 2 \minigap}{\ptmp + \qtmp - 2\ptmp \qtmp  + 2 \minigap (\minigap + \qtmp - \ptmp)}  \right) 
\end{align*}
where inequality $(i)$ holds due to monotonicity of the expression over $\dpp$ and $\dqq$ and lower bounds $\dpp > \minigap$, $\dqq > \minigap$.

Optimizing the last expression over $p \in (\minigap, 1]$ and $q \in [0, 1 - \minigap)$, we obtain
\begin{align*}
    2\prob{\arraymajentry = 1} - 1  \ge  \frac{\minigap^2}{\minigap^2 + (1 - \minigap)^2},
\end{align*}
and hence $\prob{\arraymajentry = 1} \ge \frac{1}{2} + \frac{1}{2} \frac{\minigap^2}{\minigap^2 + (1 - \minigap)^2} = \frac{1}{2} + \gamma $. Similarly, we can show that $\prob{\arrayminentry = 1} \le \frac{1}{2} - \frac{1}{2} \frac{\minigap^2}{\minigap^2 + (1 - \minigap)^2} = \frac{1}{2} - \gamma$, where $\gamma > 0$ is a constant that depends on $\minigap$.

\smallskip

\noindent \underline{Step 3. Permutation.}

\noindent At this point we are guaranteed that vectors $\arraymin$ and $\arraymaj$ constructed in Step~\ref{step:perm:majority} of the \permutation{} test, with probability $1 - \frac{\err}{2}$, contain at least $\constant \sample \minigap^2$ elements and their entries are independent Bernoulli random variables. Moreover, the entries of $\arraymaj$ have expectations larger than $\nicefrac{1}{2} + \gamma$ and entries of $\arraymin$  have expectations smaller than $\nicefrac{1}{2} - \gamma$, where $\gamma$ is independent of $\sample$.  

Conditioned on $\min\{|\arraymin|, |\arraymaj|\} > \constant \sample \minigap^2$, notice that as $\sample$ grows, the permutation test for exchangeablility of entries of $\arraymin$ and $\arraymaj$ has power growing to 1. Hence, there exists $\sample^*$ such that if $\sample > \sample^*$, then the permutation test rejects the null with probability at least $1 - \frac{\err}{2}$.

Finally, taking union bound over (i) probability that either of $\arraymaj$ and $\arraymin$ has cardinality smaller than $\constant \sample \minigap^2$ and (ii) probability that the permutation test fails to reject the null given $\min\{|\arraymin|, |\arraymaj|\} > \constant \sample \minigap^2$, we deduce that conditioned on $\testassignment$, the requirement of non-trivial power is satisfied. It now remains to notice that the established fact holds for any $\testassignment$ that is constructed by Procedure~\ref{algo:experiment} and hence Theorem~\ref{prop:absolute}(a) holds. 

\subsubsection{Proof for \counting{} test}

Similar to the proof for the\permutation{} test, the proof for the \counting{} test consists of two parts --- control over Type-I error and non-trivial power. As in the proof for the \permutation{} test, we prove both parts conditioned on the assignment $\testassignment$ computed in Step~\ref{step:assign:selection} of Procedure~\ref{algo:experiment}. The unconditional statement of the theorem then follows from the law of total probability.

\smallskip

\noindent \textbf{Control over Type-I error}

\noindent Let $\dbmatrix$ and $\sbmatrix (= \dbmatrix)$ be arbitrary matrices that fall under the definition of the null hypothesis in Problem~\ref{problem:absolute}. Consider arrays $\arraymaj$ and $\arraymin$ constructed in Step~\ref{step:count:allocation} of the \counting{} test. If any of them is empty, the test keeps the null and hence does not commit the Type-I error. Now without loss of generality assume that both $\arraymaj$ and $\arraymin$ are non-empty. By construction, conditioned on the assignment $\testassignment$, entries of arrays $\arraymaj$ and $\arraymin$ are mutually independent and bounded by 1 in absolute value. Moreover, conditioned on $\testassignment$ the size of arrays $\arraymaj$ and $\arraymin$ is fixed and is not a random variable. Next, we can show that expectation of any entry of arrays $\arraymaj$ and $\arraymin$ is zero. Indeed, consider any arbitrary entry $\arraymajentry \in \arraymaj$ and let $(j, i_1, i_2)$ be a corresponding paper and reviewers assigned to this paper in assignment $\testassignment$. Then:
\begin{align*}
    \expectation{\arraymajentry} = \frac{1}{2} \left(\sbpacc_{i_1 j} - \dbpacc_{i_2 j} \right) + \frac{1}{2} \left(\sbpacc_{i_2 j} - \dbpacc_{i_1 j} \right) = 0,
\end{align*}
where two terms correspond to two equiprobable allocations of reviewers $i_1$ and $i_2$ to conditions and the last equality follows from the fact that under the null hypothesis $\sbmatrix = \dbmatrix$. Hence, we conclude that the expectation of test statistic $\statc$ equals 0. Independence and boundedness of entries of arrays $\arraymin$ and $\arraymaj$ ensure that the test statistic $\statc$ is sub-Gaussian random variable with noise parameter $\noise$ given by
\begin{align*}
    \noise^2 = |\arraymaj|^{-1} + |\arraymin|^{-1}.
\end{align*}

Finally, applying Hoeffding's inequality we deduce that
\begin{align*}
    \prob{|\statc| > \sqrt{2 \left(|\arraymaj|^{-1} + |\arraymin|^{-1} \right) \log \nicefrac{2}{\level}}} \le 2 \exp\left\{- \frac{2 \left(|\arraymaj|^{-1} + |\arraymin|^{-1} \right) \log \nicefrac{2}{\level}}{2\left(|\arraymaj|^{-1} + |\arraymin|^{-1} \right)} \right\} = \level,
\end{align*}
which concludes the proof.

\noindent \textbf{Non-trivial power}

\noindent Consider any fixed choice of $\minigap > 0$ and $\err > 0$ in the definition of non-trivial power. The goal now is to show that there exists $\sample > 0$ such that if $\min\{|\indset|, |\noindset| \} > \sample$, then for any matrices $\dbmatrix$ and $\sbmatrix$ that satisfy the alternative hypothesis in Problem~\ref{problem:absolute} with margin $\minigap$, the \counting{} test coupled with Procedure~\ref{algo:experiment} rejects the null hypothesis with probability at least $1 - \varepsilon$. Throughout the proof we use $\constant$ to denote a universal constant and allow its value to change from line to line due to multiplications by some other universal constants. Recall that problem parameters $\papload, \revload$ and $\level$ are treated as constants. For concreteness, suppose that there is a bias in favor of papers that satisfy the property of interest.

We now consider an arbitrary instance of the bias testing problem with matrices $\sbmatrix$ and $\dbmatrix$ that fall under the definition of non-trivial power. First, Observation~\ref{observation:size} ensures that the set $\settuple$ passed to the \counting{} algorithm is such that the resulting vectors $\arraymaj$ and $\arraymin$ contain at least $\constant n_0$ elements each. Next, let $\statc_1 = \frac{1}{|\arraymaj|} \sum\limits_{\arraymajentry \in \arraymaj} \arraymajentry$ and $\statc_2 = \frac{1}{|\arraymin|} \sum\limits_{\arrayminentry \in \arraymin} \arrayminentry$, in this notation the test statistic is defined as $\statc = \statc_1 - \statc_2$. Conditioned on the assignment $\testassignment$, we have:
\begin{align*}
    \expectation{\statc_1} = \frac{1}{|\arraymaj|} \sum\limits_{\arraymajentry \in \arraymaj} \expectation{\arraymajentry} \ge \minigap.
\end{align*}
Indeed, for any arbitrary entry $\arraymajentry$ of array $\arraymaj$ let $(j, i_1, i_2)$ be corresponding paper and reviewers assigned to this paper in assignment $\testassignment$. Then requirement of non-trivial power guarantees that
\begin{align*}
    \expectation{\arraymajentry} = \frac{1}{2} \left(\sbpacc_{i_1j} - \dbpacc_{i_2 j} \right) + \frac{1}{2} \left(\sbpacc_{i_2j} - \dbpacc_{i_1 j} \right) \ge \minigap + \frac{1}{2} \left(\dbpacc_{i_1j} - \dbpacc_{i_2 j} \right) + \frac{1}{2} \left(\dbpacc_{i_2j} - \dbpacc_{i_1 j} \right) = \minigap.
\end{align*}
Similarly,
\begin{align*}
    \expectation{\statc_2} = \frac{1}{|\arraymin|} \sum\limits_{\arrayminentry \in \arraymin} \expectation{\arrayminentry} \le - \minigap.
\end{align*}

Applying Hoeffding's inequality we obtain:
\begin{align*}
    \prob{\statc_1 - \statc_2  < \minigap} \le \prob{\statc_1 - \statc_2  < \expectation{\statc_1 - \statc_2} - \minigap} \le  \exp\left(- \frac{\minigap^2}{2\left(|\arraymin|^{-1} + |\arraymaj|^{-1}  \right)} \right) \le \exp\left( -\constant \minigap^2 n_0 \right).
\end{align*}
On the other hand, the threshold for rejecting the null is such that 
\begin{align*}
    \sqrt{2 \left(|\arraymaj|^{-1} + |\arraymin|^{-1} \right) \log \nicefrac{2}{\level}} \le \constant \sqrt{\frac{1}{n_0}}. 
\end{align*}
Finally, setting $n_0 = \constant \frac{\log \nicefrac{1}{\err}}{\minigap^2}$, we ensure that if $\min\{|\indset|, |\noindset|\} > n_0$, then the \counting{} algorithm with probability at least $1 - \err$ rejects the null for any matrices $\sbmatrix$, $\dbmatrix$ that satisfy alternative hypothesis with margin $\minigap$. 

%%%%%%%%%%%%%%%%%%%%%%%%%%%%%%%%%%%%%%%%%%%%%%%%%%%%%%%%%%%

\subsection{Proof of Theorem~\ref{prop:relative}}

We prove Theorem~\ref{prop:relative} separately for the \permutation{} test and for the \counting{} tests. 

\subsubsection{Proof for \permutation{} test}

Again, the proof is presented in two parts: control over Type-I error and non-trivial power. The conceptual difference from the proof of the corresponding result for absolute bias problem is that now the parametric relationships~\eqref{eqn:genlogregmodel} and~\eqref{eqn:genlogregmodelsb} allow us to avoid conditioning on the assignment $\testassignment$.

\smallskip

\noindent \textbf{Control over Type-I error}

\noindent Let $\dbmatrix$ and $\sbmatrix$ be arbitrary matrices generated from the generalized logistic model under the absence of bias.  Consider arrays $\arraymaj$ and $\arraymin$ constructed in Step~\ref{step:perm:majority} of the \permutation{} test from the set of tuples $\settuple$ passed to the test by Procedure~\ref{algo:experiment}. If any of them is empty, the test keeps the null and hence does not commit the Type-I error. Now without loss of generality assume that both arrays $\arraymaj$ and $\arraymin$ are non-empty. 
Following the idea of the proof of Theorem~\ref{prop:absolute}, we need to show that entries of arrays $\arraymaj$ and $\arraymin$ are exchangeable random variables. First, the mutual independence follows from construction of the set $\settuple$. Second, using equations~\eqref{eqn:genlogregmodel} and~\eqref{eqn:genlogregmodelsb}, we deduce that for any paper $j \in [\numpap]$ and for any reviewer $i \in [\numrev]$:
\begin{align*}
    \log \frac{\sbpacc_{ij} (1 - \dbpacc_{ij})}{\dbpacc_{ij} (1 - \sbpacc_{ij})} = \logshift.
\end{align*}
Noticing that $\sbpacc_{ij}$ and $\dbpacc_{ij}$ under the generalized logistic model are independent of reviewer's identity, we drop index $i$ from the above equation. Now we consider any entry $\arraymajentry$ of array $\arraymaj$ together with a corresponding tuple $\tuple = (j_{\tuple}, \sbdecision_{j_{\tuple}}, \dbdecision_{j_{\tuple}}, \ind_{j_{\tuple}})$ and conclude that:
\begin{align}
    \nonumber \prob{\arraymajentry = 1 } &= \prob{\sbdecision_{j_{\tuple}} = 1 | \sbdecision_{j_{\tuple}} \ne \dbdecision_{j_{\tuple}}} \\ \nonumber &= \frac{\sbpacc_{j_{\tuple}} (1 - \dbpacc_{j_{\tuple}})}{\sbpacc_{j_{\tuple}} (1 - \dbpacc_{j_{\tuple}}) + \dbpacc_{j_{\tuple}} (1 - \sbpacc_{j_{\tuple}})} \\& \nonumber = \frac{1}{1 + \frac{\dbpacc_{j_{\tuple}} (1 - \sbpacc_{j_{\tuple}})}{\sbpacc_{j_{\tuple}} (1 - \dbpacc_{j_{\tuple}})}} \\ & = \label{eqn:nopaprep}\frac{1}{1 + e^{-\logshift}}. 
\end{align}

Importantly, the value of the paper representation $\paprepresentation_j$ does not appear in equation~\eqref{eqn:nopaprep}, implying that entries of array $\arraymaj$ are identically distributed. Applying the same argument to entries of array $\arraymin$ we deduce that entries of arrays $\arraymaj$ and $\arraymin$ are exchangeable random variables and hence the permutation test with the test statistic $\statp$ defined in Step~\ref{step:perm:decision} of Test~\ref{test:permutation} is guaranteed to control for the Type-I error rate at any given significance level $\level \in (0, 1)$ which concludes the proof. 

%%%%%%%%%%%%%%%%%%%%%%%%%%%%%%%%%%%%%%%%%%%%%%%%%%%%%%%%%%%

\smallskip

\noindent \textbf{Non-trivial power}

\noindent Consider any fixed choice of $\minigap > 0$ and $\err > 0$ in the definition of non-trivial power. The goal now is to show that there exists $\sample = \sample(\err, \minigap)$ such that if $\min\{|\indset|, |\noindset| \} > \sample$, then for any matrices $\dbmatrix$ and $\sbmatrix$ generated from the generalized logistic model that satisfy the alternative hypothesis in Problem~\ref{problem:relative} with margin $\minigap$, the \permutation{} test coupled with Procedure~\ref{algo:experiment} is guaranteed to reject the null hypothesis with probability at least $1 - \err$. Throughout the proof we use $\constant$ to denote a universal constant and allow its value to change from line to line due to multiplications by some other universal constants. Recall that problem parameters $\papload, \revload$ and $\level$ are treated as constants. For concreteness, throughout the proof we assume that the bias is in favor of papers from $\indset$. The same argument can be repeated in case of bias against papers from $\indset$. 

\smallskip

\noindent \underline{Step 1. Cardinality of $\arraymaj$ and $\arraymin$.}

\noindent Consider any matrices $\sbmatrix$ and $\dbmatrix$ generated from the generalized logistic model that satisfy the alternative hypothesis in Problem~\ref{problem:relative} with margin $\minigap$. First, we notice that scores $\paprepresentation_j, j \in [\numpap]$, and coefficients $\coef_0, \coef_1$ are bounded in absolute value by some constant $\bound$, and hence using equation~\eqref{eqn:genlogregmodel} we conclude that for all $(i, j) \in [\numpap] \times [\numrev]$
\begin{align}
\label{eqn:boundaaa}
    \dbpacc_{ij} \in (\lb, \ub)  \quad \forall j \in [\numpap],
\end{align}
where $0 < \lb < \ub < 1$ and values of $\lb$ and $\ub$ are determined by $\bound$. Now consider any tuple $\tuple = (j_{\tuple}, \sbdecision_{i_1 j_{\tuple}}, \dbdecision_{i_2 j_{\tuple}}, \ind_{j_{\tuple}})$ from the set of tuples $\settuple$. Then
\begin{align*}
    \prob{\sbdecision_{j_{\tuple}} \ne \dbdecision_{j_{\tuple}}} & = \sbpacc_{i_1 j_{\tuple}}(1 - \dbpacc_{i_2 j_{\tuple}}) + \dbpacc_{i_2 j_{\tuple}}(1 - \sbpacc_{i_1 j_{\tuple}}) \\ &\ge \min\{\dbpacc_{i_2 j_{\tuple}}, 1 - \dbpacc_{i_2 j_{\tuple}} \} \left( \sbpacc_{i_1 j_{\tuple}} + 1 - \sbpacc_{i_1 j_{\tuple}} \right)  \\ &=
    \min\{\dbpacc_{i_2 j_{\tuple}}, 1 - \dbpacc_{i_2 j_{\tuple}} \} \\ & \ge \min\{\lb, 1 - \ub \}, 
\end{align*}
where the last inequality follows from equation~\eqref{eqn:boundaaa}. Applying Hoeffding's inequality in the same way as we did in the proof of Theorem~\ref{prop:absolute} to get the bound~\eqref{eqn:cardinality}, we deduce that with probability at least $ 1 - \frac{\err}{2}$, cardinalities of arrays $\arraymaj$ and $\arraymin$ are at least $\constant \sample$ for some constant $\constant$ that may depend on $\minigap$ and $\bound$.

\smallskip

\noindent \underline{Step 2. Distribution}

\noindent By definition of non-trivial power requirement, it must be the case that for all $(i, j) \in [\numrev] \times [\numpap]$ we have $|\sbpacc_{ij} - \nullshift(\dbpacc_{ij})| > \minigap$, where function $\nullshift$ belongs to class $\logregclass$ defined in~\eqref{eqn:logmember} and for all $(i, j) \in [\numrev] \times [\numpap]$ satisfies: 
\begin{subequations}
\begin{align}
\label{eqn:anticipated}
    \log \frac{\nullshift(\dbpacc_{ij})}{1 - \nullshift(\dbpacc_{ij})} &= \coef_0 + \logshift + \coef_1 \paprepresentation_j \\ &= \label{anticipated2} \log \frac{\dbpacc_{ij}}{1 - \dbpacc_{ij}} + \logshift,
\end{align}
\end{subequations}
for some value of $\logshift \in (-\bound, \bound)$. Observe that values $\coef_0, \logshift, \coef_1, \paprepresentation_j$ in the RHS of equation~\eqref{eqn:anticipated} are bounded in absolute value by constant $\bound$. Next, recall that the definition of the non-trivial power requirement ensures that for each reviewer $i \in [\numrev]$ it must be the case that (a) for each paper $j \in \indset$ we have $\nullshift(\dbpacc_{ij}) < 1 - \minigap$ and (b) for each paper $j \in \noindset$ we have $\nullshift(\dbpacc_{ij}) > \minigap$. Finally, we are guaranteed that for any pair of reviewer $i \in [\numrev]$ and paper $j \in [\numpap]$ we have
\begin{align*}
    \nullshift(\dbpacc_{ij}) \in
    \begin{cases}
         \left(\lb', \min\{\ub', 1 - \minigap\}\right) & \text{if } j \in \indset \\
         \left(\max\{\lb', \minigap\}, \ub'\right) & \text{if } j \in \noindset.
    \end{cases}
\end{align*}
Notice that constants $\lb'$ and $\ub'$ are such that  $0 < \lb' < \ub' < 1$ and may be different from $\lb$ and $\ub$, because in equation~\eqref{eqn:anticipated} we have additional term $\logshift$ which is absent in~\eqref{eqn:genlogregmodel}.

Let us now define two quantities $\upinc$ and $\lrinc$ as 
\begin{subequations}
\begin{align}
\label{eqn:infa}
    \upinc &= \inf\limits_{t \in \left(\lb', \min\{\ub', 1 - \minigap\}\right)} \left(\log \frac{t + \minigap}{1 - (t + \minigap)} - \log \frac{t}{1 - t}\right) \\ \label{eqn:infb}
    \lrinc &= \inf\limits_{t \in \left(\max\{\lb', \minigap\}, \ub'\right)} \left(\log \frac{t}{1 - t} - \log \frac{t - \minigap}{1 - (t - \minigap)}\right). 
\end{align}
\end{subequations}
Notice that both quantities $\upinc$ and $\lrinc$ are some functions of $\minigap$ and $\bound$ and are strictly positive, because function $\log\frac{x}{1-x}$ is strictly increasing on the interval $(0, 1)$ with its derivative being lower bounded by $\constant > 0$, where $\constant$ is independent of problem parameters.

Putting together equations~\eqref{eqn:anticipated} - \eqref{eqn:infb}, we now show that for each reviewer $i \in [\numrev]$ the definition of non-trivial power requirement ensures that for each paper $j \in \indset$ 
\begin{align}
\label{eqn:i1}
    \log \frac{\sbpacc_{ij}}{1 - \sbpacc_{ij}} \ge \log \frac{\dbpacc_{ij}}{1 - \dbpacc_{ij}} + \logshift + \upinc,
\end{align}
and for each paper $j \in \noindset$
\begin{align}
\label{eqn:i2}
    \log \frac{\sbpacc_{ij}}{1 - \sbpacc_{ij}} \le \frac{\dbpacc_{ij}}{1 - \dbpacc_{ij}} + \logshift - \lrinc.
\end{align}

Consider any arbitrary entry $\arraymajentry$ of array $\arraymaj$ and the corresponding tuple $(j_{\tuple}, \sbdecision_{i_1 j_{\tuple}}, \dbdecision_{i_2 j_{\tuple}}, \ind_{j_{\tuple}})$. Then,
\begin{align*}
    \prob{\arraymajentry = 1} &= \prob{\sbdecision_{i_1 j_{\tuple}} = 1 | \sbdecision_{i_1 j_{\tuple}} \ne \dbdecision_{i_2 j_{\tuple}}} \\ &= \frac{ \sbpacc_{i_1 j_{\tuple}} (1 - \dbpacc_{i_2 j_{\tuple}}) }{\sbpacc_{i_1 j_{\tuple}} (1 - \dbpacc_{i_2 j_{\tuple}}) + \dbpacc_{i_2 j_{\tuple}} (1 - \sbpacc_{i_1 j_{\tuple}})} \\ &= \frac{1}{1 + \frac{\dbpacc_{i_2 j_{\tuple}} (1 - \sbpacc_{i_1 j_{\tuple}})}{\sbpacc_{i_1 j_{\tuple}} (1 - \dbpacc_{i_2 j_{\tuple}})}} \\ &\ge \frac{1}{1 + e^{-\logshift - \upinc}},
\end{align*}
where the last inequality follows from~\eqref{eqn:i1}. Similarly, using~\eqref{eqn:i2} we show that for each entry $\arrayminentry$ of array $\arraymin$
\begin{align*}
    \prob{\arrayminentry = 1} \le \frac{1}{1 + e^{-\logshift + \lrinc}}.
\end{align*}

\noindent \underline{Step 3. Permutation.}

\noindent At this point we are guaranteed that vectors $\arraymin$ and $\arraymaj$ constructed in Step~\ref{step:perm:majority} of the \permutation{} test, with probability $1 - \frac{\err}{2}$, contain at least $\constant \sample$ elements and their entries are independent Bernoulli random variables. Moreover, the entries of $\arraymaj$ have expectations larger than $\frac{1}{1 + e^{-\logshift}} + \gamma$ and entries of $\arraymin$  have expectations smaller than $\frac{1}{1 + e^{-\logshift}} - \gamma$, where $\gamma$ is independent of $\sample$, but depends on $\minigap$ and $\bound$. 

Conditioned on $\min\{|\arraymin|, |\arraymaj|\} > \constant \sample $, notice that as $\sample$ grows, the permutation test for exchangeablility of entries of $\arraymin$ and $\arraymaj$ has power growing to 1. Hence, there exists $\sample^*$ such that if $\sample > \sample^*$, then the permutation test rejects the null with probability at least $1 - \frac{\err}{2}$.

Finally, taking union bound over (i) probability that either of $\arraymaj$ and $\arraymin$ has cardinality smaller than $\constant \sample$ and (ii) probability that the permutation test fails to reject the null given $\min\{|\arraymin|, |\arraymaj|\} > \constant \sample$, we deduce that the requirement of non-trivial power is satisfied.

\smallskip

\subsubsection{Proof for \counting{} test}

We give a proof for an extended version of the generalized linear model in which for each $(i, j) \in [\numpap] \times [\numrev]$ we substitute $\paprepresentation_j$ with $\paprepresentation_{ij}$, thus allowing subjectivity of reviewers. In the proof we will be using two observations we made in the beginning of Appendix~\ref{appendix:proof:absolute}. As in the proof of Theorem~\ref{prop:absolute}, we prove the result conditioned on the assignment $\testassignment$ constructed in Step~\ref{step:assign:selection} of Procedure~\ref{algo:experiment}. The unconditional statement of the theorem then follows from the law of total probability. 

\smallskip

\noindent \textbf{Control over Type-I error}

\noindent Let $\dbmatrix$ and $\sbmatrix$ be arbitrary matrices generated under the generalized linear model that fall under the null hypothesis in Problem~\ref{problem:relative}. Consider arrays $\arraymaj$ and $\arraymin$ constructed in Step~\ref{step:count:allocation} of the \counting{} test. If any of them is empty, the test keeps the null and hence does not commit the Type-I error. Now without loss of generality assume that both $\arraymaj$ and $\arraymin$ are non-empty. By construction, conditioned on the assignment $\testassignment$, entries of arrays $\arraymaj$ and $\arraymin$ are mutually independent and bounded by 1 in absolute value. Moreover, conditioned on $\testassignment$ the size of arrays $\arraymaj$ and $\arraymin$ is fixed and is not a random variable. Next, for any arbitrary entry $\arraymajentry \in \arraymaj$ let $(j, i_1, i_2)$ be a corresponding paper and reviewers assigned to this paper in assignment $\testassignment$. Then,
\begin{align*}
    \expectation{\arraymajentry} &= \frac{1}{2} \left(\sbpacc_{i_1 j} - \dbpacc_{i_2 j} \right) + \frac{1}{2} \left(\sbpacc_{i_2 j} - \dbpacc_{i_1 j} \right) \\ &= \frac{1}{2} \left(\paprepresentation_{i_1j} + \shift - \paprepresentation_{i_2 j} \right) + \frac{1}{2} \left(\paprepresentation_{i_2j} + \shift - \paprepresentation_{i_1 j} \right) \\ &= \shift.
\end{align*}
Similarly, it follows that for any arbitrary entry $\arrayminentry \in \arraymin$:
\begin{align*}
    \expectation{\arrayminentry} = \shift.
\end{align*}
Hence, we conclude that the expectation of the test statistic $\statc$ equals 0. Independence and boundedness of entries of arrays $\arraymin$ and $\arraymaj$ ensure that the test statistic $\statc$ is sub-Gaussian random variable with noise parameter $\noise$ given by
\begin{align*}
    \noise^2 = |\arraymaj|^{-1} + |\arraymin|^{-1}.
\end{align*}

Finally, applying Hoeffding's inequality we deduce that
\begin{align*}
    \prob{|\statc| > \sqrt{2 \left(|\arraymaj|^{-1} + |\arraymin|^{-1} \right) \log \nicefrac{2}{\level}}} \le 2 \exp\left\{- \frac{2 \left(|\arraymaj|^{-1} + |\arraymin|^{-1} \right) \log \nicefrac{2}{\level}}{2\left(|\arraymaj|^{-1} + |\arraymin|^{-1} \right)} \right\} = \level,
\end{align*}
which concludes the proof.
\medskip

\noindent \textbf{Non-trivial power}

\noindent Consider any fixed choice of $\minigap > 0$ and $\err > 0$ in the definition of non-trivial power. The goal now is to show that there exists $\sample > 0$ such that if $\min\{|\indset|, |\noindset| \} > \sample$, then for any matrices $\dbmatrix$ and $\sbmatrix$ generated under the generalized linear model that satisfy the alternative hypothesis in Problem~\ref{problem:relative} with margin $\minigap$, the \counting{} test coupled with Procedure~\ref{algo:experiment} rejects the null hypothesis with probability at least $1 - \varepsilon$. Throughout the proof we use $\constant$ to denote a universal constant and allow its value to change from line to line due to multiplications by some other universal constants. Recall that problem parameters $\papload, \revload$ and $\level$ are treated as constants. For concreteness, suppose that there is a bias in favor of papers that satisfy the property of interest.

We now consider an arbitrary instance of the bias testing problem with matrices $\sbmatrix$ and $\dbmatrix$ that fall under the definition of non-trivial power. First, Observation~\ref{observation:size} ensures that the set $\settuple$ passed to the \counting{} algorithm is such that resulting vectors $\arraymaj$ and $\arraymin$ contain at least $\constant n_0$ elements each. Next, let $\statc_1 = \frac{1}{|\arraymaj|} \sum\limits_{\arraymajentry \in \arraymaj} \arraymajentry$ and $\statc_2 = \frac{1}{|\arraymin|} \sum\limits_{\arrayminentry \in \arraymin} \arrayminentry$, in this notation the test statistic is defined as $\statc = \statc_1 - \statc_2$. Conditioned on the assignment $\testassignment$, we have:
\begin{align*}
    \expectation{\statc_1} = \frac{1}{|\arraymaj|} \sum\limits_{\arraymajentry \in \arraymaj} \expectation{\arraymajentry} \ge \shift + \minigap.
\end{align*}
Indeed, for any arbitrary entry $\arraymajentry$ of array $\arraymaj$ let $(j, i_1, i_2)$ be corresponding paper and reviewers assigned to this paper in assignment $\testassignment$. Then the definition of the non-trivial power guarantees that
\begin{align*}
    \expectation{\arraymajentry} &= \frac{1}{2} \left(\sbpacc_{i_1j} - \dbpacc_{i_2 j} \right) + \frac{1}{2} \left(\sbpacc_{i_2j} - \dbpacc_{i_1 j} \right) \\ &\ge \frac{1}{2} \left(\paprepresentation_{i_1 j} + \shift + \minigap - \paprepresentation_{i_2 j} \right) + \frac{1}{2} \left(\paprepresentation_{i_2 j} + \shift + \minigap - \paprepresentation_{i_1 j} \right) \\ &= \shift + \minigap.
\end{align*}
Similarly,
\begin{align*}
    \expectation{\statc_2} = \frac{1}{|\arraymin|} \sum\limits_{\arrayminentry \in \arraymin} \expectation{\arrayminentry} \le \shift - \minigap.
\end{align*}

Applying Hoeffding's inequality we obtain:
\begin{align*}
    \prob{\statc_1 - \statc_2  < \minigap} \le \prob{\statc_1 - \statc_2  < \expectation{\statc_1 - \statc_2} - \minigap} \le  \exp\left(- \frac{\minigap^2}{2\left(|\arraymin|^{-1} + |\arraymaj|^{-1}  \right)} \right) \le \exp\left( -\constant \minigap^2 n_0 \right).
\end{align*}
On the other hand, the threshold for acceptance is such that 
\begin{align*}
    \sqrt{2 \left(|\arraymaj|^{-1} + |\arraymin|^{-1} \right) \log \nicefrac{2}{\level}} \le \constant \sqrt{\frac{1}{n_0}}. 
\end{align*}
Finally, setting $n_0 = \constant \frac{\log \nicefrac{1}{\err}}{\minigap^2}$, we ensure that if $\min\{|\indset|, |\noindset|\} > n_0$, then the \counting{} algorithm with probability at least $1 - \err$ rejects the null for any matrices $\sbmatrix$, $\dbmatrix$ that satisfy the alternative hypothesis with margin $\minigap$. 

%%%%%%%%%%%%%%%%%%%%%%%%%%%%%%%%%%%%%%%%%%%%%%%%%%%%%%%%%%%

\subsection{Proof of Theorem~\ref{prop:richness}}

Assume that the premises of Theorem~\ref{prop:richness} are satisfied, that is, there exist functions $g, h \in \subclass$ and values $0 \le x_1 < x_2 \le 1$ such that $g(x_1) < h(x_1)$ and $g(x_2) > h(x_2)$. 

The high-level idea of the proof is to construct matrices $\dbmatrix$ and $\sbmatrix$ which simultaneously satisfy the null hypothesis of Problem~\ref{problem:relative} specified by some function $\nullshift \in \subclass$ and the alternative hypothesis of Problem~\ref{problem:relative} specified by another function $\nullshift' \in \subclass$ with margin $\minigap > 0$. If such matrices exist, then there exist two instances of a bias testing problem --- one with presence of bias and the other with absence of bias --- such that the distributions of the reviewers' decisions for these two instances coincide. Hence, any test that uniformly controls for the Type-I error rate at the level $\level$ for every $\nullshift \in \subclass$ must under the second instance have power upper bounded by $\level$ and thus violate the requirement of non-trivial power over the class of functions $\subclass$.

We begin with building a matrix $\dbmatrix$. For any reviewer $i \in [\numrev]$ and for any paper $j \in [\numpap]$ we let
\begin{align*}
    \dbpacc_{ij} = \begin{cases}
                        x_1 & \text{if } \ind_j = 1 \\
                        x_2 & \text{if } \ind_j = -1.
                   \end{cases}
\end{align*}
Next, we define $\sbmatrix$ as follows. For any reviewer $i \in [\numrev]$ and for any paper $j \in [\numpap]$
\begin{align*}
    \sbpacc_{ij} = h(\dbpacc_{ij}).
\end{align*}

By construction matrices $\sbmatrix$ and $\dbmatrix$ satisfy the null hypothesis specified by function $h \in \subclass$. On the other hand, notice that for each paper $j \in \indset$  we have
\begin{align*}
    \sbpacc_{ij} = h(x_1) > g(x_1) = g(\dbpacc_{ij}),
\end{align*}
and for each paper $j \in \noindset$ we have
\begin{align*}
    \sbpacc_{ij} = h(x_2) < g(x_2) = g(\dbpacc_{ij}).
\end{align*}

Hence, matrices $\dbmatrix$ and $\sbmatrix$ also satisfy the alternative hypothesis specified by function $g$. Moreover, $\dbmatrix$ and $\sbmatrix$ satisfy this alternative with margin $\minigap = \min\{|h(x_1) - g(x_1)|, |h(x_2) - g(x_2)| \} > 0$. We now conclude the proof by noting that our construction holds for any choice of parameters $\papload, \revload, \numpap, \numrev$ and hence the requirement of non-trivial power must be violated by any testing algorithm that controls for Type-I error at the level $\level \in (0, 1)$.

\subsection{Proof of Corollary~\ref{prop:joint}}

\newcommand{\altnullshift}{\widetilde{\nullshift}}

To prove Corollary~\ref{prop:joint}, we consider any choice of parameters $\gap \in (0, 0.5)$ and $\bound > 0$ and construct two functions $\nullshift$ and $\altnullshift$ together with two numbers $0 \le x_1 < x_2 \le 1$ such that 
\begin{enumerate}[itemsep=3pt, leftmargin=*, label=(\roman*)]
    \item Functions $\nullshift$ and $\altnullshift$ describe the behavior of reviewers under the absence of bias under the generalized linear and generalized logistic models respectively, that is,  $\nullshift \in \linmodel, \ \altnullshift \in \logregclass$, where class $\linmodel$ is defined by equation~\eqref{eqn:linmember} and class $\logregclass$ is specified in equation~\eqref{eqn:logmember}
    
    \item Values $x_1$ and $x_2$ are such that:
    \begin{enumerate}[itemsep=1pt, label=(\alph*)]
        \item One can select parameters $\tscore_j, j \in [\numpap]$, that fall under the definition of the generalized linear model such that matrix $\dbmatrix$ generated according to the equation~\eqref{eqn:genlinregmodel} satisfies the following equation:
        \begin{align}
        \label{eqn:cndii}
        \dbpacc_{ij} = \begin{cases}
                                x_1 &\text{if } j \in \indset \\
                                x_2 &\text{if } j \in \noindset.
                           \end{cases}
        \end{align}
        
        \item One can select parameters $\tscore_j, j \in [\numpap],$ and $\coef_0, \coef_1$ that fall under the definition of the generalized logistic model such that matrix $\dbmatrix$ generated according to the equation~\eqref{eqn:genlogregmodel} satisfies the equation~\eqref{eqn:cndii}.
    \end{enumerate}

    \item Functions $\nullshift$ and $\altnullshift$ are such that
    \begin{align*}
        \text{sign}\left(\nullshift(x_1) - \altnullshift(x_1)\right) \times \text{sign}\left(\nullshift(x_2) - \altnullshift(x_2)\right) = -1,
    \end{align*}
    where $\text{sign}(\cdot)$ is the sign function. That is, at $x_1$ the function $\nullshift$ is strictly larger than $\altnullshift$ and at $x_2$ the function $\nullshift$ is strictly smaller than $\altnullshift$, or vice versa.
\end{enumerate}

Assume for the moment that conditions (i)-(iii) are satisfied and consider the matrix $\dbmatrix$ whose entries are given by equation~\eqref{eqn:cndii}. Then one can select values of papers' representations $\paprepresentation_j, j \in [\numpap],$ such that $\dbmatrix$ satisfies the model of DB reviewers in the generalized linear model~\eqref{eqn:genlinregmodel}. Similarly, there exists another choice of papers' representations $\paprepresentation'_j, j \in [\numpap],$ and parameters $\coef_0, \coef_1$, such that the same matrix $\dbmatrix$ satisfies the model of DB reviewers in the generalized logistic model~\eqref{eqn:genlogregmodel}. Now define matrix $\sbmatrix_1$ whose entries for each $(i, j) \in [\numrev]\times[\numpap]$ are given by:
\begin{align*}
    \sbpacc_{ij} = \nullshift(\dbpacc_{ij})
\end{align*}
and matrix $\sbmatrix_2$ whose entries for each $(i, j) \in [\numrev]\times[\numpap]$ are given by:
\begin{align*}
    \sbpacc_{ij} = \altnullshift(\dbpacc_{ij}).
\end{align*}
Matrices $\dbmatrix, \sbmatrix_1$ satisfy the null hypothesis under the generalized \emph{linear} model specified by the function $\nullshift$. Moreover, condition (iii) ensures that they \emph{simultaneously} satisfy the alternative hypothesis under the generalized \emph{logistic} model specified by the function $\altnullshift$ with margin $\minigap = \min\{|\nullshift(x_1) - \altnullshift(x_1)|, |\nullshift(x_2) - \altnullshift(x_2)| \}$. Hence, if the testing procedure $\procedure_2$ (which has a non-trivial power under the generalized logistic model) is given decisions of SB and DB reviewers sampled according to the pair of matrices $\dbmatrix, \sbmatrix_1$, then it will reject the null hypothesis with probability that goes to 1 as the minimum of $|\indset|$ and $|\noindset|$ grows. Finally, given that matrices $\dbmatrix$ and $\sbmatrix$ solely determine the distribution of observed reviewers' decisions, our construction implies that under the generalized linear model procedure $\procedure_2$ does not control for the Type-I error rate at any level $\level < 1$.

A similar argument applies to the pair of matrices $\dbmatrix, \sbmatrix_2$, and it follows that under the generalized logistic model the procedure $\procedure_1$ does not control for the Type-I error rate at any level $\level < 1$.

To conclude the proof it remains to find $x_1, x_2, \nullshift, \altnullshift$ that satisfy aforementioned conditions (i)-(iii). To this end, let us define quantities $\gamma_1, \gamma_2$:
\begin{align*}
    \gamma_1 &= \max\left\{ \gap, \left(1 + \exp\{\bound + \bound^2\} \right)^{-1} \right\} \\
    \gamma_2 &= \min\left\{ 1 - \gap, \left(1 + \exp\{-\bound - \bound^2\} \right)^{-1} \right\}.
\end{align*}
Notice that the value of $\gamma_1$ is by definition smaller than 0.5. Moreover, for each pair $(i, j) \in [\numrev] \times [\numpap]$ it gives a lower bound on the value $\dbpacc_{ij}$ that can be generated from both the generalized linear (with parameter $\gap$) and generalized logistic (with parameter $\bound$) models. Likewise, the value of $\gamma_2$ is at least 0.5 and gives the corresponding upper bound. Hence, any values of $x_1, x_2$ such that $\gamma_1 < x_1 < x_2 < \gamma_2$ satisfy the condition~(ii).

Next, find values $\shift \in (0, \gap)$ and $\logshift \in (0, \bound)$ such that for functions $\chng \in \linmodel$ and $\logchng \in \logmodel$ defined in equations~\eqref{eqn:linmember} and~\eqref{eqn:logmember} respectively the following equality holds:
\begin{align*}
    \chng(0.5) = \logchng(0.5).
\end{align*}
Observe that such values must exist because $\chng$ and $\logchng$ are continuous functions of $\shift$ and $\logshift$ respectively and
\begin{align*}
    \lim\limits_{\shift \to +0} \chng(0.5) = \lim\limits_{\logshift \to +0} \logchng(0.5) = 0.5.
\end{align*}
Consider now two possible cases:

\noindent \underline{Case 1.}  Functions $\chng$ and $\logchng$ are such that there exist two points $y \in (\gamma_1, 0.5)$ and  $z \in (0.5, \gamma_2)$ for which the following equation holds:
\begin{align}
\label{eqn:sign_change}
    \text{sign}\Big(\chng(y) - \logchng(y)\Big) \times \text{sign}\Big(\chng(z) - \logchng(z)\Big) = -1, 
\end{align}
%as illustrated in Figure~\ref{fig:functions}. 
Observe that in this case conditions (i)-(iii) are satisfied by the choice $\nullshift = \chng, \altnullshift = \logchng, x_1 = y, x_2 = z$ and hence the result of the theorem follows.

\smallskip

\noindent \underline{Case 2.} Functions $\chng$ and $\logchng$ are such that $\chng$ is a tangent line to $\logchng$ at $0.5$.  This case reduces to the Case 1 by setting $\shift' = \shift - \varepsilon$ for a sufficiently small $\varepsilon \in (0, \shift)$. Indeed, if $\varepsilon$ is sufficiently small, then due to strict concavity and differentiability of the function $\logchng$, by shifting the tangent line down we ensure that there exist points $y = 0.5$ and $z \in (0.5, \gamma_2)$ such that 
    \begin{align*}
        \text{sign}\Big(h_{\shift'}(y) - \logchng(y)\Big) \times \text{sign}\Big(h_{\shift'}(z) - \logchng(z)\Big) = -1.
    \end{align*}
Hence, we can satisfy conditions (i)-(iii) by setting $\nullshift = h_{\shift'}, \altnullshift = \logchng, x_1 = 0.5, x_2 = z$.

\medskip

To conclude the proof, we notice that Cases 1 and 2 are complementary, because function $\logchng$ is differentiable and strictly concave on the interval (0, 1) and function $\chng$ is a linear function.

%%%%%%%%%%%%%%%%%%%%%%%%%%%%%%%%%%%%%%%%%%%%%%%%%%%%%%%%%%%

\subsection{Proofs of auxiliary results}
\label{appendix:auxiliary}

In this section we give proofs for auxiliary results stated in appendix. 

%%%%%%%%%%%%%%%%%%%%%%%%%%%%%%%%%%%%%%%%%%%%%

\subsubsection{Proof of Proposition~\ref{prop:clearq}}

We prove Lemma~\ref{prop:clearq} by straightforward verification. First, let $\sbdecision_{ij}$ be generated from model~\eqref{eqn:glinmp:sb} and $\dbdecision_{i'j}$ be generated from model~\eqref{eqn:glinmp:db}. Then
\begin{align*}
    \expectation{\sbdecision_{ij} - \dbdecision_{i'j}} = \paprepresentation_j + \sbcoef_0 + \sum\limits_{\ell \in [\numcat]} \sbcoef_{\ell} \ind_j^{(\ell)} - \paprepresentation_j = \sbcoef_0 + \sum\limits_{\ell \in [\numcat]} \sbcoef_{\ell} \ind_j^{(\ell)}.
\end{align*}

Similarly, let $\sbdecision_{ij}$ be generated from model~\eqref{eqn:glogmp:sb} and $\dbdecision_{i'j}$ be generated from model~\eqref{eqn:glogmp:db}. Then
\begin{align*}
    \expectation{\sbdecision_{ij} | \sbdecision_{ij} \ne \dbdecision_{i'j}} &= \frac{\sbpacc_{ij}(1 - \dbpacc_{i'j})}{\sbpacc_{ij}(1 - \dbpacc_{i'j}) + \dbpacc_{i'j}(1 - \sbpacc_{ij})} \\ &= \left[1 + \frac{\dbpacc_{i'j}(1 - \sbpacc_{ij})}{\sbpacc_{ij}(1 - \dbpacc_{i'j})}\right]^{-1} \\ &= \left[1 + \exp\left\{- \left(\sbcoef_0 + \sum\limits_{\ell \in [k]} \sbcoef_{\ell + 1} \ind_{j}^{(\ell)} - \dbcoef_0 \right) \right\}\right]^{-1},
\end{align*}
and hence
\begin{align*}
    \log \frac{\expectation{\sbdecision_{ij} | \left(\sbdecision_{ij} \ne \dbdecision_{i'j}\right)}}{1 - \expectation{\sbdecision_{ij} | \left(\sbdecision_{ij} \ne \dbdecision_{i'j}\right)}} = \sbcoef_0 - \dbcoef_0 + \sum\limits_{\ell \in [k]} \sbcoef_{\ell + 1} \ind_{j}^{(\ell)}.
\end{align*}

%%%%%%%%%%%%%%%%%%%%%%%%%%%%%%%%%%%%

\subsubsection{Proof of Lemma~\ref{lemma:exact}}

Consider any assignment of papers to SB reviewers that satisfy $(\papload, \revload)-$constraint with $\papload > \revload$. Then pick any subset of papers $\papset \subseteq [\numpap]$ and denote a set of SB reviewers who are assigned to at least one paper from $\papset$ as $\sbrevset$. Then one can notice that
\begin{align*}
    |\sbrevset| \ge \frac{\papload |\papset|}{\revload} \ge |\papset|, 
\end{align*}
and hence by Hall's theorem there exists a matching that maps each paper to one reviewer such that each reviewer is matched to at most one paper. This matching is computed in Step~\ref{step:exact:hungary} of Algorithm~\ref{alg:exact}.

The same argument applies to DB reviewers and hence, joining these two matchings, the algorithm in Step~\ref{step:exact:final} constructs a set of tuples $\settuple$ where for each paper $j \in [\numpap]$ there exists a tuple that corresponds to this paper. 

%%%%%%%%%%%%%%%%%%%%%%%%%%%%%%%%%%%%%%%%%%%%%

\subsubsection{Proof of Lemma~\ref{lemma:greedy}}

\newcommand{\fraction}{\gamma}

Consider any assignments of papers to SB and DB reviewers that satisfy $(\papload, \revload)-$constranints. Let $\fraction$ be a maximum integer that satisfies inequality
\begin{align*}
    \fraction \le \min \left\{ \frac{|\indset|}{4 \revload}, \frac{|\noindset|}{4 \revload} \right\}.
\end{align*}

Without loss of generality, assume that $\fraction > 1$. Given that $\revload$ and $\papload$ are treated as constants and that we only need to proof the result for large enough $\min\{|\indset|, |\noindset|\}$, we ignore the cases when $\min\{|\indset|, |\noindset| \}$ is small. 

Consider a graph $G$ before the first iteration of Steps~\ref{step:greedy:find} 
- \ref{step:greedy:iterate} of Algorithm~\ref{alg:greedy}. Each paper in this graph is connected to $\papload$ SB and $\papload$ DB reviewers such that each reviewer is connected to at most $\revload$ papers. 

Now let $(i_1, j, i_2)$ and $(i_1', j', i_2')$ be triples found in the first iteration of the algorithm. These triples exists provided that $\fraction > 1$. Then in Step~\ref{step:greedy:iterate} we remove reviewers $i_1, i_1', i_2, i_2'$ and corresponding edges from graph $G$. One can see that these reviewers are connected to at most $4 \revload$ papers in total and hence before the second iteration of Steps ~\ref{step:greedy:find} 
- \ref{step:greedy:iterate} graph $G$ will have at least $|\indset| - 4 \revload \ge 4 \revload (\fraction - 1)$ papers from $\indset$ and $|\noindset| - 4 \revload \ge 4 \revload (\fraction - 1)$ papers from $\noindset$ that are connected to $\papload$ SB and $\papload$ DB remaining reviewers and each of the remaining reviewers (there must be at least $8\papload(\fraction - 1)$ SB and $8\papload(\fraction - 1)$ DB reviewers) will be connected to at most $\revload$ papers.

By induction we can show that in the first $\fraction$ iterations of Steps ~\ref{step:greedy:find} - \ref{step:greedy:iterate} the greedy algorithm will be able to find non-empty triples in Steps~\ref{step:greedy:find} and \ref{step:greedy:find2}. Hence the resulting set of tuples $\settuple$ will contain at least $\fraction$ tuples that correspond to papers from $\indset$ and at least $\fraction$ tuples that correspond to papers from $\noindset$. We then conclude the proof noticing that $\fraction = \constant \min\{|\indset|, |\noindset| \}$, where $\constant$ is a constant that depends only on $\revload$.

%%%%%%%%%%%%%%%%%%%%%%%%%%%%%%%%%%%%%%%%%%%%%

\subsubsection{Proof of Proposition~\ref{prop:tomkins}}

\newcommand{\tripleset}{\mathcal{C}}
\newcommand{\triplefam}{\mathbb{C}}
\newcommand{\tuplefam}{\mathbb{T}}
\newcommand{\triple}{c}

The proof of Proposition~\ref{prop:tomkins} follows the idea of the proof of Theorem~\ref{prop:absolute} with some changes which we now discuss. Consider any set of triples $\tripleset$ such that (i) each triple $\triple \in \tripleset$ is of the form $(j, i_1, i_2)$ (one paper and two reviewers) and (ii) each reviewer $i \in [\numrev]$ appears in at most one triple. Let $\triplefam$ denote a collection of all such sets of triples. Then any set of tuples $\settuple$ passed to the \permutation{} or \counting{} tests as input corresponds to one member of $\triplefam$ which is constructed as follows: for each $\tuple \in \settuple$ let $(j_{\tuple}, i_{\tuple}, i_{\tuple}')$ be a corresponding paper, SB reviewer and DB reviewer assigned to this paper, then $\tripleset = \bigcup\limits_{\tuple \in \settuple} (j_{\tuple}, i_{\tuple}, i_{\tuple}') $. Conversely, each member $\tripleset \in \triplefam$ gives rise to a family of sets of tuples $\tuplefam(\tripleset)$ which contains $2^{|\tripleset|}$ elements and each element corresponds to a different allocation of reviewers in each triple $(j, i_1, i_2) \in \tripleset$ to SB and DB conditions. For example, let $\tripleset = \{(j, i_1, i_2), (j', i_1', i_2') \}$, then the family $\tuplefam(\tripleset)$ consists of four sets of tuples:
\begin{align*}
    \settuple_1 &= \{(j, \sbdecision_{i_1 j}, \dbdecision_{i_2 j}, \ind_{j}), (j', \sbdecision_{i_1 j'}, \dbdecision_{i_2 j'}, \ind_{j'}) \} \\
    \settuple_2 &= \{(j, \sbdecision_{i_2 j}, \dbdecision_{i_1 j}, \ind_{j}), (j', \sbdecision_{i_1 j'}, \dbdecision_{i_2 j'}, \ind_{j'}) \} \\
    \settuple_3 &= \{(j, \sbdecision_{i_1 j}, \dbdecision_{i_2 j}, \ind_{j}), (j', \sbdecision_{i_2 j'}, \dbdecision_{i_1 j'}, \ind_{j'}) \} \\
    \settuple_4 &= \{(j, \sbdecision_{i_2 j}, \dbdecision_{i_1 j}, \ind_{j}), (j', \sbdecision_{i_2 j'}, \dbdecision_{i_1 j'}, \ind_{j'}) \}
\end{align*}

Next, for concreteness assume that $\papload \ge \revload$, that is, Algorithm~\ref{alg:exact} is used to construct a set $\settuple$. Then conditioned on the fact that the set of tuples $\settuple$ constructed by the algorithm belongs to $\tuplefam(\tripleset)$, the randomness of the allocation of reviewers to conditions, the random assignment procedure used to assign reviewers to papers in each condition and randomness in the tie-breaking in the matching algorithm ensure that $\settuple \in \uniform\left[\tuplefam(\tripleset)\right]$, that is, all elements of $\tuplefam(\tripleset)$ are equally likely to be constructed and no other set of tuples can be constructed. 

For each member $\tripleset \in \triplefam$, let $\prob{\tripleset}$ be probability that Algorithm~\ref{alg:exact} constructs a set of tuples that belongs to $\tuplefam(\tripleset)$. Notice that for some $\tripleset \in \triplefam$ we have $\prob{\tripleset} = 0$ which happens for example when $|\tripleset| < \numpap$, because Lemma~\ref{lemma:exact} ensures that $|\settuple| = \numpap$. Now, conditioning on any set $\tripleset$ with $\prob{\tripleset} > 0$ (instead of conditioning on $\testassignment$) and using Lemma~\ref{lemma:exact} (instead of Observation~\ref{observation:size}), we repeat the proof of Theorem~\ref{problem:absolute} for both \permutation{} and \counting{} tests. The unconditional result then follows from the law of total probability. The same argument applies to the case when $\papload < \revload$ and hence we conclude the proof.

%%%%%%%%%%%%%%%%%%%%%%%%%%%%%%%%%%%%%%%%%%%%%

\subsubsection{Proof of Corollary~\ref{prop:extgll}}

The high-level idea of the proof is to construct matrices $\dbmatrix$ and $\sbmatrix$ that simultaneously (for different choices of $\sbcoef_0$ and $\sbcoef_1$ coefficients) satisfy the null and the alternative hypotheses under the extended model given by equations~\eqref{eqn:extlogregdb} and~\eqref{eqn:extlogregsb}.

We begin our construction from specifying values of $\paprepresentation_j, j \in [\numpap]$. For each paper $j \in [\numpap]$, let 
\begin{align*}
    \paprepresentation_j = \begin{cases}
                            -1 & \text{if } \ \ind_j = 1 \\
                            0 & \text{if } \ \ind_j = -1. 
                           \end{cases}
\end{align*}
Then $\dbmatrix$ is generated from model~\eqref{eqn:extlogregdb} with $\dbcoef_0 = 0$ and $\dbcoef_1 = 1$. In this way, for any reviewer $i \in [\numrev]$ and for any paper $j \in [\numpap]$, probability of acceprance $\dbpacc_{ij}$ satisfies:
\begin{align*}
   M_0: \ \log \frac{\dbpacc_{ij}}{1 - \dbpacc_{ij}} = \paprepresentation_j.
\end{align*}
That is, for any reviewer $i \in [\numrev]$ and for any paper $j \in [\numpap]$ we have
\begin{align*}
     \dbpacc_{ij} = \begin{cases}
                        \frac{1}{1 + e} & \text{if } \ind_j = 1 \\
                        0.5             & \text{if } \ind_j = -1.
                    \end{cases}
\end{align*}

We now consider two different choices of coefficients for SB reviewers which result into two different models of behaviour of SB reviewers under the absence of bias:
\begin{align*}
    {M_1 \ (\sbcoef_0 = 1, \sbcoef_1 = 1):}& \ \ \log \frac{\sbpacc_{ij}}{1 - \sbpacc_{ij}} = 1 + \paprepresentation_j \\
    {M_2 \ (\sbcoef_0 = \nicefrac{3}{2}, \sbcoef_1 = 2):} & \ \ \log \frac{\sbpacc_{ij}}{1 - \sbpacc_{ij}} = \frac{3}{2} + 2\paprepresentation_j 
\end{align*}
Consider a matrix $\sbmatrix$ whose components for each $i \in [\numrev]$ and $j \in [\numpap]$ are defined as follows:
\begin{align*}
    \sbpacc_{ij} = \begin{cases}
                        0.5 & \text{if } \ind_j = 1 \\
                        \frac{1}{1 + e^{-1}} & \text{if } \ind_j = -1,
                   \end{cases}
\end{align*}
it is not hard to see that (i) entries of matrix $\sbmatrix$ satisfy the model $M_1$ and (ii) for each paper $j \in \indset$ corresponding entries of matrix $\sbmatrix$ are larger than prescribed by model $M_2$ by $\minigap > 0$ and for each paper $j \in \noindset$ corresponding entries are smaller than those prescribed by $M_2$ by $\minigap > 0$, where $\minigap$ is some universal constant. Hence, depending on which model of SB reviewer under the absence of bias ($M_1$ or $M_2$) is correct, pair of matrices $(\dbmatrix, \sbmatrix)$ corresponds to the absence or presence of bias.

Given that matrices $\dbmatrix$ and $\sbmatrix$ solely determine a distribution of reviewers' decisions,  we have shown that reviewers' decisions are identically distributed under both null and alternative hypotheses under the extended version of the generalized logistic model. Hence, we conclude the proof by declaring that any algorithm that operates on reviewers' decision and keeps Type-I error below $\level$ must have power at most $\level$ under the alternative specified by models $M_0, M_2$ and matrices $\sbmatrix, \dbmatrix$ for all values of $\min\{|\indset|, |\noindset| \}$ and hence violates the non-trivial power requirement.

\end{document}